\documentclass[letterpaper]{article} %
\usepackage{aaai25}  %
\usepackage{times}  %
\usepackage{helvet}  %
\usepackage{courier}  %
\usepackage[hyphens]{url}  %
\usepackage{graphicx} %
\usepackage{natbib}  %
\usepackage{caption} %
\usepackage{algorithm}
\usepackage{algorithmic}
\usepackage{amsmath}
\usepackage{newfloat}
\usepackage{listings}
\DeclareCaptionStyle{ruled}{labelfont=normalfont,labelsep=colon,strut=off} %
\floatstyle{ruled}
\newfloat{listing}{tb}{lst}{}
\floatname{listing}{Listing}
\title{The Impact of Pseudo-Science in Financial Loans Risk Prediction}
\author{
    Bruno Scarone\textsuperscript{\rm 1},
    Ricardo Baeza-Yates\textsuperscript{\rm 2}
}
\affiliations{
    \textsuperscript{\rm 1}Khoury College of Computer Sciences, Northeastern University, Boston, USA\\
    \textsuperscript{\rm 2}Barcelona Supercomputing Center, Barcelona, Spain\\
    scarone.b@northeastern.edu, rbaeza@acm.org
}

\usepackage{caption}
\usepackage{subcaption}

\def\ie{\emph{i.e., }}
\def\eg{\emph{e.g., }}

\begin{document}

\maketitle

\begin{abstract}
We study the societal impact of pseudo-scientific assumptions for predicting the behavior of people in a straightforward application of machine learning to risk prediction in financial lending. This use case also exemplifies the impact of survival bias in loan return prediction. 
We analyze the models in terms of their accuracy and social cost, showing that the socially optimal model may not imply a significant accuracy loss for this downstream task. Our results are verified for commonly used learning methods and datasets.
Our findings also show that there is a natural dynamic when training models that suffer survival bias where accuracy slightly deteriorates, and whose recall and precision improves with time. These results act as an illusion, leading the observer to believe that the system is getting better, when in fact the model is suffering from increasingly more unfairness and survival bias. 
\end{abstract}

\maketitle

\section{Introduction}
\label{sec:intro}

Nowadays many important decisions that affect the lives of people are being taken with the support of predictive machine learning (ML) models with little human intervention. Some typical use cases include hiring, loans, scholarships, government subsidies, or even justice, among others. In this context, these tools may discriminate or favor people, having an important but unknown social impact. For this reason, in the new EU regulation for the use of AI \cite{AIAct}, most of these tools are considered high-risk. However, we do not really know well how to measure social risk and its overall impact, as well as other related issues. %

In addition, these predictive tools are based on several assumptions that weaken their scientific validity. First, there is the belief that the available data represents reasonably well the problem being solved (dataism \cite{dataism2013}), which in many cases is arguable and also depends on every instance (that is, a given person). This creates a so-called {\em reality gap} \cite{hildebrandt2021}. Second, for a new person, in some of the use cases there is the assumption that her or his behavior can be predicted from data of ``similar'' people. This may work on average, but certainly does not work for all individuals, moreover considering that the similarity metric is another proxy for reality. Third, normally we evaluate the success of the system ({\eg accuracy), which implies that all errors have the same impact, which in most cases is false. Finally, this type of predictive tools have many technical errors in their design and implementation as shown in \cite{predopt2024,SnakeOil}. Hence, in many ways these applications are closer to pseudoscience and alchemy rather than science and engineering, triggering many ethical issues \cite{pseudo2024}. Hence, the main goal of this work is to expose the impact of these questionable scientific and technical assumptions.

Given the problems above, consider the following hypothetical scenario. Imagine that a bank has built a ML model based on historical data to predict if a new client will return a loan or not. That is, we are assuming that from other people can be used to predict the behavior of a new client. This hypothesis is pseudo-science as it may work many people, but certainly does not work for all people. If the prediction is positive, the person obtains a loan, otherwise it is rejected. After a certain period of time, as the bank has more data, we can retrain the model periodically, with the expectation of improving it. However, the new data suffers from survival or survivorship bias on the model errors. That is, we know what happens with all the false positives (favored people that represent the survival bias) and with none of the false negatives (harmed people). This scenario in addition to neglect the bias issue above, combines the three problems mentioned above: the model errors are a good proxy for the size of the reality gap and the prediction failures allow us to estimate the social impact.

In this work, using real data, we simulate this bank lending scenario and measure the information lost in time due to survival bias (false negatives), its impact on the model performance, and an estimation of the social impact of the errors. For this we compare across time the model with an oracle that always knows all the available data and hence makes less mistakes. We also propose another model that minimizes the estimated social impact instead of maximizing the accuracy, mitigating the harm of the behavioral prediction hypothesis. 

Our main contributions are:
\begin{itemize}
    \item Using the socially optimal models does not always imply a significant accuracy loss and there are models for which the loss is particularly small ($<1.3\%$), which further encourages the use of this fairness optimization criterion in a model.
    \item The identified dynamic to retrain ML models, which intuitively seems to improve the models from the perspective of an external observer, actually deteriorates them as the process is introducing more survival bias in the data.
    \item A simple methodology to simulate a dynamic version of a predictive problem when enough data is available.
\end{itemize}
Our results are valid for two widely used non-linear ML algorithms and three datasets with different imbalance ratios.

The rest of the paper is organized as follows. We first cover related work and then explain our methodology. Then we present the results, ending with the conclusions.

\section{Related Work}
\label{sec:related}

\subsection{Legitimacy of Machine Learning}

The use of ML in high-stakes scenarios has been repeatedly questioned, both from a methodological (scientific legitimacy and functionality) and an ethical perspective. Andrews {\it et al.} \cite{pseudo2024} reviews a large number of works, undertaken by academic research groups, private firms and government agencies, where the use of ML has received harsh criticism on both ethical and epistemic grounds (referred to as ``AI pseudoscience''). The authors analyze which characteristics of the applied ML field make it a fertile ground for this type of pseudoscience and how it can lend itself to social harms. Among the reasons they list: the meta-narratives and cycles of hype surrounding ML, the unprincipled and purely quantitative nature of the evaluation used in the field, as well as the ease of use of modern ML tools. They also question the idea of applied ML being paradigmatically objective, remarking that in the case of supervised learning tasks, what is being predicted is the judgment of the labeler of the data (a good example is emotion recognition \cite{le2021professional}). This is related to the problem of ML practitioners paying less attention to how the examples from which the model should learn are created, which is also highlighted in the paper. Stark deems many uses of applied ML as fundamentally conjectural, meaning that they are used to draw conclusions based on incomplete information \cite{stark2023artificial}. In particular, the author states that ``individual human actions cannot be reliably aggregated into general and repeatable empirical rules'' and thus such techniques should not be used to predict future human activity. Barbierato and Gatti \cite{barbierato2024challenges} also question the validity of the field by stating that although ML has a mathematical/statistical foundation, it cannot be strictly regarded as science from a methodological perspective. The main reason for this is the lack of causal explanations for the predictions of the models. They also state that the process of learning is more complex and that statistical methods on their own cannot achieve this goal. Raji and Kumar {\it et al.} list a large number of high stakes scenarios where AI systems do not functionally work or do not provide benefits, yet such products are still being used \cite{raji2022fallacy}. They not only include corporations and the technology press under this claim, but also scholars and policymakers. They summarize the harms of different types of functionality failures in a taxonomy of known AI functionality issues. The authors highlight the need for AI system developers to more honestly understand and articulate the limits of their products prior to their release into the market or public use, as well as understanding data validation as a critical step in the construction of an ML system.

\subsection{Loan Default Prediction}

Credit scoring and the estimation of default risk are central activities in financial lending \cite{he2018novel}. An increasingly popular method for this consists of ML based techniques \cite{suhadolnik2023machine}. Some works on financial lending \cite{jagtiani2017fintech, jagtiani2018roles} have highlighted that when lending institutions (\eg banks) consider nontraditional alternative information sources to collect soft information about creditworthiness, customers end up achieving ``better'' loan grades. The nature and quality of the data used in this application domain have also been questioned \cite{ippoliti2017dark}. The datasets available are oftentimes imbalanced, and there seems to be a consensus that models are better on the balanced versions \cite{markov2022credit,alam2020investigation}.

In terms of the models being used, tree-based models tend to be a popular option because of their interpretability \cite{markov2022credit,alam2020investigation,suhadolnik2023machine}. However, when systematically reviewing credit scoring methods, Markov {\it et al.} \cite{markov2022credit} remarks that ``the literature does not provide a single answer to what method is the most efficient means of credit scoring: various authors appear to hold different, if not opposite, opinions on the issue''. We consider a wide range of models, which we train on three widely used datasets. We select the best performing ones in terms of accuracy, precision and recall, both being tree-based models. We balance two of the datasets using under-sampling.

\subsection{Survival Bias in Loan Default Prediction}

An extensive list of works has highlighted the diverse types of bias that can affect the ML pipeline \cite{mehrabi2021survey,hellstrom2020bias}. We consider survival bias, a form of selection bias in which one focuses on a subset of a population that passes (``survives'') a given selection criterion. The existence of survival bias is known in the context of mutual funds \cite{wermers1997momentum,elton2015,rohleder2011survivorship}, where many datasets only include those that are currently in operation, ignoring the ones that were unsuccessful. Piovoso {\it et al.}~\cite{weinblat2018forecasting} comment that the self-reporting nature of some dataset makes it particularly vulnerable to this type of bias, since firms with poor performance are less likely to report their data voluntarily. In all these cases, the survival bias in the data is not originated by the use of ML models, but rather by data reported by the companies. Survival bias mitigation techniques, also called reject inference techniques, has been addressed in several papers \cite{banasik2007reject, liu2018delayed, mancisidor2020deep, ehrhardt2021reject, anderson2023monte}. They will be considered in future work.

\newpage

\subsection{Our Work}

We study a natural dynamic to retrain ML models, which intuitively seems to improve the models but which in reality deteriorates them, given that it introduces a form of survival bias in the data. We believe that this type of phenomena is often overlooked by ML practitioners due to the unprincipled nature of the field, as well as the nature of the evaluations of the models, both already highlighted in the literature. We also operationalize the social cost of a model to illustrate the trade-off between this notion and the accuracy of the models. We propose this for two reasons: first, to highlight the need to take this notion into account when deploying ML tools, but also as a concrete (but insufficient) initial approach to account for it in their design process. Of course, this should be considered in conjunction with the other social and ethical considerations stated in the literature. 

\section{Methodology and Data}
\label{sec:method}

In this section, we introduce the methodology and the datasets used.

\subsection{Predictive Task}

We consider the task of loan default prediction using a binary classifier, where the goal is to predict whether a loan applicant will default on their loan or repay it successfully. This classification problem is essential in the financial sector, as it helps lending institutions manage risk and optimize their decision-making processes \cite{aslam2019empirical,madaan2021loan}. The model's performance can significantly impact both the profitability of the lender institution and the accessibility of loans for applicants, making accuracy, fairness, and robustness crucial considerations in its design and implementation \cite{bhatore2020machine}.

In this setting, we consider a bank that uses ML to grant loans to their clients. We evaluate the models based on classical ML performance metrics and on their social cost,\ie how much do the predictions of the model affect society.  

On the one hand, we assess the trade-off between the most accurate and socially optimal instances of the models. On the other hand, we consider a dynamic setting in which the set of clients of the bank evolves over time. The bank has $C(t)$ clients at time $t$, 
and trains an ML model based on the historical data of the clients, which we denote by $M(t)$. Our objective in this context is to evaluate how does the evolution of the data affect the classical ML performance metrics of the models. 

\subsection{Data}

For the study, we consider three datasets that have already been used in the literature \cite{he2018novel,markov2022credit}. The Default dataset~\cite{default_of_credit_card_clients_350}, which contains information about customers' default payments in Taiwan, the PPDai dataset obtained from customer data from a Chinese P2P lending platform named PaiPaiDai \cite{ppdai} and a dataset derived from LendingClub ~\cite{lendingclube2007_2020Q3}, an American peer-to-peer lending company. The binary label of all datasets indicates whether a person defaulted on their loan payment. The datasets are described in Table \ref{tab:datasets}.

\begin{table*}[t]
    \centering
    \begin{tabular}{|c|c|c|c|c|}
        \hline
        Name & Nr. of rows & Nr. of features & Imbalance Ratio\\
        \hline
        Default & 30,000 & 23 & 77.88\%\\
        \hline
        PPDai & 55,596 & 29 & 87.08\%\\
        \hline
        PPDai down-sampling (ppdai\_bal\_imbr77) & 31,389 & 29 & 77.01\%\\
        \hline
        lc2017 & 443,579 & 142 & 84.89\%\\
        \hline
        lc2017 down-sampling (lc17\_bal\_imbr50) & 121,974 & 36 & 50\%\\
        \hline
    \end{tabular}
    \caption{Descriptions of the datasets used in this paper.}
    \label{tab:datasets}
\end{table*}

Regarding preprocessing, for the LendingClub (LC) dataset, we started by selecting the data from the year 2017. Then we discarded the top-10 columns with the highest number of Nulls, all having more that 314,212 of Null values. %
We then proceeded to drop the tuples having Null values. Lastly, we performed feature selection: we computed the correlation matrix of the dataset, selected the top-50 attributes that have the highest correlation with the label. From this subset, we then considered pairs of features that have a correlation higher than $0.85$ and discarded the one from the pair that has the least correlation with the label attribute. Both the PPDai and LendingClub datasets were down-sampled to achieve the imbalance ratios shown in Table \ref{tab:datasets}.

\subsection{Selected Models}

We initially considered eight binary classification models. These models were trained on the three datasets (Table \ref{tab:datasets}) by randomly splitting the data with 80\% for training and 20\% for evaluation. Accuracy, precision and recall were computed for each one. All measures were averaged over 10 iterations and we discarded linear models due to the complexity of the problem. The results for the Default, PPDai and LC datasets are shown in Tables \ref{tab:performance_default},  \ref{tab:performance_ppdaibal_lcbal}. Among the four best performing models for the three datasets, we selected two widely used tree based models: Random Forest (RF) and Gradient Boosting Decision Trees (GBDT). 

We use the open source ML library Scikit-learn~\cite{sklearn_gral} to implement our learning algorithms. The default decision threshold used in Scikit-learn is $\tau=0.5$, %
a value that does not always maximize accuracy. For example, for the Default dataset and the Random Forest model, $\tau=0.5$ gives an accuracy of $0.822$, while $\tau=0.461$ results in $0.824$.

\subsection{Decision Threshold}

A binary classifier typically outputs a probability score (which can we interpreted as a measure of confidence) that represents how likely a given input belongs to the positive class \cite{murphy2012machine}. The decision threshold $\tau$ is the cutoff value applied to this probability to decide the predicted class. If the predicted probability is greater than or equal to the threshold, the instance is assigned to the positive class or otherwise, to the negative class. The choice of the decision threshold affects the trade-off between recall (also known as True Positive Rate,\ie fraction of actual positive instances that the model successfully identifies) and specificity (also known as True Negative Rate,\ie how good a model is at identifying negative instances). Raising the threshold increases specificity (fewer false positives), but may result in more false negatives. Lowering the threshold increases sensitivity (more true positives), but may lead to more false positives. We denote the probability of a tuple with features $x$ to be assigned label $y\in\{0,1\}$ by $\Pr(y|x)$, and use $\hat{y}$ to denote the given label. Using this notation, if $\Pr(y=1|x)\geq\tau$ then we assign $\hat{y}=1$ and $\hat{y}=0$ otherwise.

\begin{table}[t]
    \centering
    \begin{tabular}{|c|c|c|c|}
        \hline
        Model & Accuracy & Precision & Recall \\
        \hline
        Ada Boost & 0.8240 & 0.6916 & 0.3410 \\
        \hline
        GBDT & 0.8168 & 0.6739 & 0.3673 \\
        \hline
        Extra Trees & 0.8103 & 0.6545 & 0.3663 \\
        \hline
        Random Forest & 0.8083 & 0.6491 & 0.3508 \\
        \hline
        MLP & 0.7858 & 0.6197 & 0.0338 \\
        \hline
        kNN & 0.7535 & 0.3877 & 0.1890 \\
        \hline 
    \end{tabular}
    \caption{Performance measures for the best six models on the Default Dataset. All measures are averaged over ten runs.}
    \label{tab:performance_default}
\end{table}

\begin{table*}[t]
    \centering
    \begin{tabular}{|c|c|c|c|p{.5cm}|c|c|c|c|}
        \cline{1-4} \cline{6-9} 
        Model & Accuracy & Precision & Recall & & Model & Accuracy & Precision & Recall\\
        \cline{1-4} \cline{6-9}
        Random Forest & 0.7751 & 0.5581 & 0.1195 & & GBDT & 0.9702 & 0.9759 & 0.9645 \\
        \cline{1-4} \cline{6-9}
        GBDT & 0.7751 & 0.5217 & 0.0844 & & Random Forest & 0.9724 & 0.9787 & 0.9658\\
        \cline{1-4} \cline{6-9}
        Ada Boost & 0.7748 & 0.6043 & 0.0951 & & Extra Trees & 0.9662 & 0.9661 & 0.9656 \\
        \cline{1-4} \cline{6-9}
        Extra Trees & 0.7709 & 0.5333 & 0.0935 & & Ada Boost & 0.9572 & 0.9590 & 0.9555\\
        \cline{1-4} \cline{6-9}
        kNN & 0.7300 & 0.3113 & 0.159 & & MLP & 0.9424 & 0.9725 & 0.9106 \\
        \cline{1-4} \cline{6-9}
        MLP & 0.7209 & 0.3364 & 0.1703 & & kNN & 0.7939 & 0.7668 & 0.8357\\
        \cline{1-4} \cline{6-9}
    \end{tabular}
    \caption{Performance measures for the best six models on the balanced PPDai (left) and LendingClub (right) datasets, averaged over ten runs.}
    \label{tab:performance_ppdaibal_lcbal}
\end{table*}

\subsection{Modeling the Bank Customers}
\label{sec:client_data}

We use the recurrence defined in Equation \ref{eq:clients_filtered} to determine the set of bank customers for the iterations of the process. 
\begin{equation}
\label{eq:clients_filtered}
    \begin{cases}
        C(0)  = & Unif(U,n_0)\\
        C(i>0)= & C(i-1)\cup \{p\in Samp : M(i-1,p)=1 \} \\ 
            & \text{ with } Samp \leftarrow Unif(U\setminus C(i-1),n')
    \end{cases}
\end{equation}

Initially (base case, iteration zero), we take a uniform sample of size $n_0$ from our complete dataset $U$. Since this constitutes the initial data collection step, the bank does not have data to train an ML model yet. This case models either the scenario where the bank starts by accessing an initial dataset (open data or in house data) or where it starts giving loans using non-ML techniques to a representative set of clients to generate an initial dataset.

For the recursive case, we first take a uniform random sample of size $n'$ from the remaining data. Then the current model (the one trained with $C(i-1)$) is used to filter the sample, only accepting clients who are believed to be able to repay the loan using the latest model. %

\subsection{Determining the Initial Sample Size}
\label{sec:nzero}

We propose a data-driven way of determining the initial batch size of the process, \ie the value of $n_0$ used in Equation \ref{eq:clients_filtered}.
Suppose we have a dataset of size $n$ and we consider a step size $k<n$. 
Then we train instances of a given model with uniform random samples of size $\{k,2k,\dots ,\big\lfloor\frac{n}{k}\big\rfloor k,n\}$ and determine the size up from which the accuracy stabilizes
, which we measure by looking at the slope of the trend line of the curve. The full list of $n_0$ values are shown in Table \ref{tab:n0s}. The plots for the Default dataset both for the RF and GBDT models are shown in Figure \ref{fig:n0_default_rf_gbdt}.

\begin{table}[]
    \centering
    \begin{tabular}{|c|c|c|}
        \hline
        Dataset & Model & $n_0$ \\
        \hline
        Default & Random Forest & 10,000\\
        \hline
        Default & GBDT & 6,000\\
        \hline
        ppdai\_bal\_imbr77 & Random Forest & 10,000\\
        \hline
        ppdai\_bal\_imbr77 & GBDT & 15,000\\
        \hline 
        lc17\_bal\_imbr50 & Random Forest & 60,000\\
        \hline
        lc17\_bal\_imbr50 & GBDT & 60,000\\
        \hline
    \end{tabular}
    \caption{Initial batch size ($n_0$) for every dataset and model.}
    \label{tab:n0s}
\end{table}

\begin{figure*}[t]
    \begin{subfigure}{.5\textwidth}
      \centering
      \includegraphics[scale=.35]{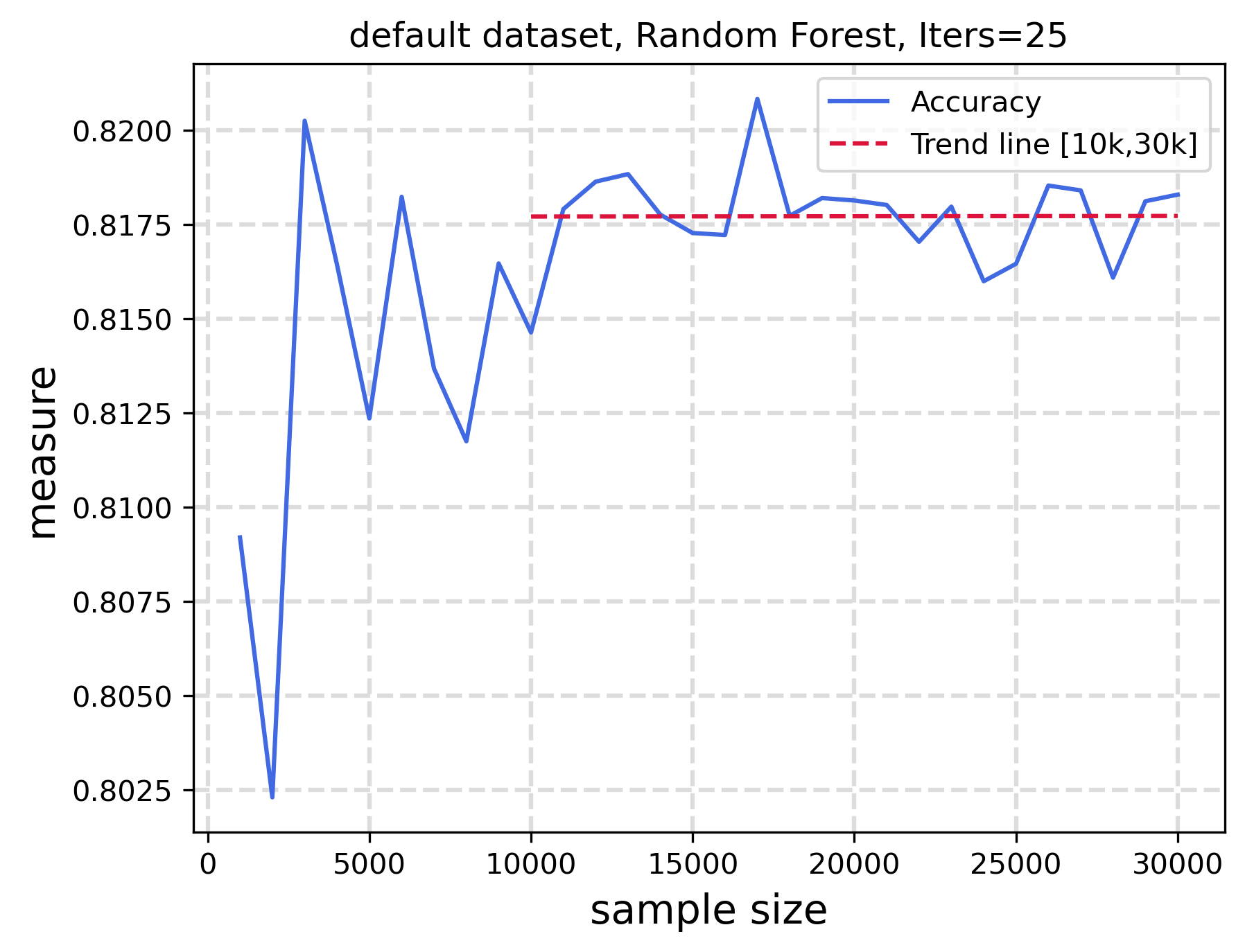}
      \caption{Random Forest}
    \end{subfigure}%
    \begin{subfigure}{.5\textwidth}
      \centering
      \includegraphics[scale=.35]{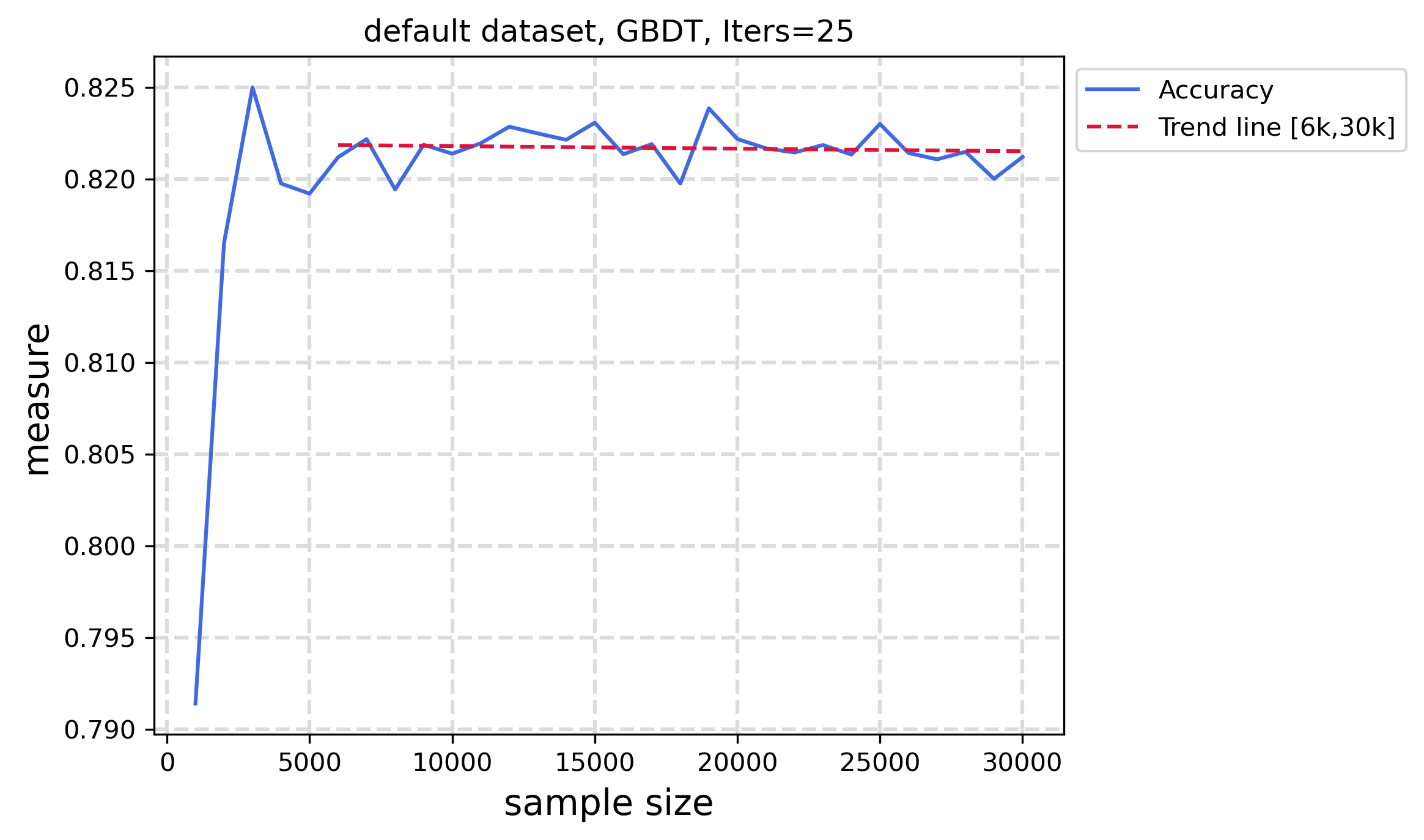}
      \caption{GBDT}
    \end{subfigure}
    \caption{Determining the initial batch size ($n_0$) based on the data - Default dataset.}
    \label{fig:n0_default_rf_gbdt}
\end{figure*}

\subsection{Model Variations}

For each model instance that we train, we are interested in measuring the following quantities:

\begin{itemize}
    \item The number of \textbf{benefited people}, \ie the ones that were given a loan and were not able to pay it back, given by the number of False Positives ($\hat{y}=1$ and $y=0$).
    \item The number of \textbf{negatively affected or harmed people}, \ie the ones that were denied a loan and would have been able to pay it back, given by the number of False Negatives ($\hat{y}=0$ and $y=1$)
    \item The accuracy, precision and recall of the model.
    \item The \textbf{social cost} of the predictions. We assume that both types of affected people negatively impact society. Thus, the social cost of a model trained with decision threshold $\tau$ is $sc(\tau)=c_{FN}\cdot FN(\tau)+c_{FP}FP(\tau)$, where $c_{FN}$ and $c_{FP}$ are the cost of a FN and a FP, respectively. For simplicity, in what follows we take $c_{FP}=1$ and use $c_{FN}$ and $c$ interchangeably. 

\end{itemize}

In this work, we consider two ways of tuning a given model, which are orthogonal to each other:
\begin{itemize}
    \item Changing the decision threshold. When varying the decision threshold of the models, we consider the following two  types of instances :%
        \begin{enumerate}
            \item Using the decision threshold that maximizes accuracy, denoted $\tau_{acc}$.
            \item Using the decision threshold that minimizes social cost, for different values of $c$ denoted by $\tau_{sc,c}$. In the experiments we consider $c\in\{1,\dots ,5\}$, but for lack of space we only show the results for $c \in \{1, 3, 5\}$. The resulting decision thresholds are given in the Additional Material. %
        \end{enumerate}
    \item Oracle vs Biased models: In order to asses how does the simulation affect the measures of interest, we compare the models resulting from the experiment (which we call Biased models) against models that have access to the full sample at each step, \ie where no filtering is applied (Oracle models). We use the name ``Biased models'', since the training dynamic of the bank introduces a form of survival bias in the training data, as discussed in the Introduction. %
\end{itemize}

\section{Results and Analysis}
\label{sec:results}

In this section, we present our experimental results.\footnote{The source code is available at \url{https://github.com/bscarone/illegitimate-ml-pub}.} In Section \ref{sec:tempevol}, we start by analyzing how the proposed training dynamic affects the performance of the model. Later in Section \ref{sec:acc_vs_sc}, we study the accuracy loss with respect to the most accurate model when considering  socially optimal models.

\subsection{Temporal Evolution}\label{sec:tempevol}

We show the temporal evolution of the Biased and Oracle models for the three classical performance measures of interest: Accuracy (Figure \ref{fig:rf_tempevol_accuracy}), Precision (Figure \ref{fig:rf_tempevol_precision}) and Recall (Figure \ref{fig:rf_tempevol_recall}), as well as for the percentage of FNs (Figure \ref{fig:rf_tempevol_perFN}) and FPs (Figure \ref{fig:rf_tempevol_perFPs}), \ie the number of harmed and benefited people, respectively. In each of the figures, we consider the following instances of the models: the one that maximizes accuracy $\tau_{acc}$ and the ones that minimize the social cost for $c_{FN}\in\{1,\dots,5\}$. The results for the GBDT model are given in the additional material.
We start by qualitatively analyzing the results for each measure. We then explain the variations in terms of how the different types of predictions (TN, TP, FN and FP) evolve over time.

In general, all measures for the Oracle remain constant or exhibit small variations. This is because, by definition, we select the values of $n_0$ such that the trained models have a stable performance. Regarding the accuracy for the Biased models, it deteriorates for both RF and GBDT in all datasets and for every threshold, except for the case of $\tau_{acc}$ for the lc17\_bal\_imbr50 dataset. We conjecture that for this last case,  the initially trained model has such a good performance that the effect of the subsequent biased data is not enough to deteriorate its performance in a significant way. In terms of precision, for the RF model, both for Default, PPDai and the LC data, the Biased models slightly improve for every threshold. For GBDT, both the Biased and the Oracle models tend to have the same behavior. In terms of recall, the measure for the Biased model increases significantly in all the cases compared to the Oracle. The percentage of FNs of the Oracles remains stable for both models, while it decreases for the bast majority of the cases for the Biased models (for all datasets and every threshold). The percentage of FPs of the Oracles remains stable for both models and every threshold for the three datasets and increases significantly for all the Biased models, regardless of the threshold.

Now, we analyze how the way in which we train the Biased models affects the different types of predictions over time and relate this with the previous analysis. Regarding the data used to train the Biased models, $C(t)$, we only have tuples with $\hat{y}=0$ in $C(0)$. In all subsequent iterations these are filtered out, which means that we are not adding neither TNs ($\hat{y}=0$ and $y=0$) nor FNs ($\hat{y}=0$ and $y=1$). This explains why both, the number of FNs and TNs, grow with the number of iterations in the Oracle models, but remain constant in the Biased models (Figure \ref{fig:oraclevsfilt_tpfptnfn}). Since the other tuples are not affected by the data filtering, each iteration increases the number of FPs and TPs roughly by the same amount, which is why these quantities grow at the same rate, both for the Oracle and Biased models in the experiments. %

\begin{figure*}[t]
    \centering
    \begin{tabular}{ccc}
        Default & ppdai\_bal\_imbr77 & lc17\_bal\_imbr50\\
        \includegraphics[scale=.25]{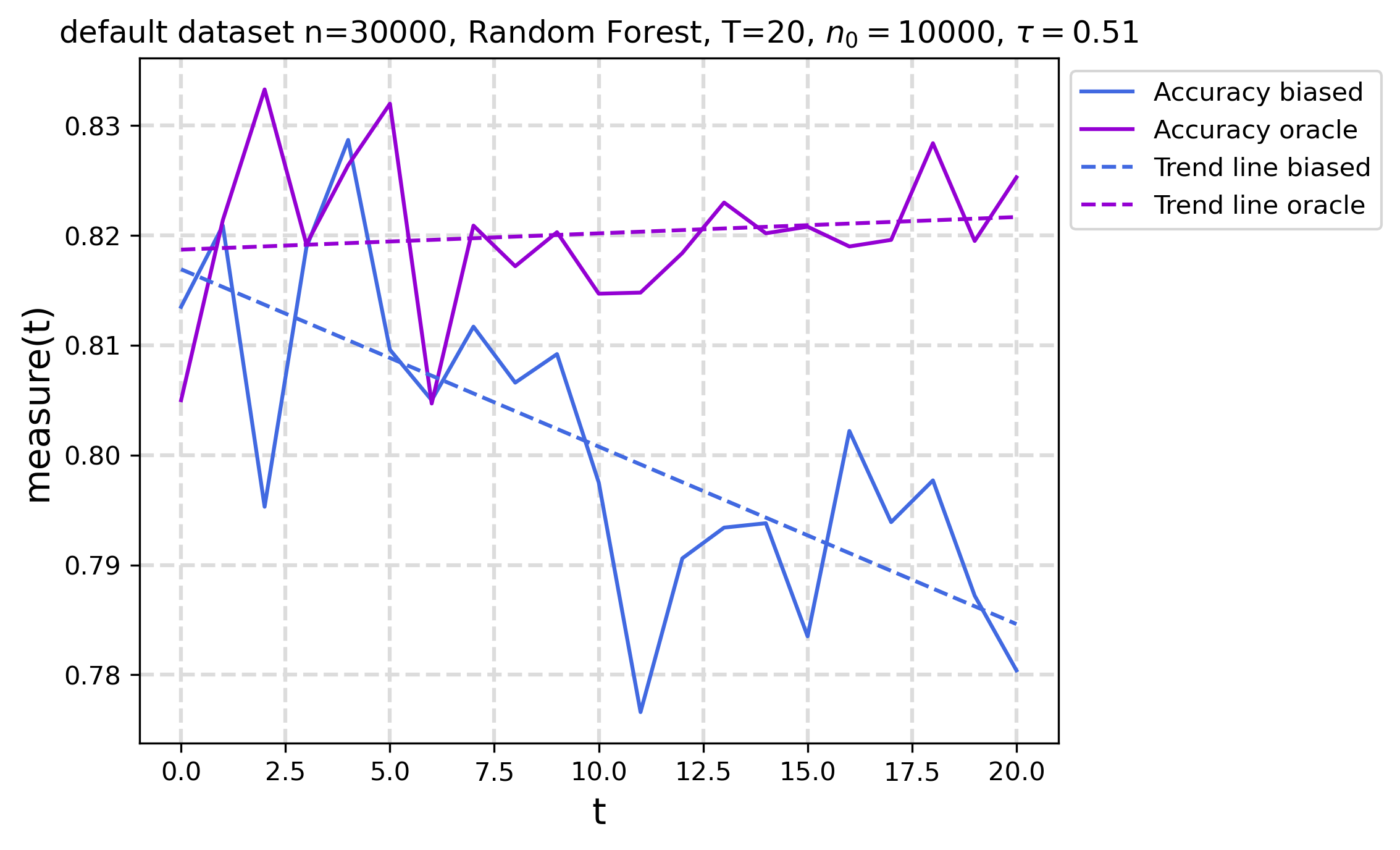} & \includegraphics[scale=.25]{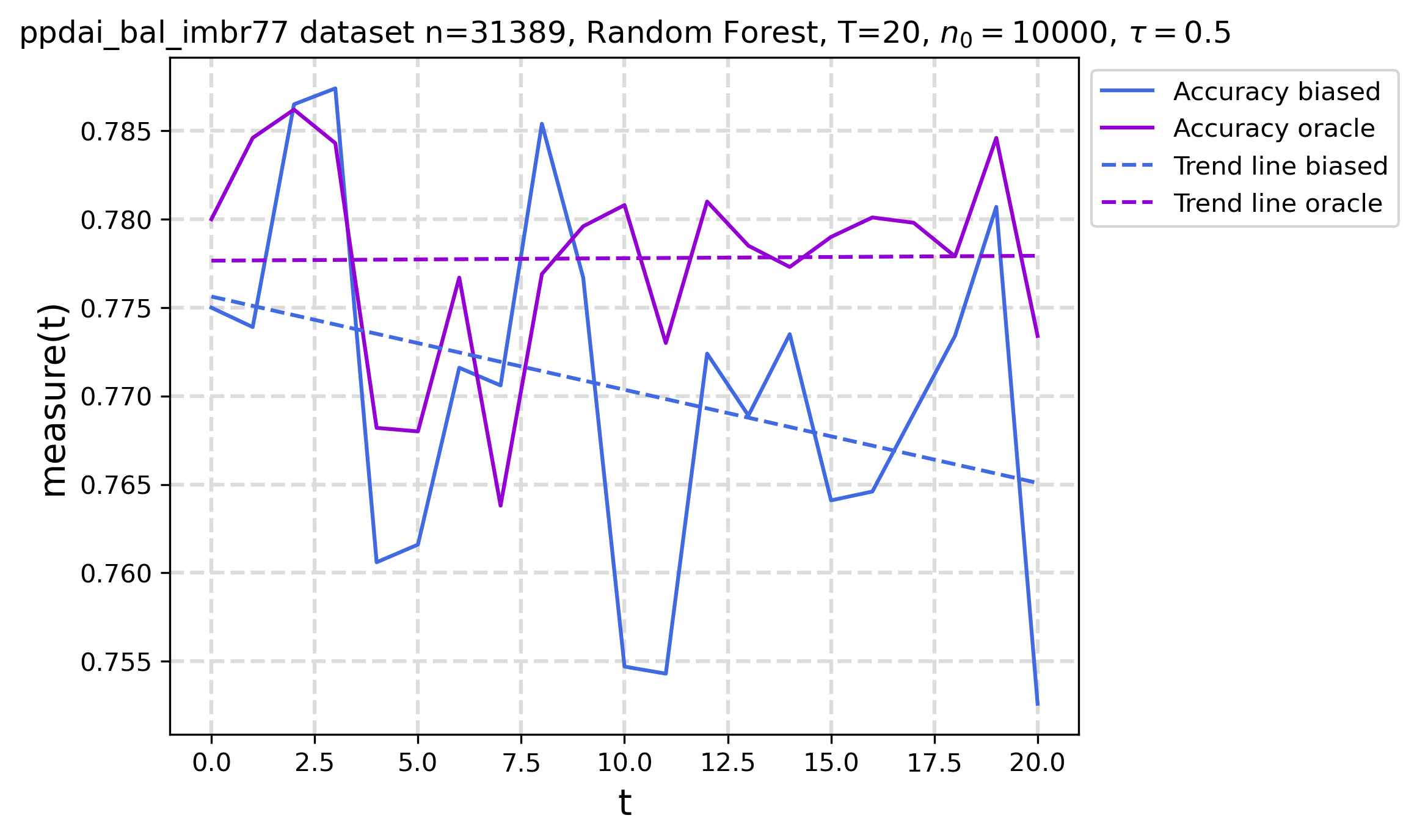} & \includegraphics[scale=.25]{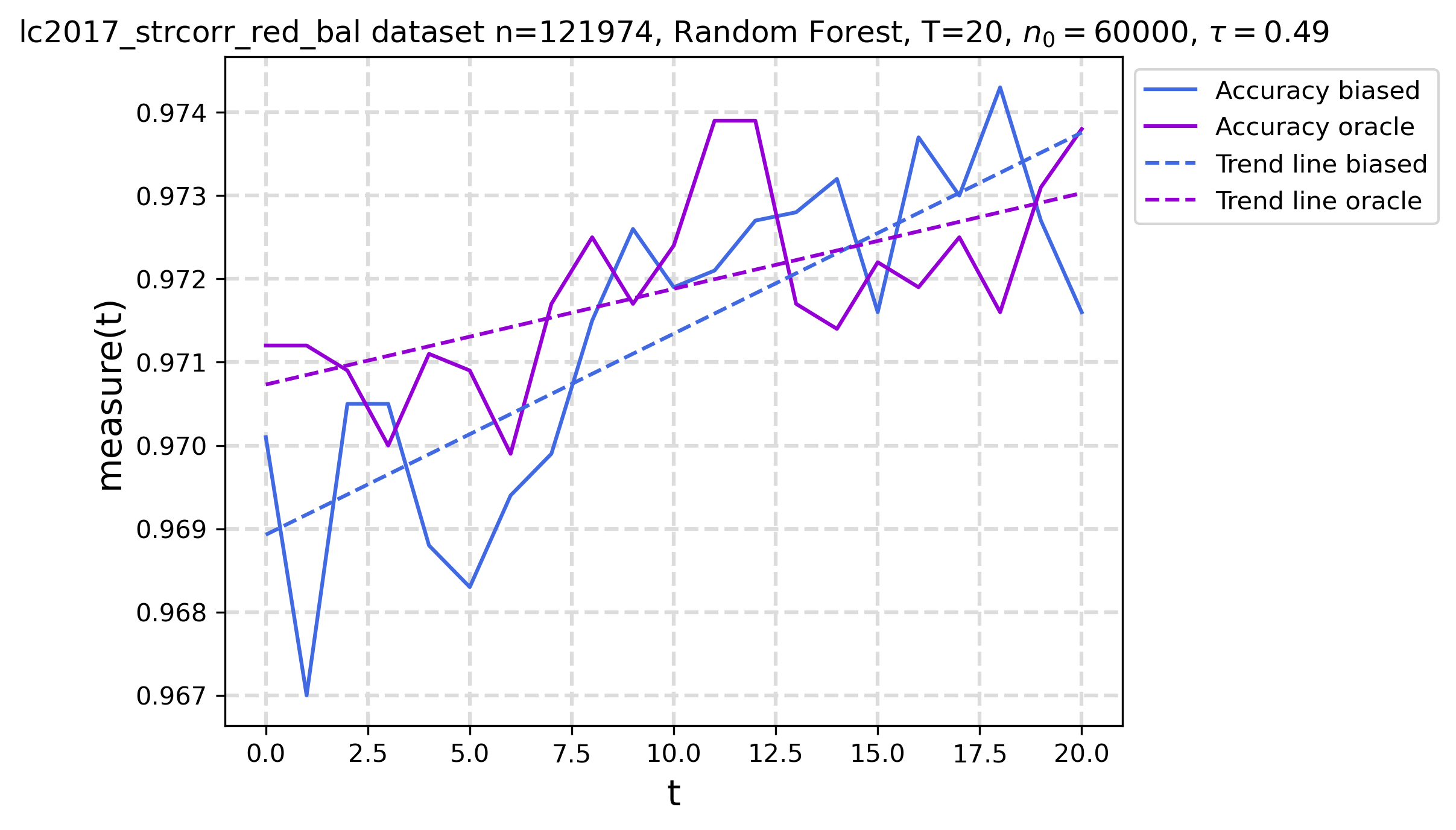}\\
        \includegraphics[scale=.25]{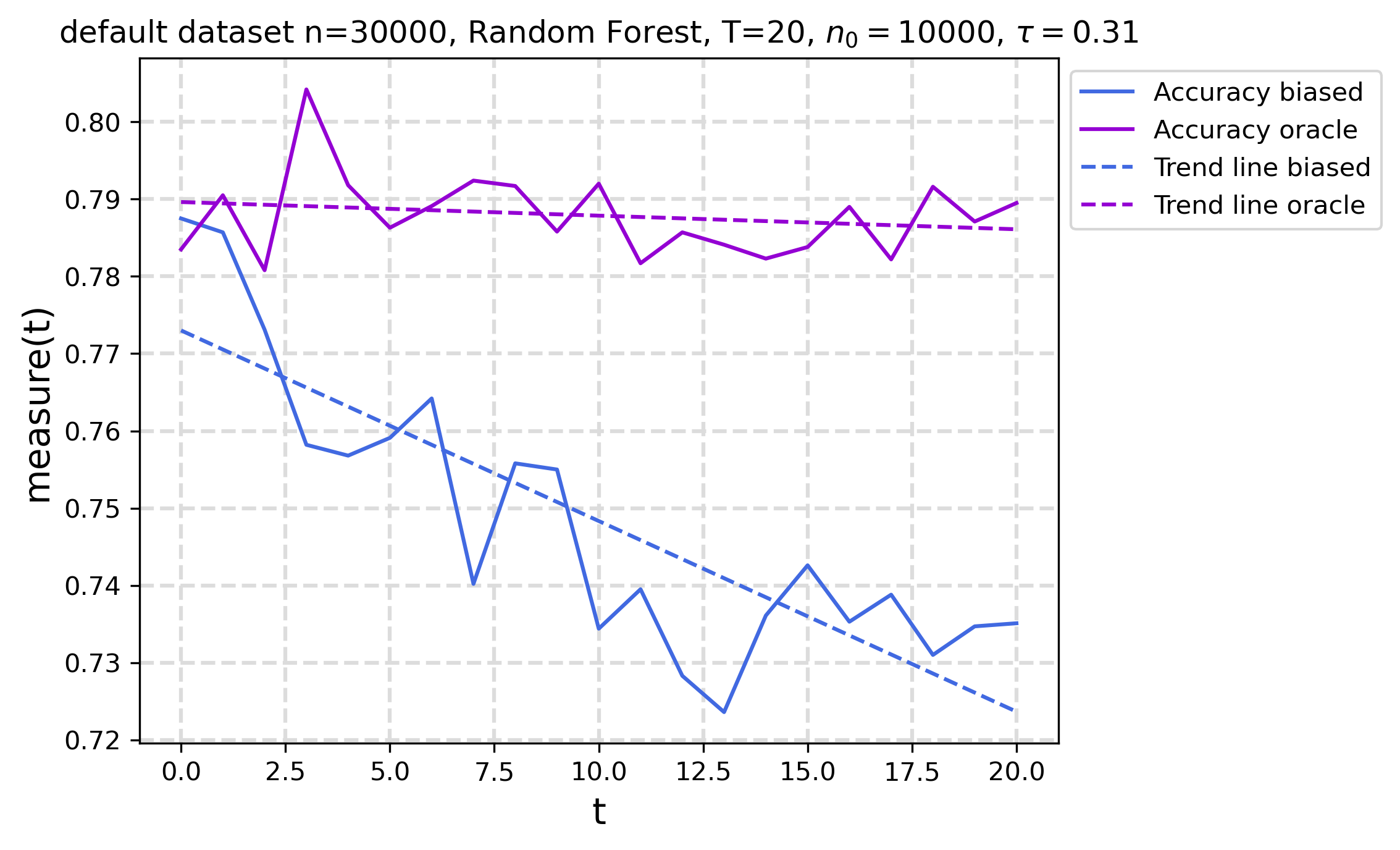} & \includegraphics[scale=.25]{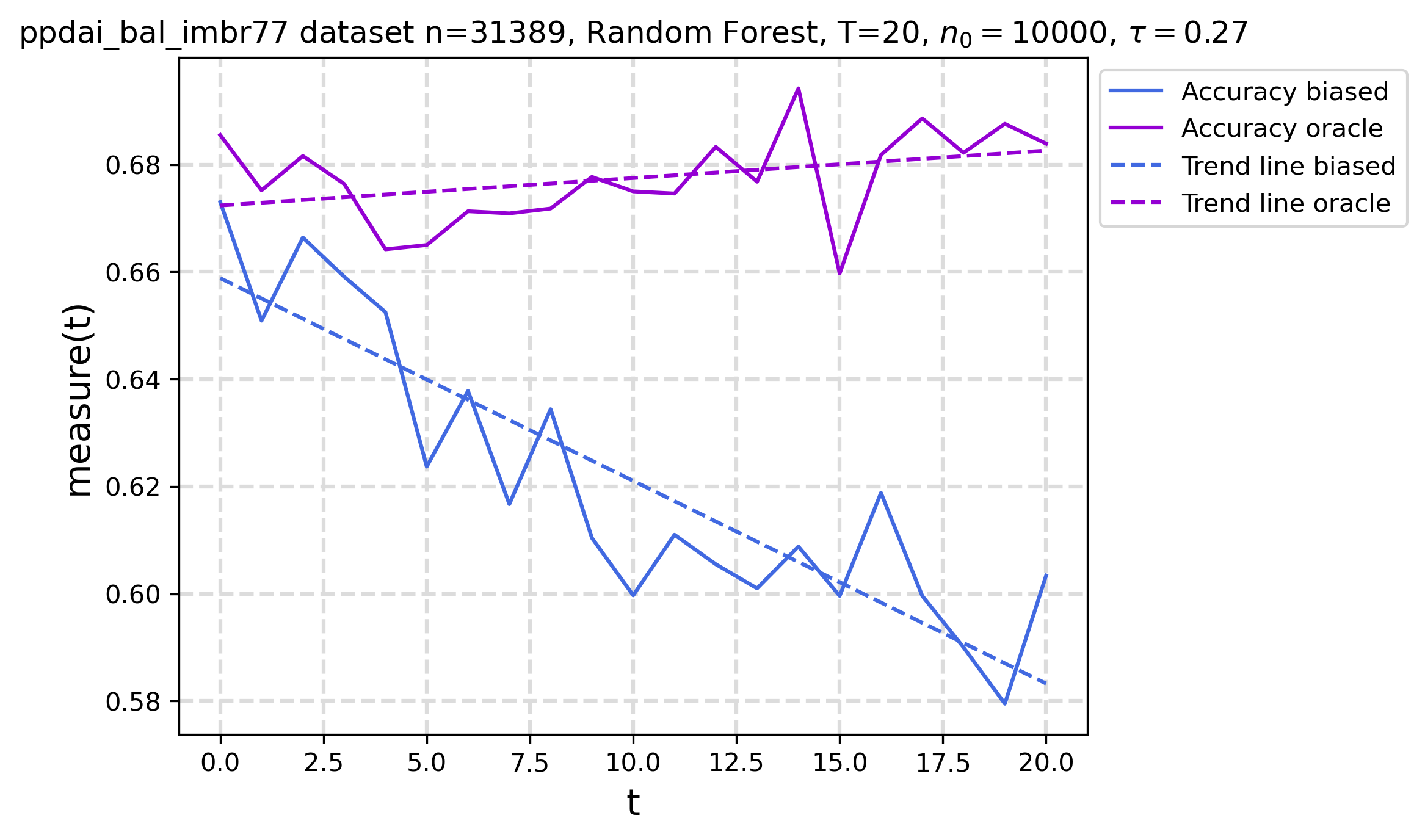} & \includegraphics[scale=.25]{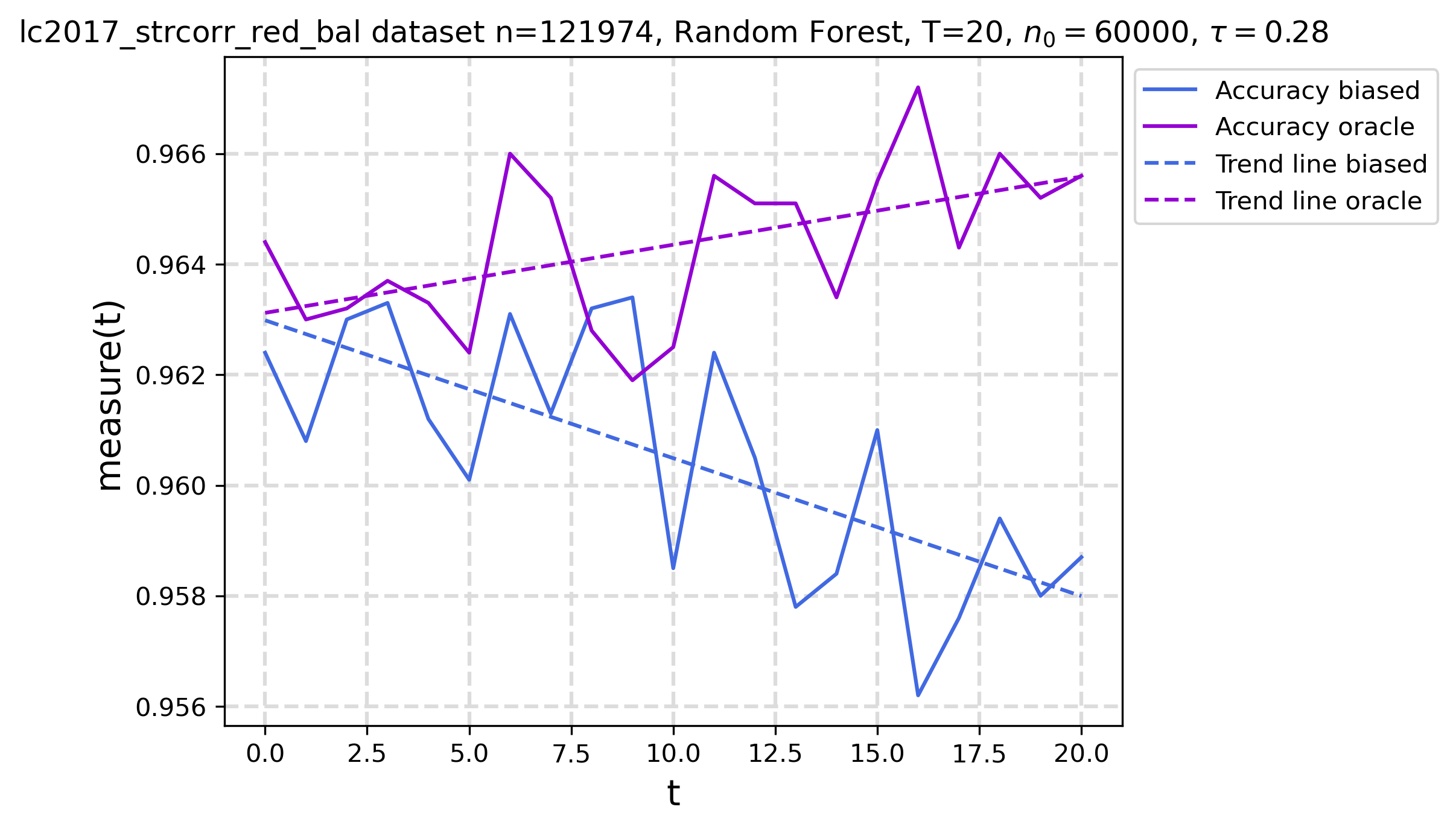}\\
        \includegraphics[scale=.25]{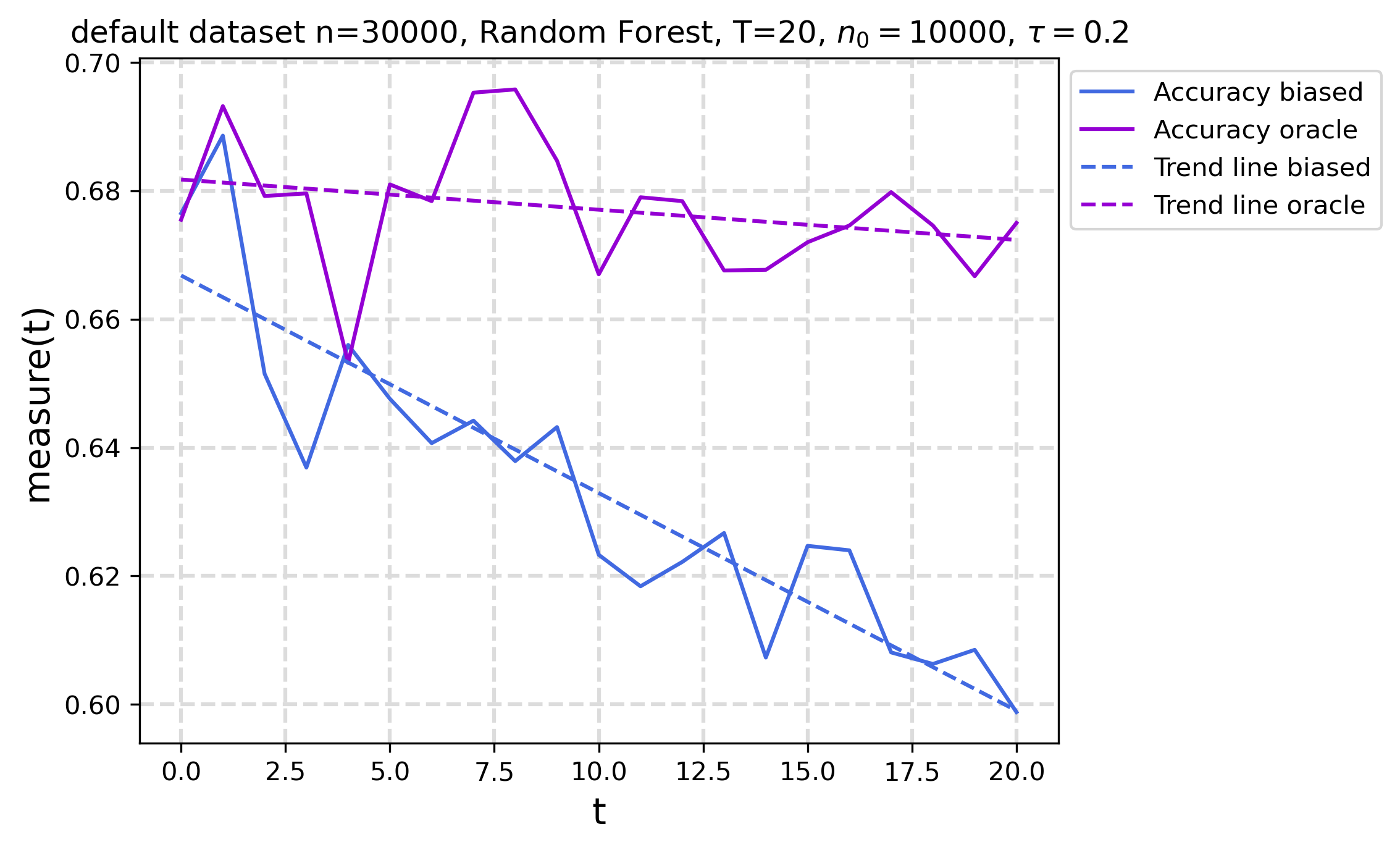} & \includegraphics[scale=.25]{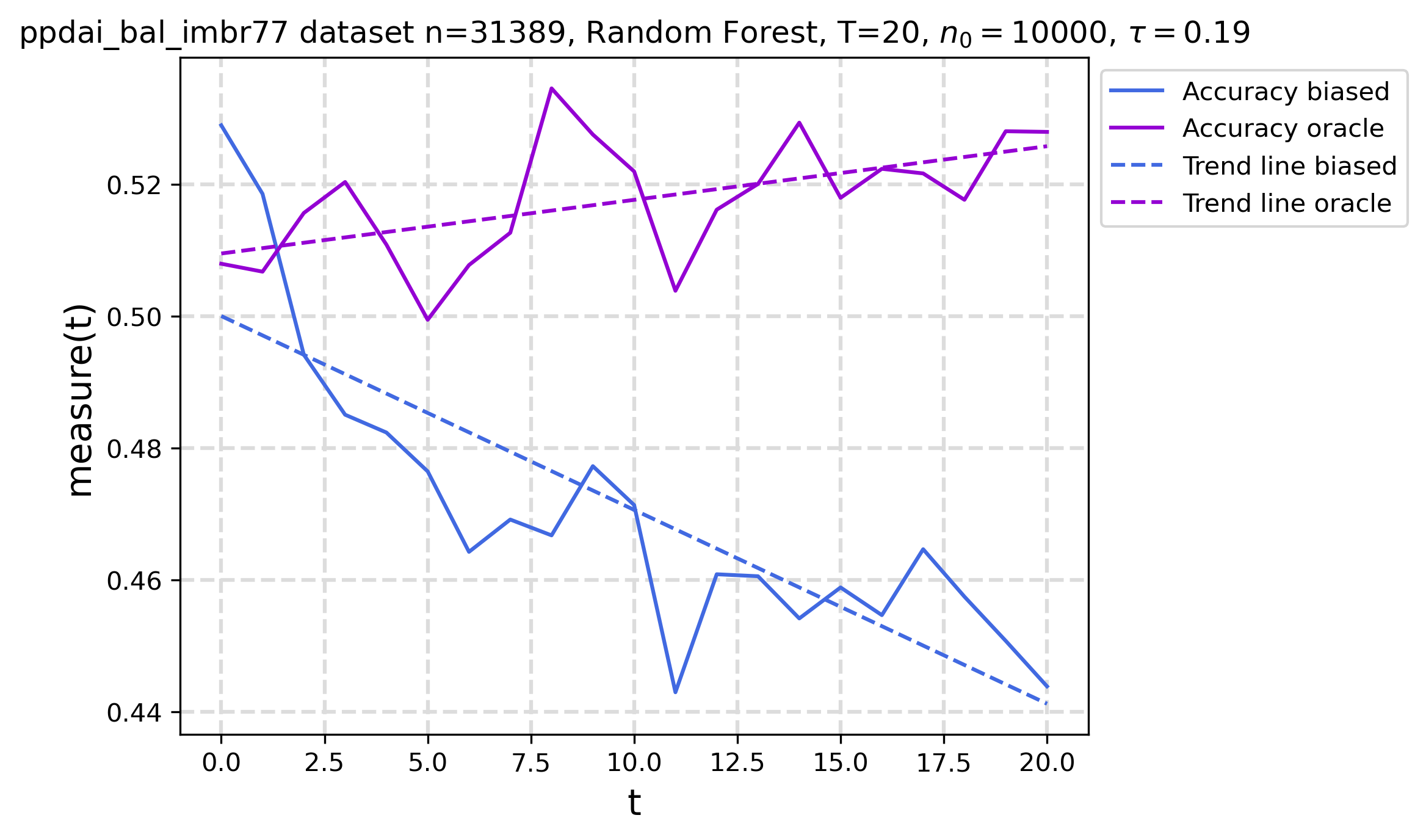} & \includegraphics[scale=.25]{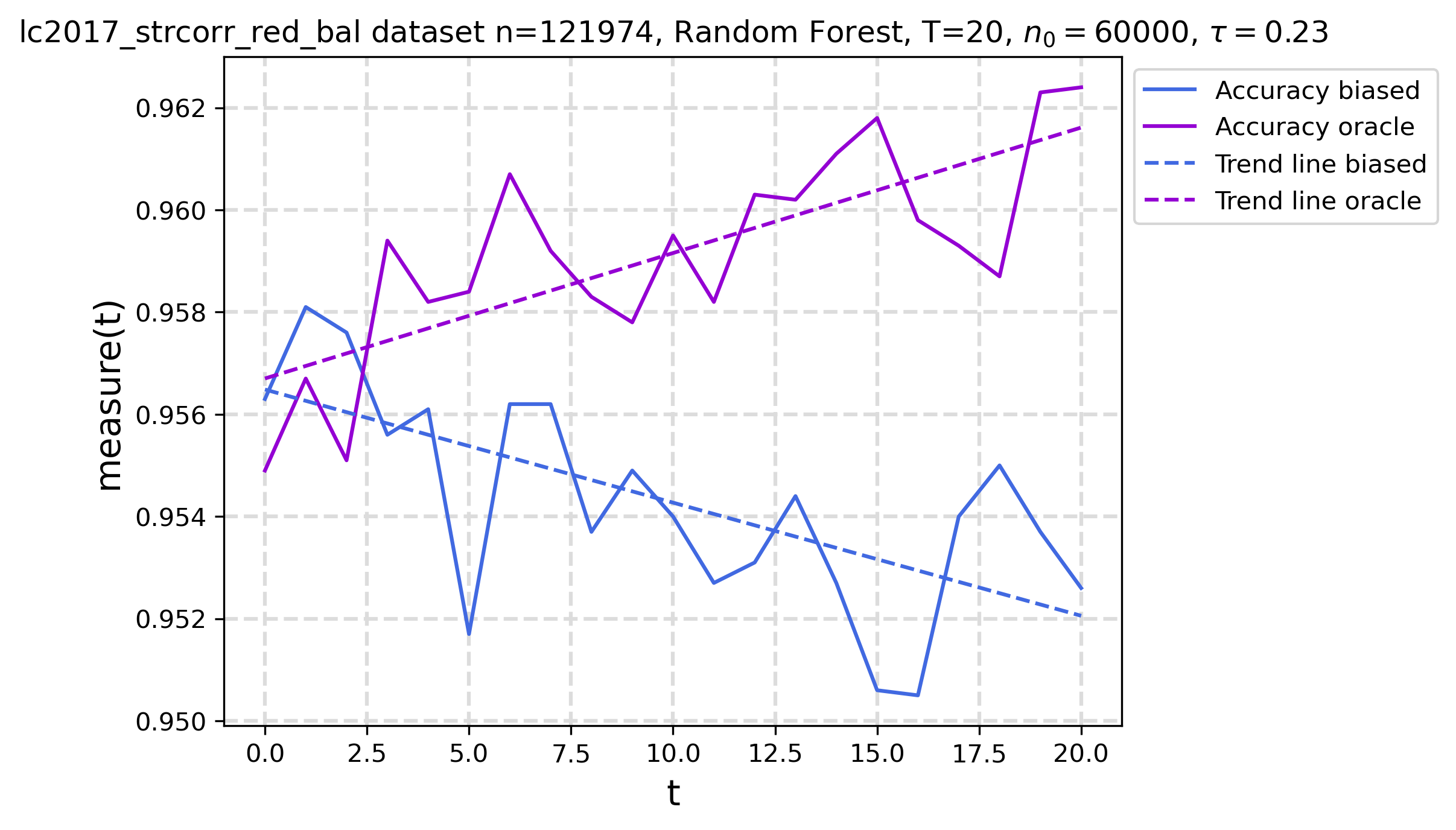}\\
    \end{tabular}
    \caption{Temporal evolution - Random Forest - Decision thresholds for $c \in \{1, 3, 5\}$ - Accuracy.}
    \label{fig:rf_tempevol_accuracy}
\end{figure*}

\begin{figure*}[t]
    \centering
    \begin{tabular}{ccc}
    Default & ppdai\_bal\_imbr77 & lc17\_bal\_imbr50\\
    \includegraphics[scale=.25]{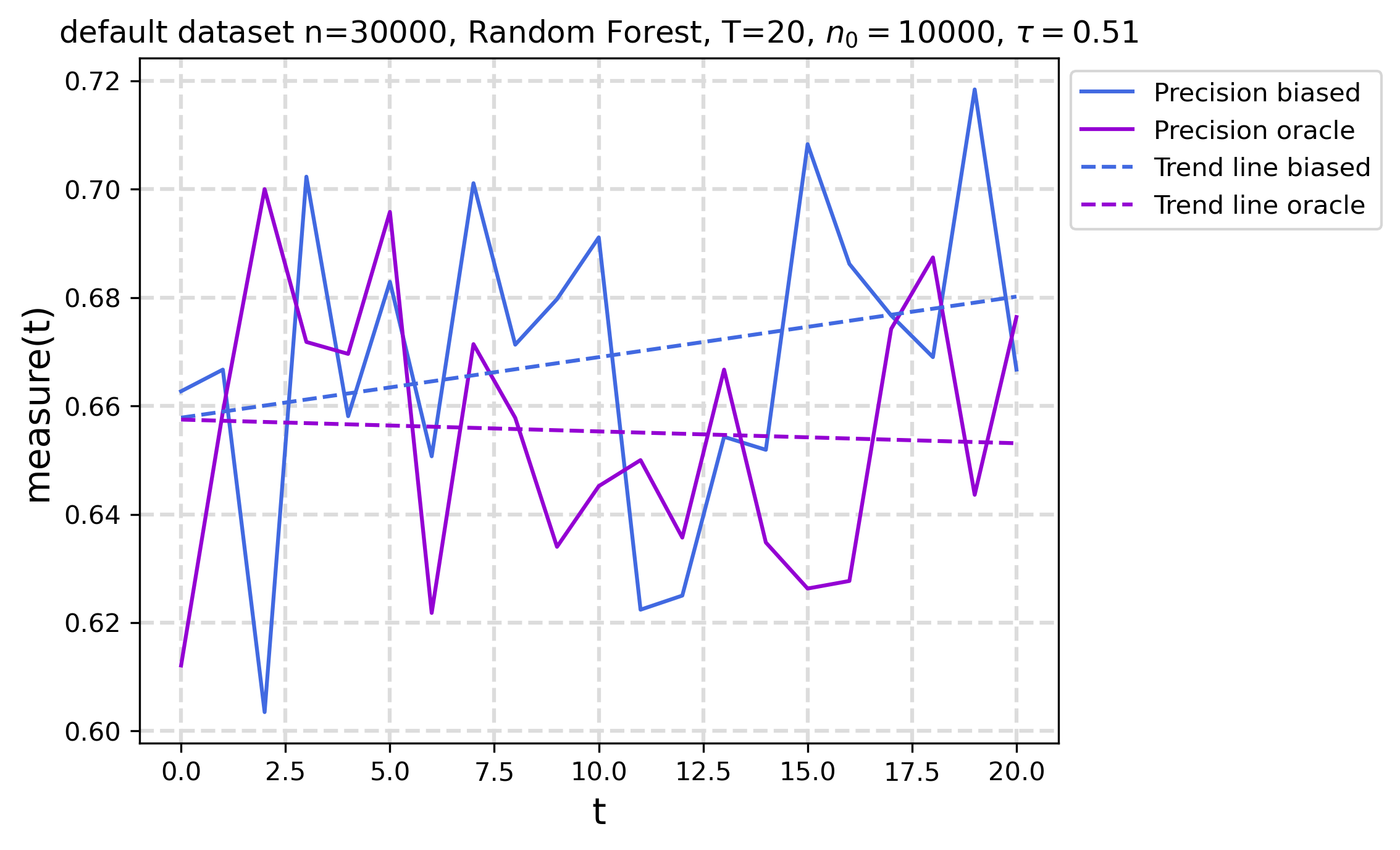} & \includegraphics[scale=.25]{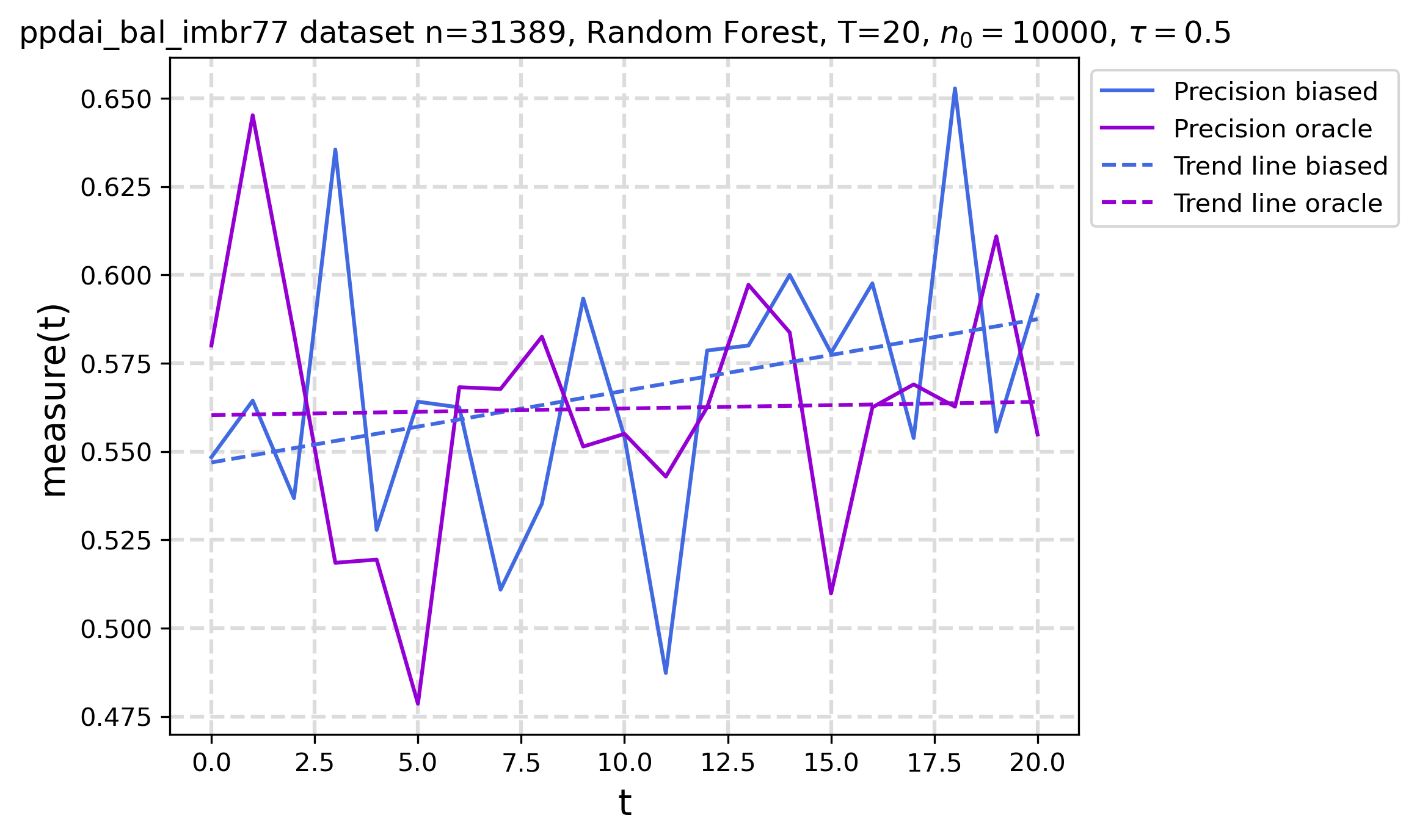} & \includegraphics[scale=.25]{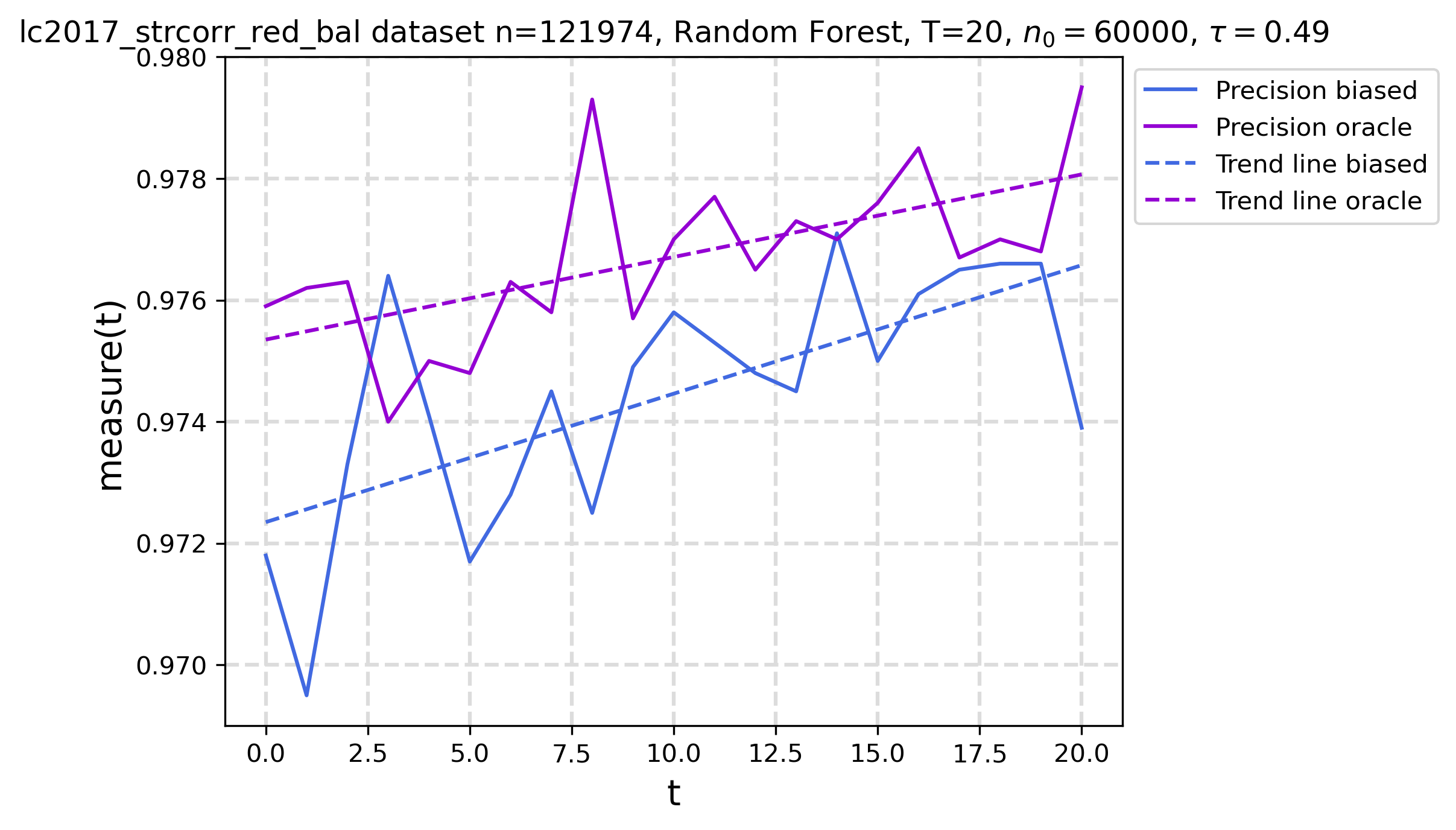}\\
  \includegraphics[scale=.25]{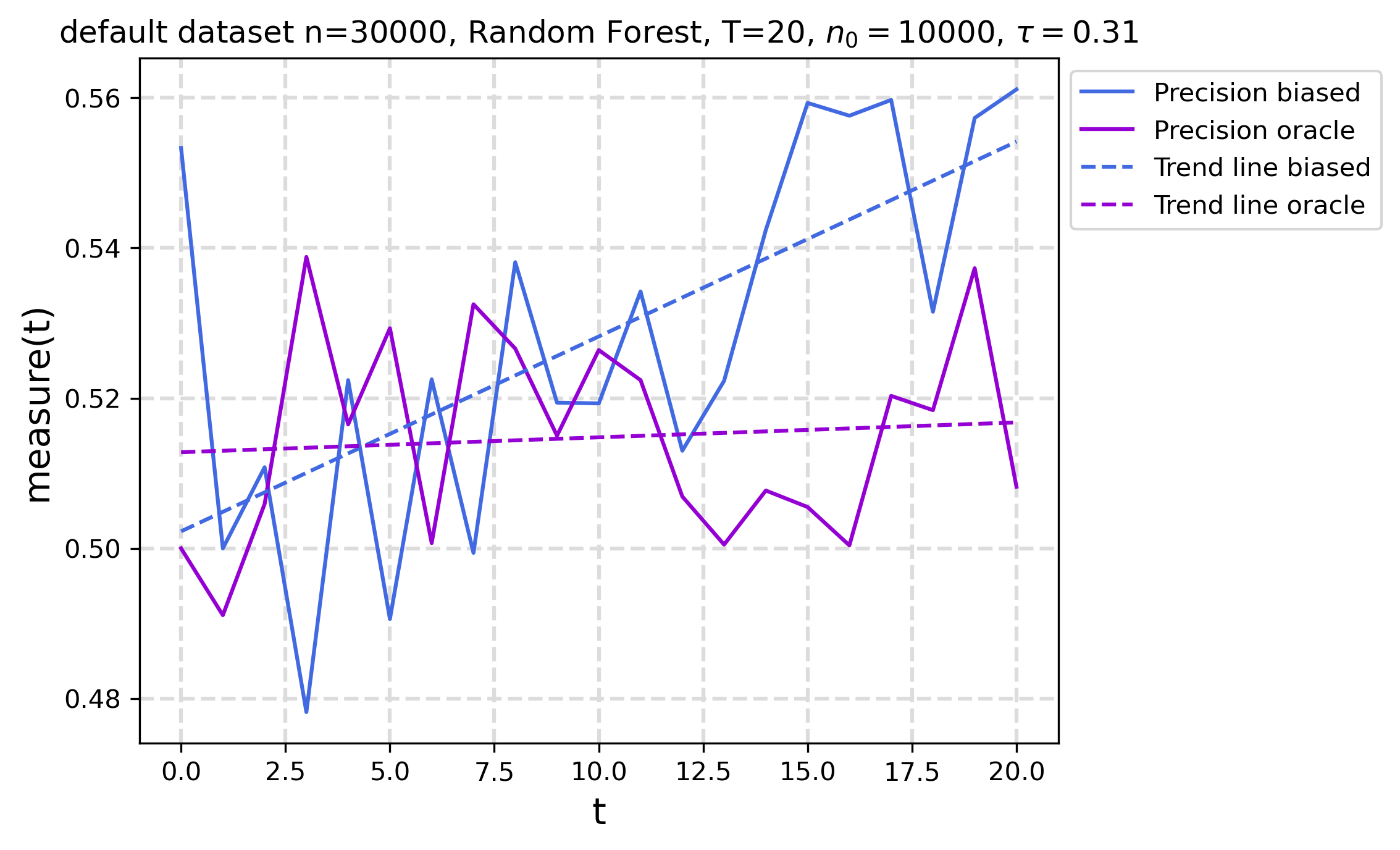} & \includegraphics[scale=.25]{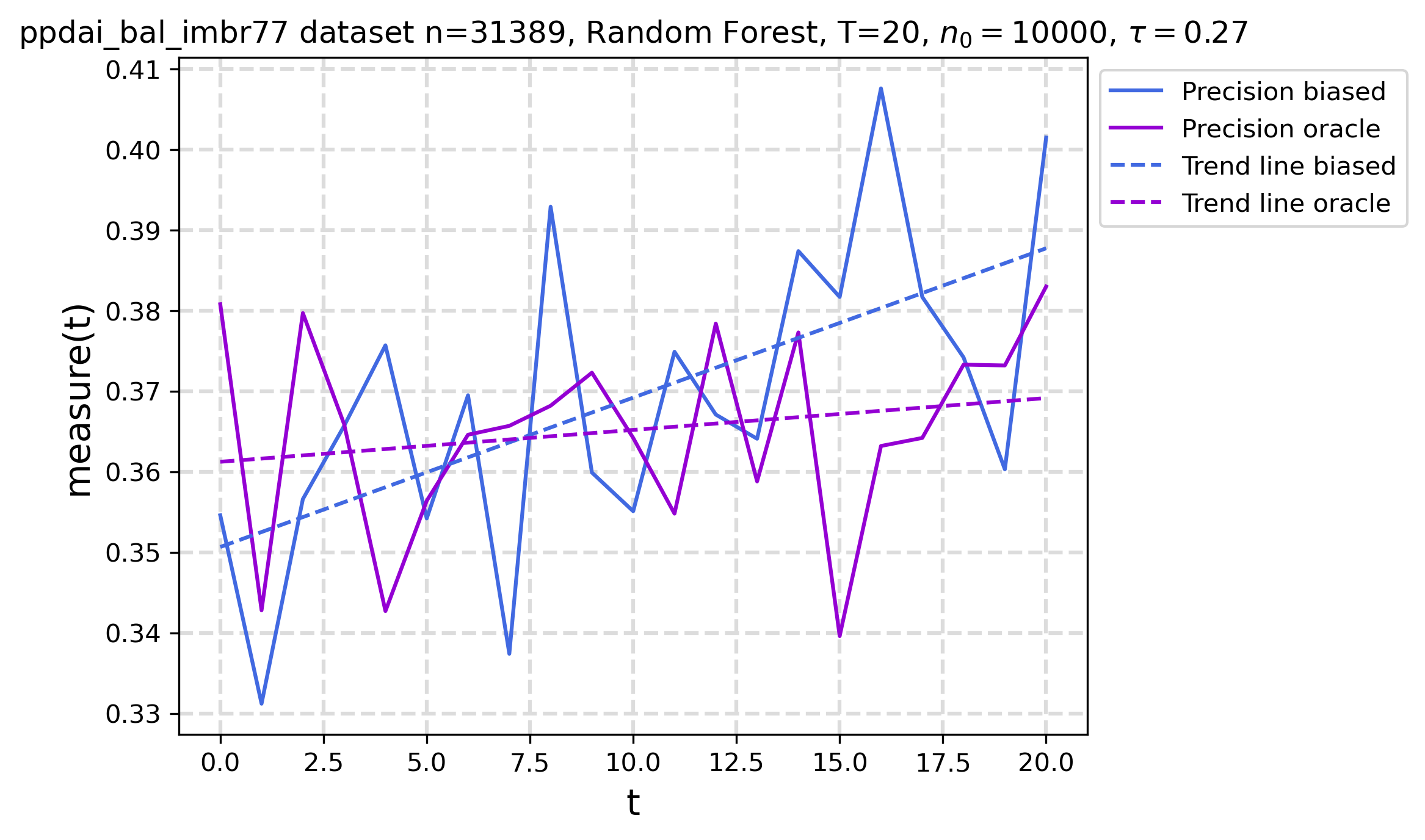} & \includegraphics[scale=.25]{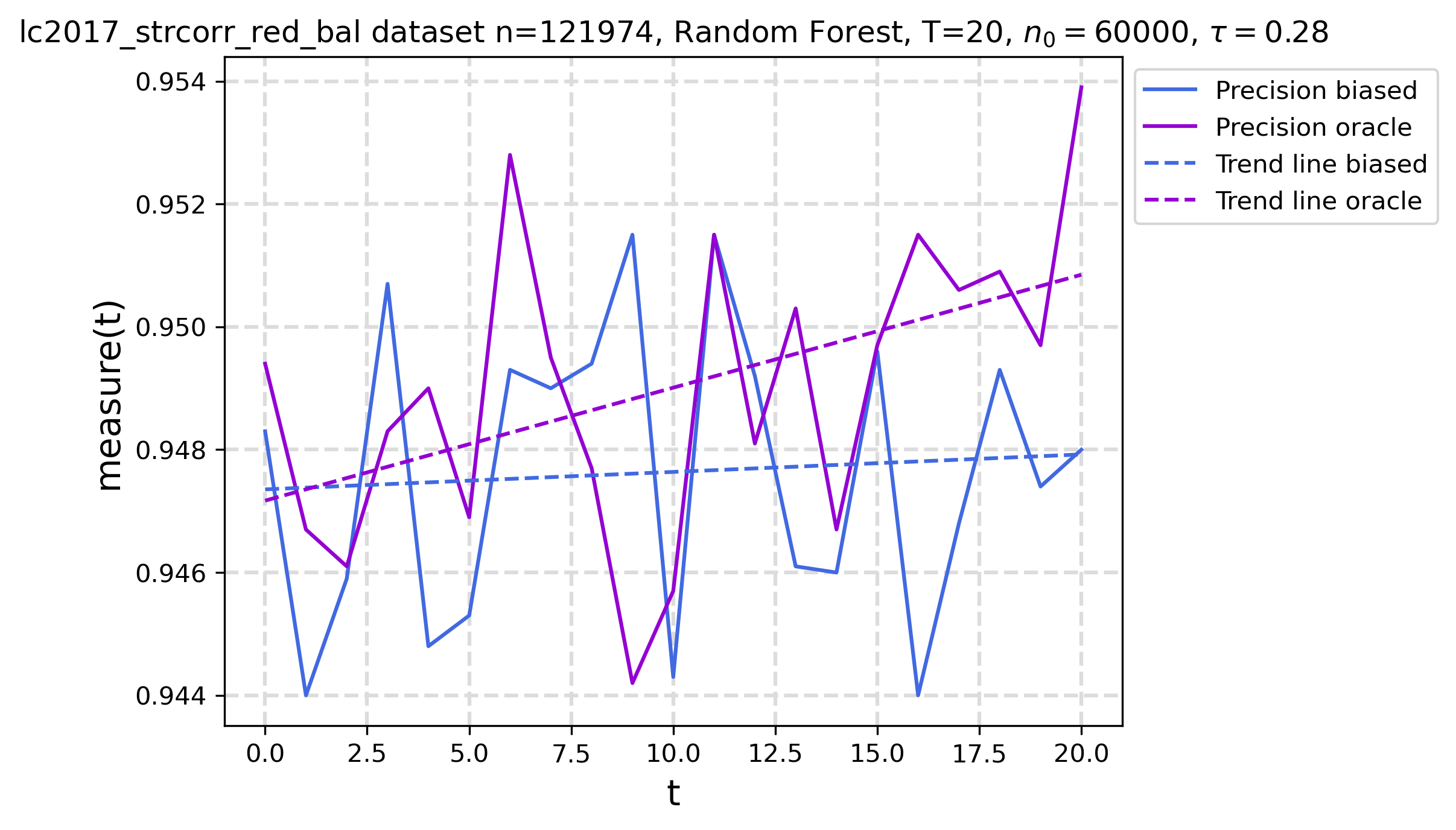}\\
    \includegraphics[scale=.25]{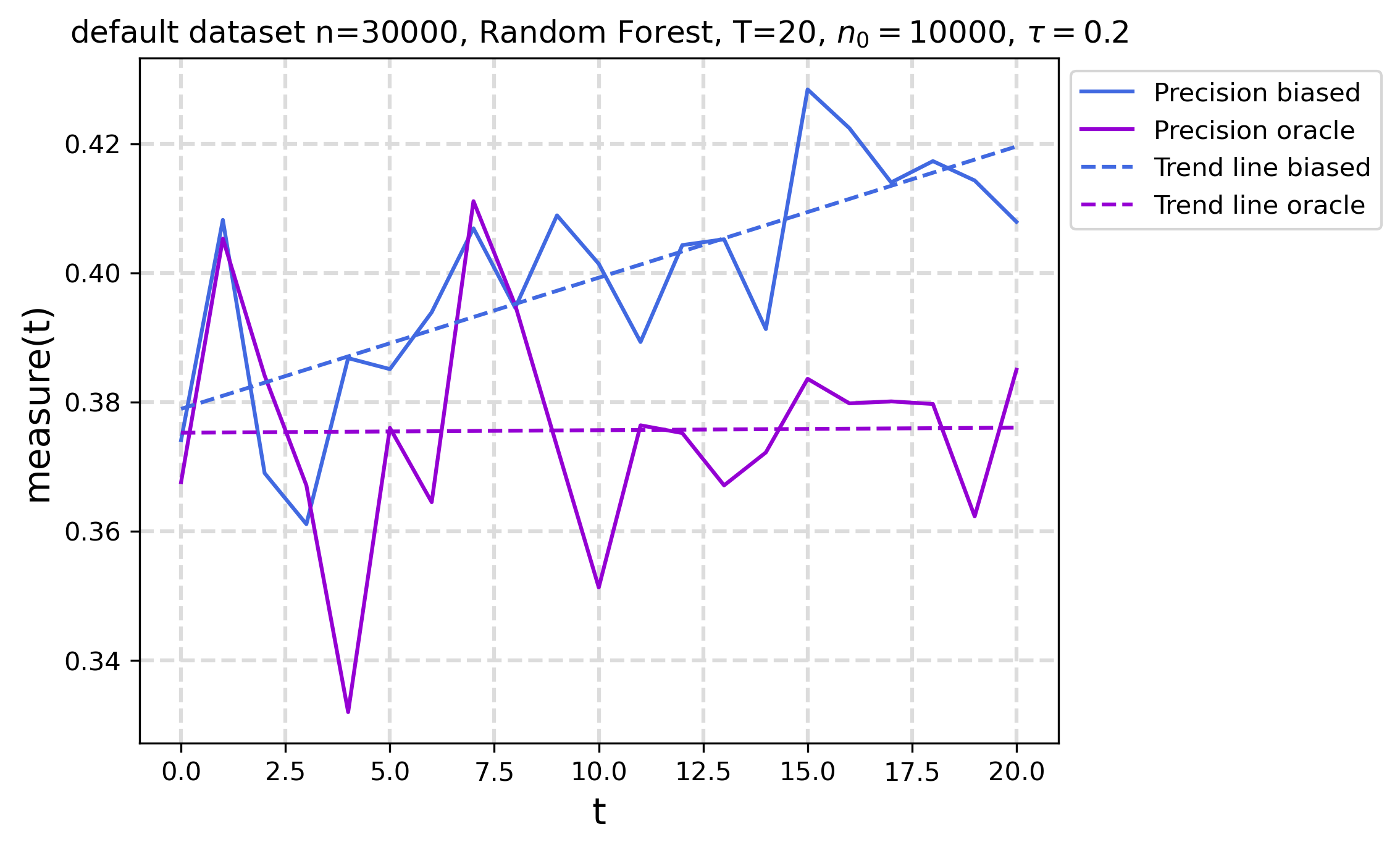} & \includegraphics[scale=.25]{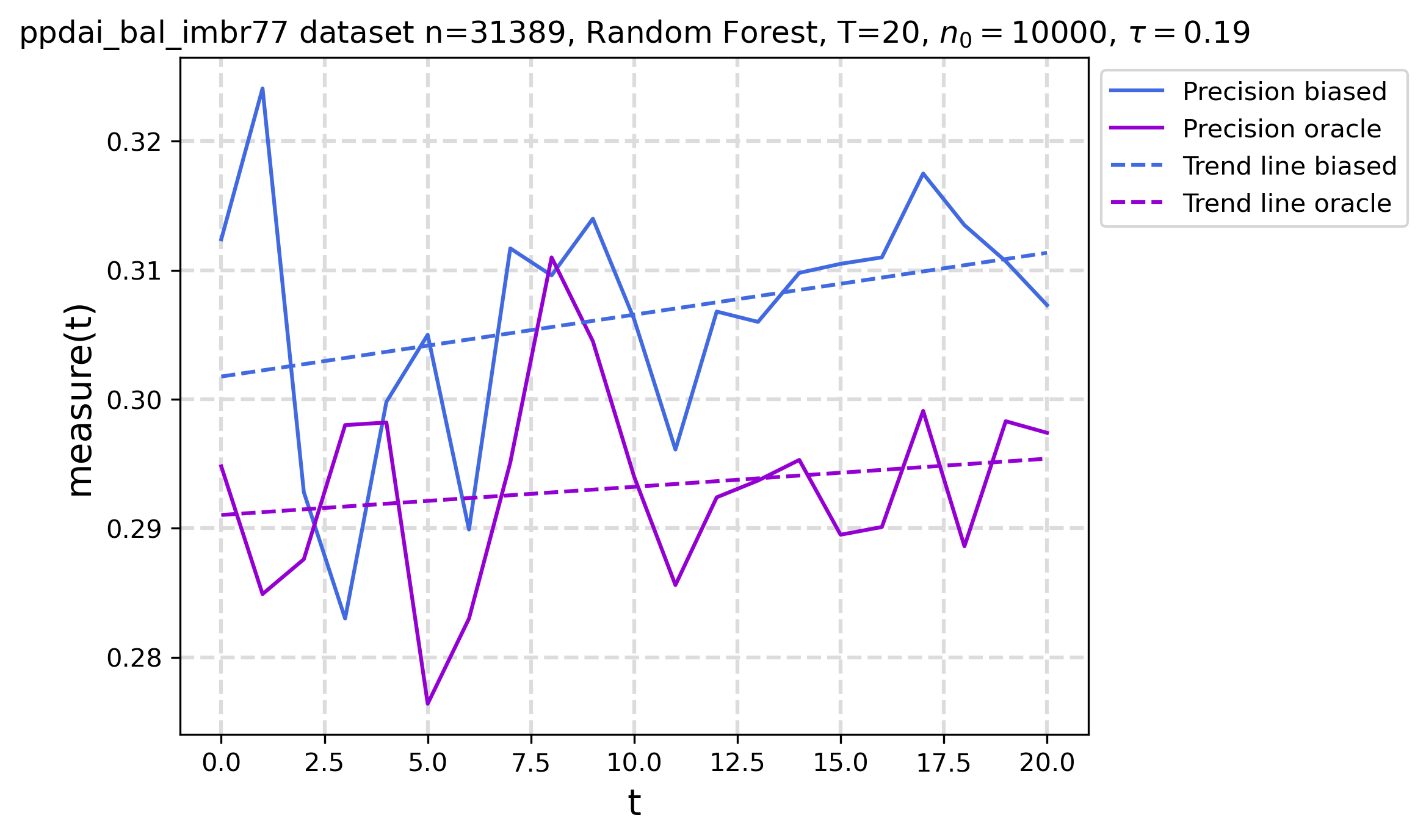} & \includegraphics[scale=.25]{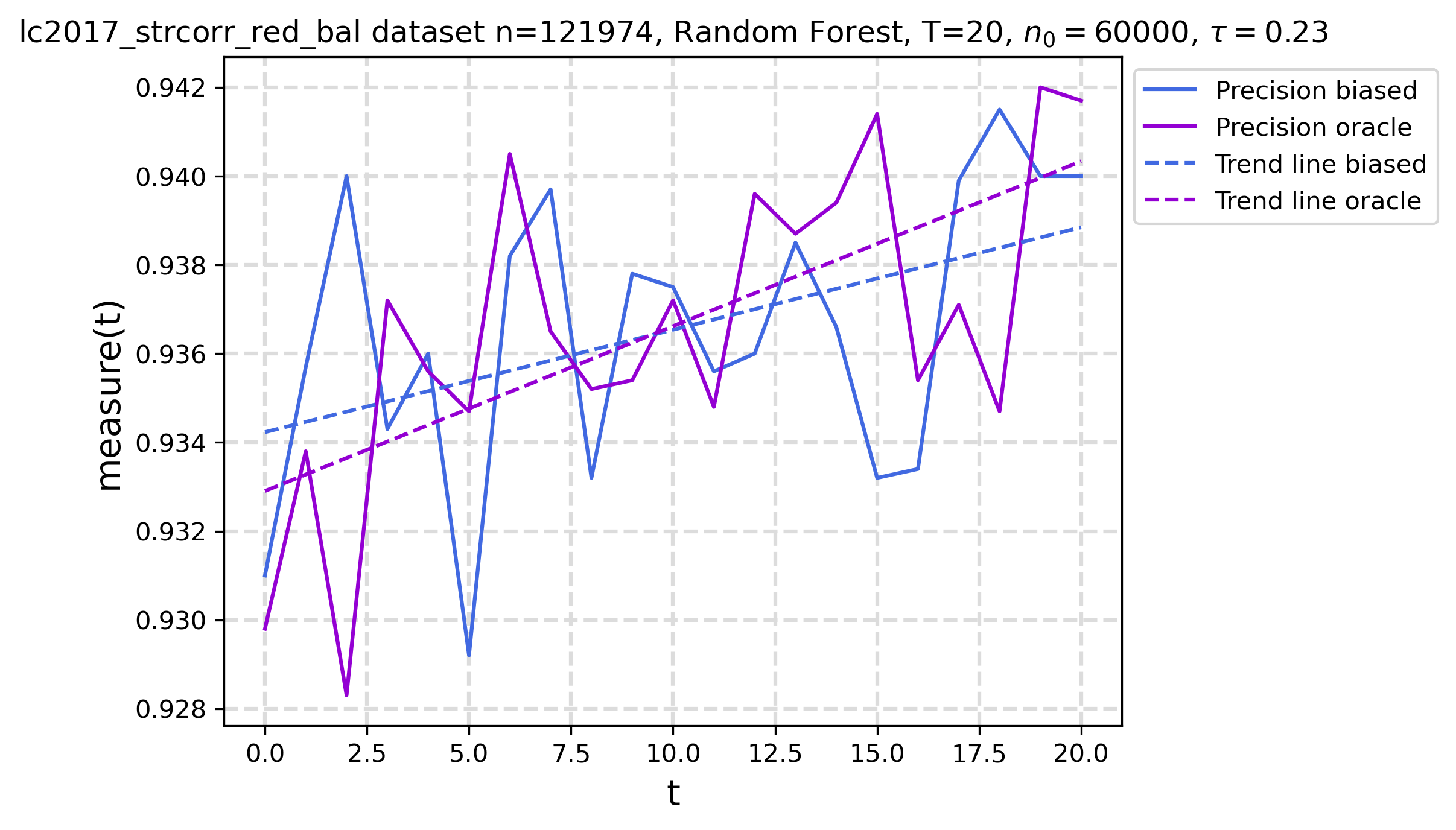}\\
    \end{tabular}
    \caption{Temporal evolution - Random Forest - Decision thresholds for $c \in \{1, 3, 5\}$ - Precision.}
    \label{fig:rf_tempevol_precision}
\end{figure*}

\clearpage

\begin{figure*}[t]
    \centering
    \begin{tabular}{ccc}
    Default & ppdai\_bal\_imbr77 & lc17\_bal\_imbr50\\
    \includegraphics[scale=.25]{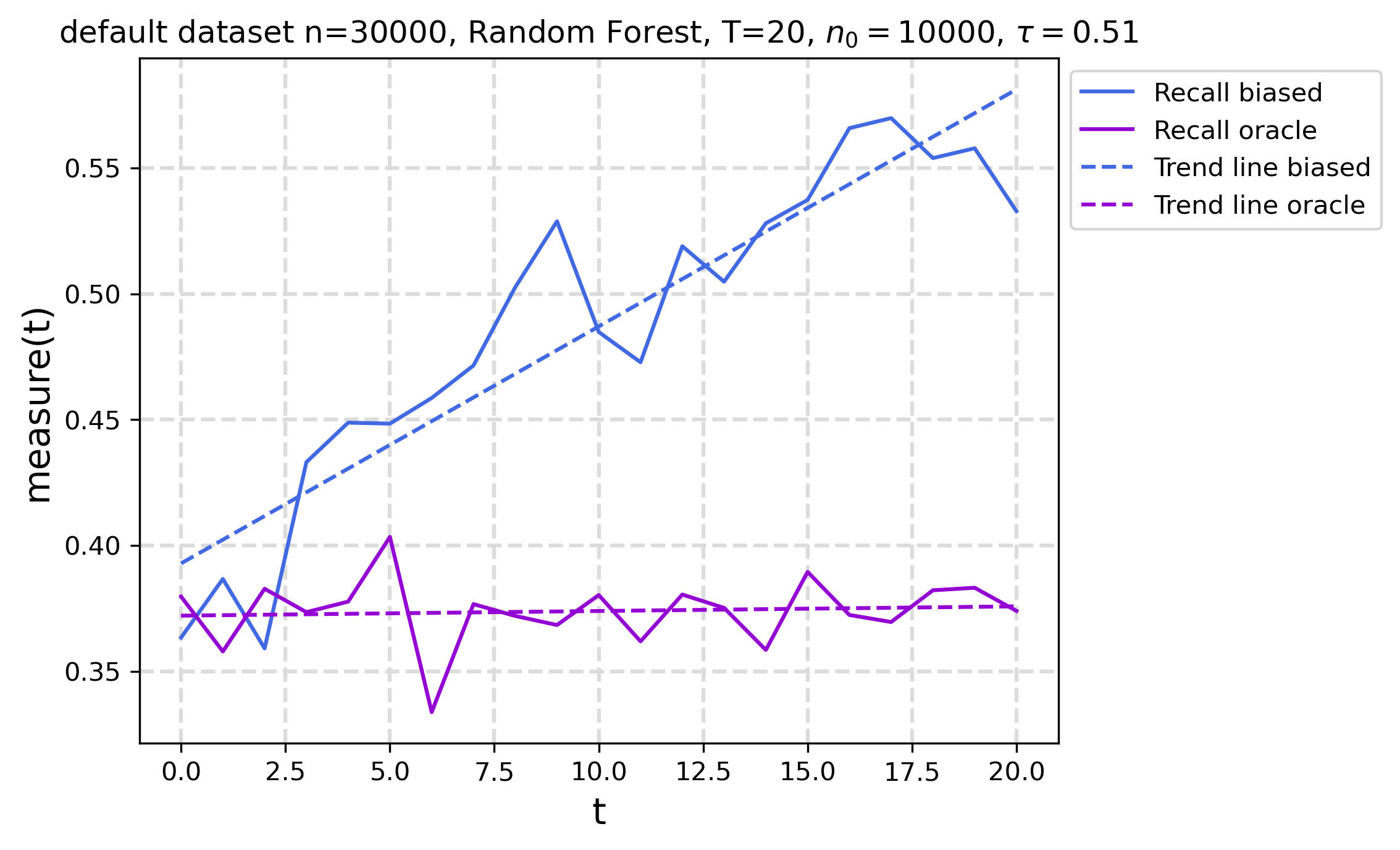} & \includegraphics[scale=.25]{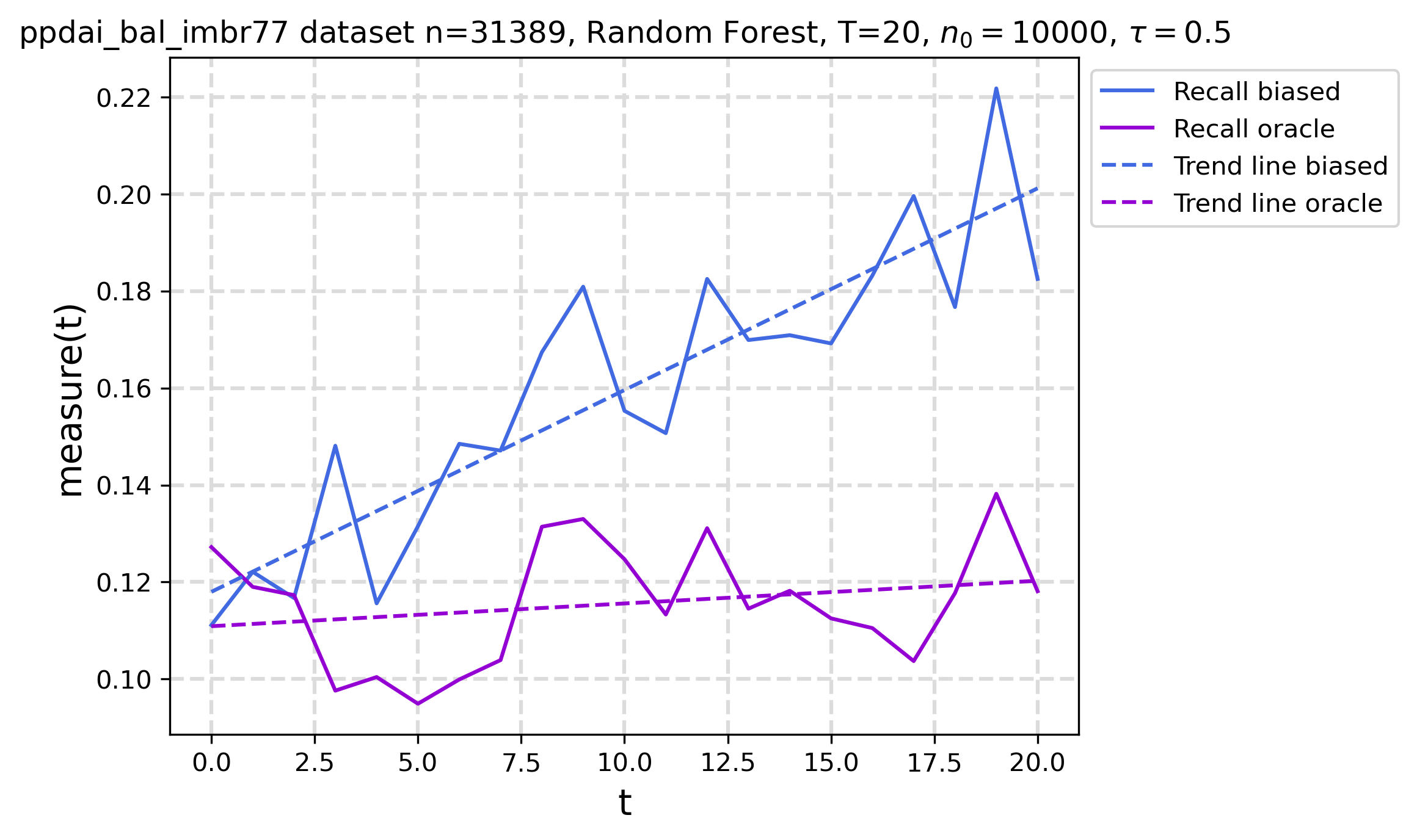} & \includegraphics[scale=.25]{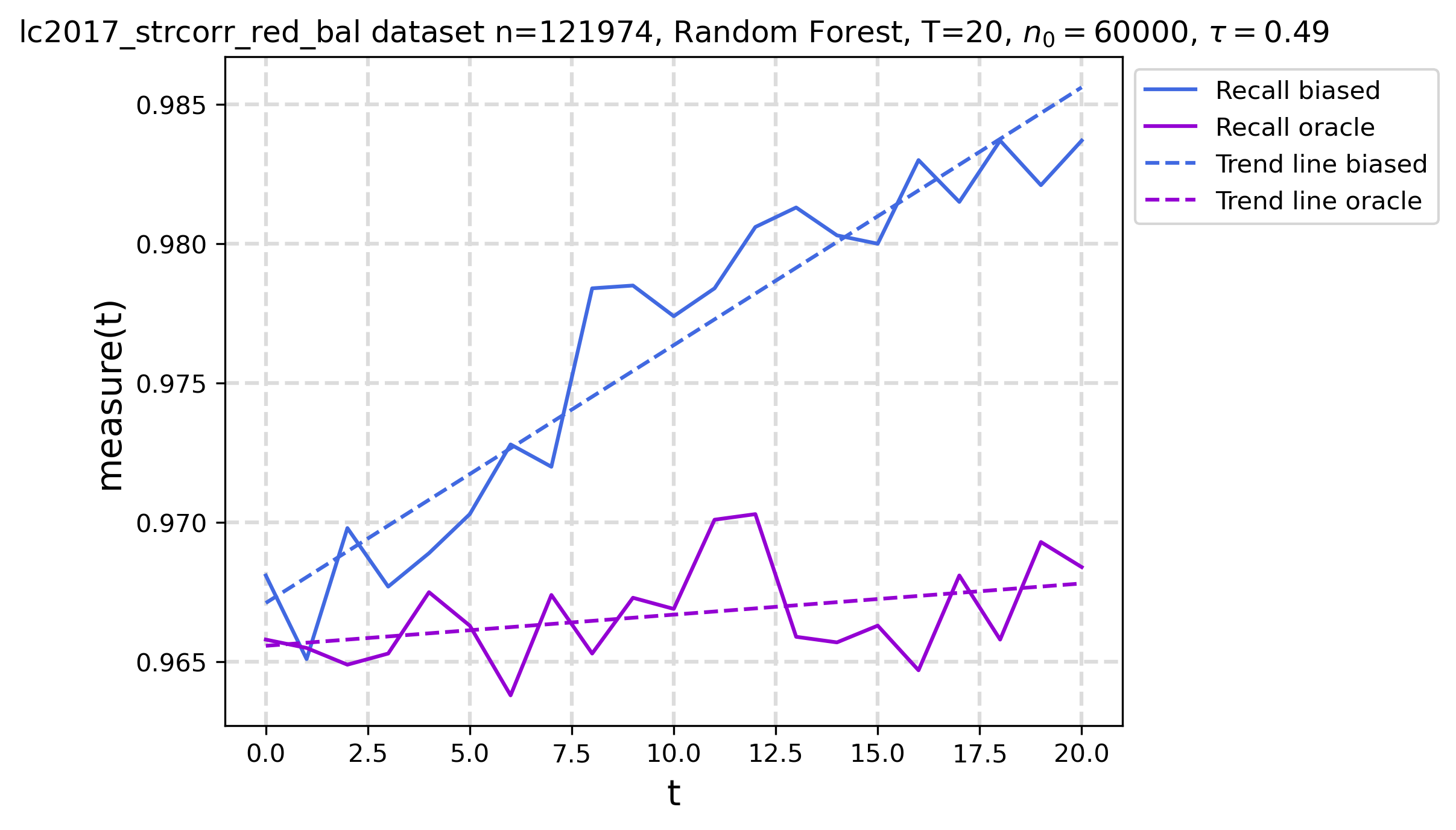}\\
   \includegraphics[scale=.25]{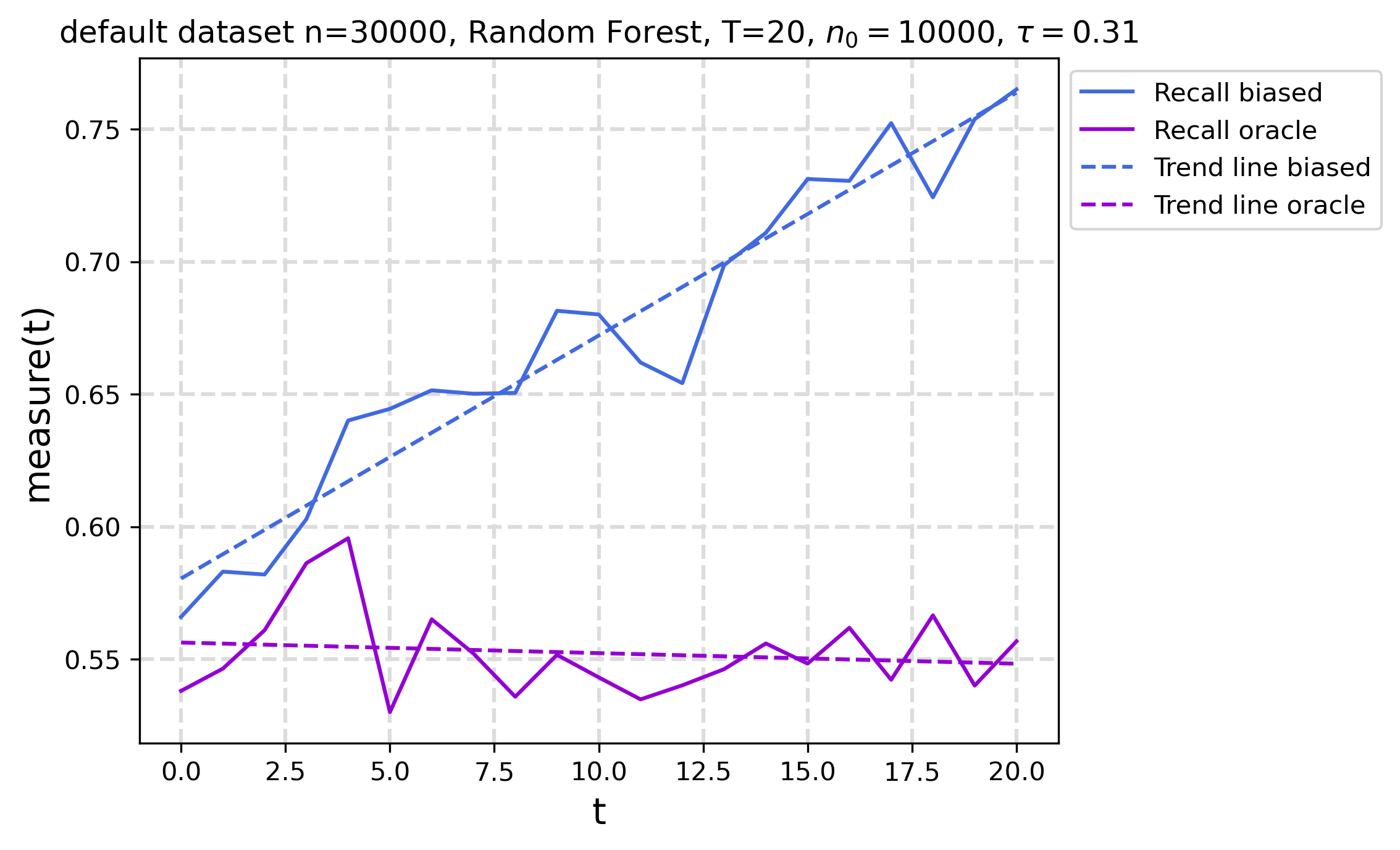} & \includegraphics[scale=.25]{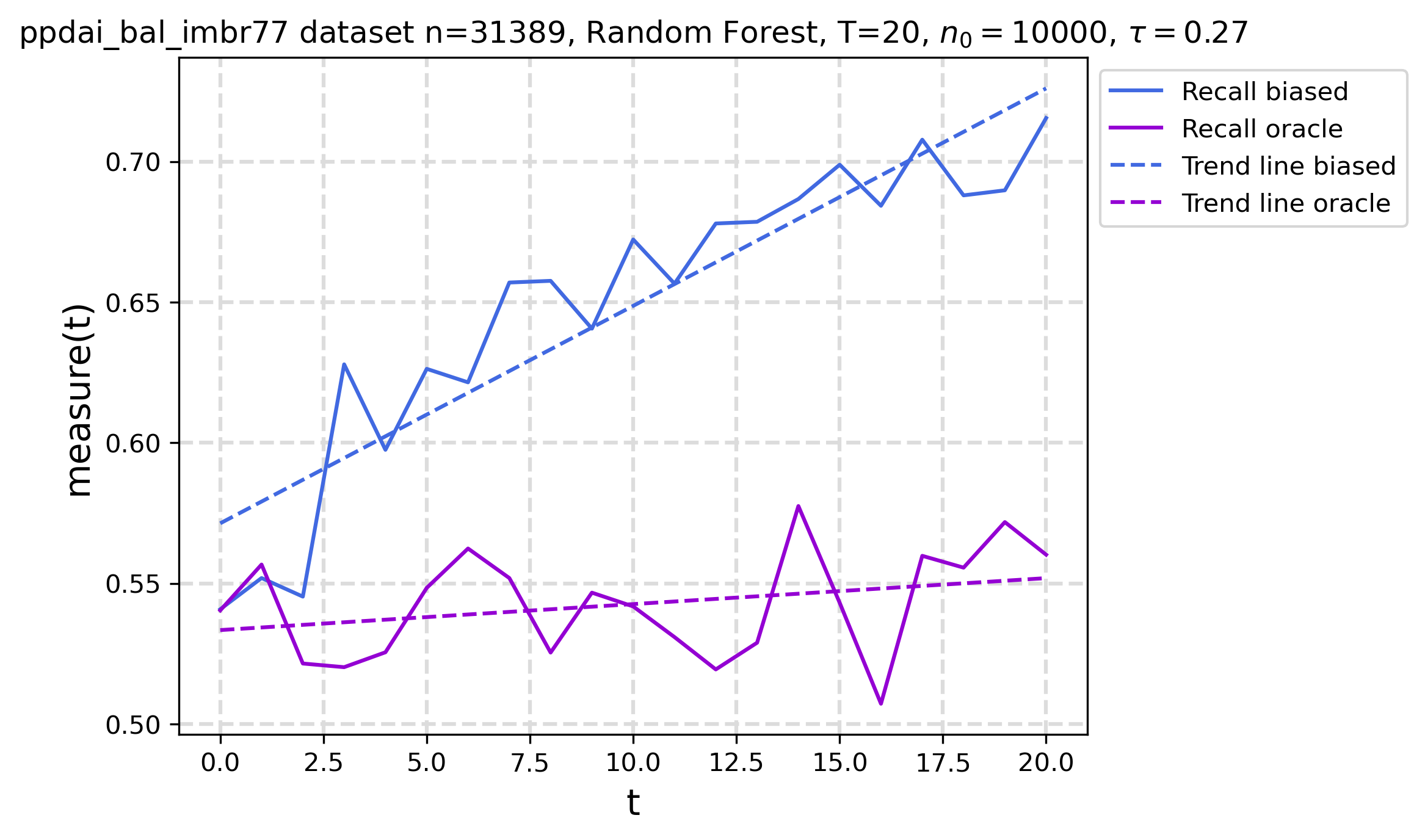} & \includegraphics[scale=.25]{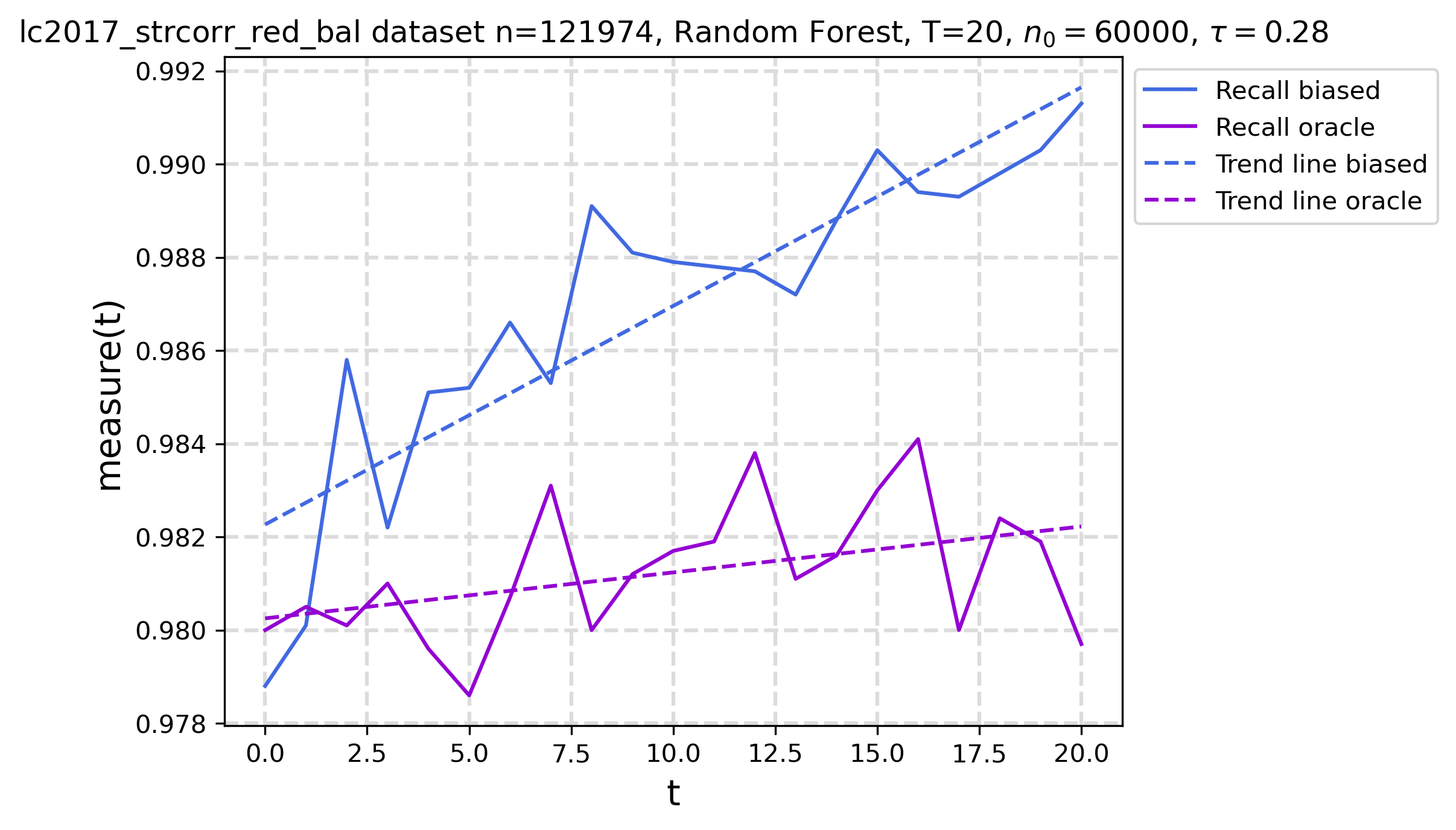}\\
    \includegraphics[scale=.25]{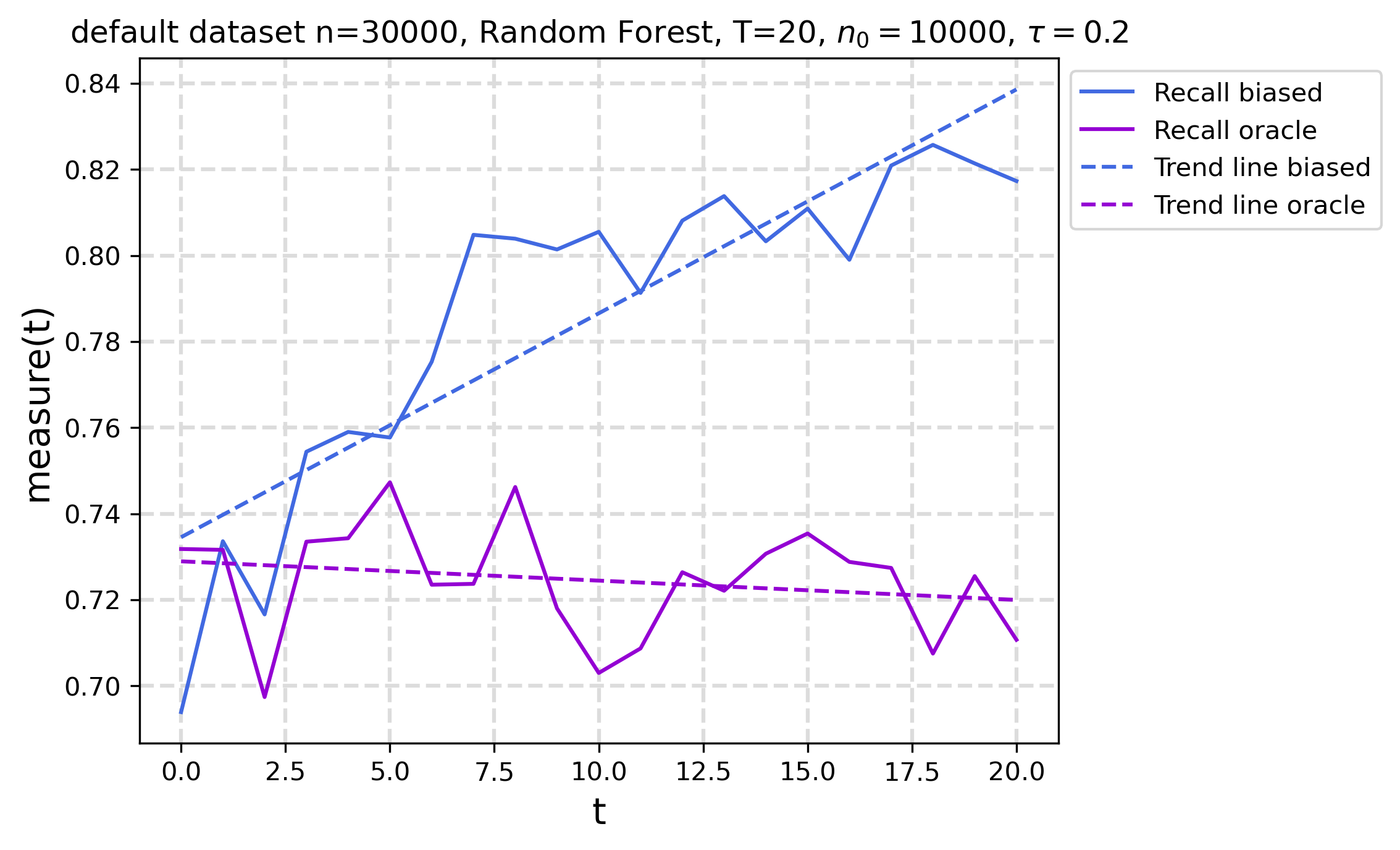} & \includegraphics[scale=.25]{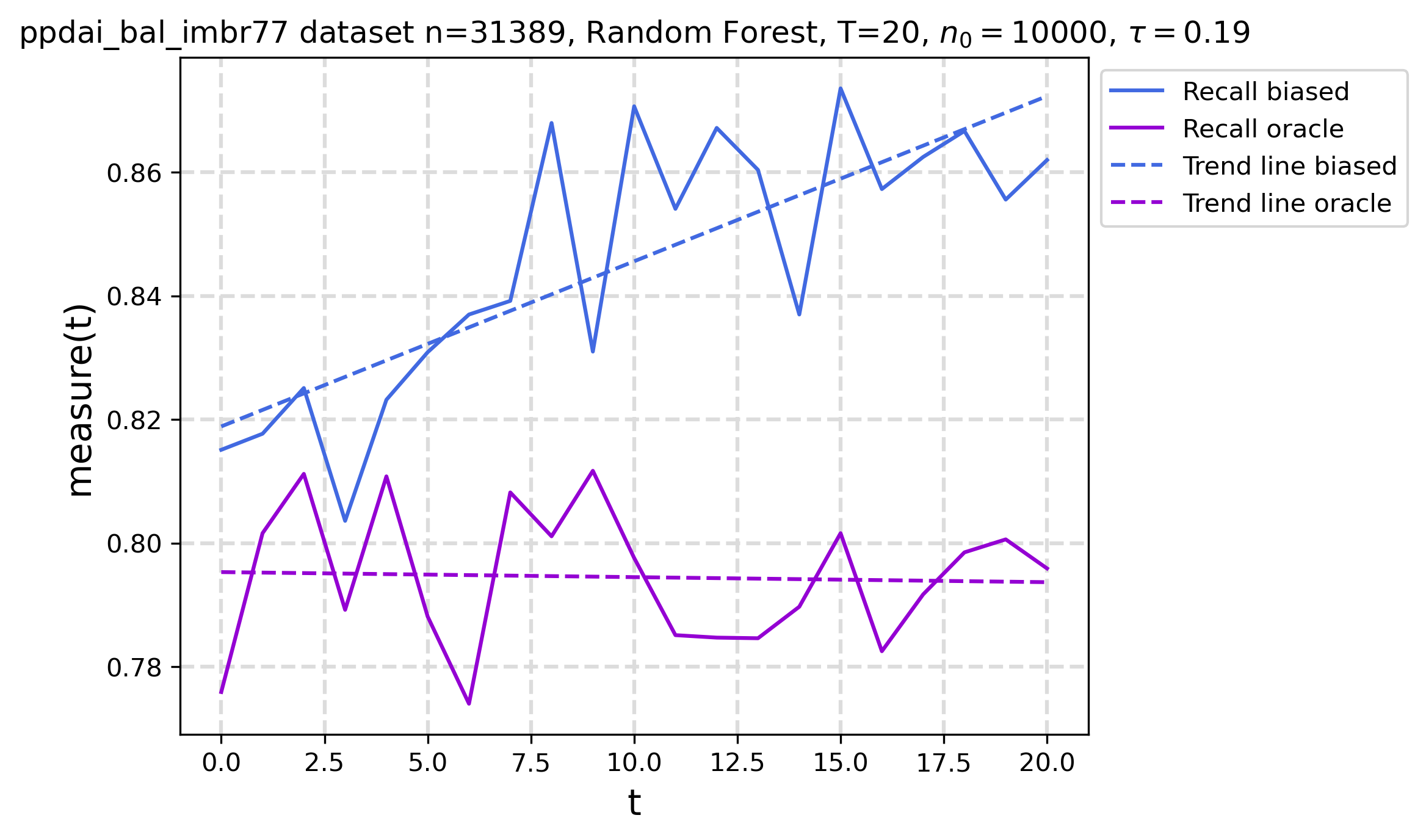} & \includegraphics[scale=.25]{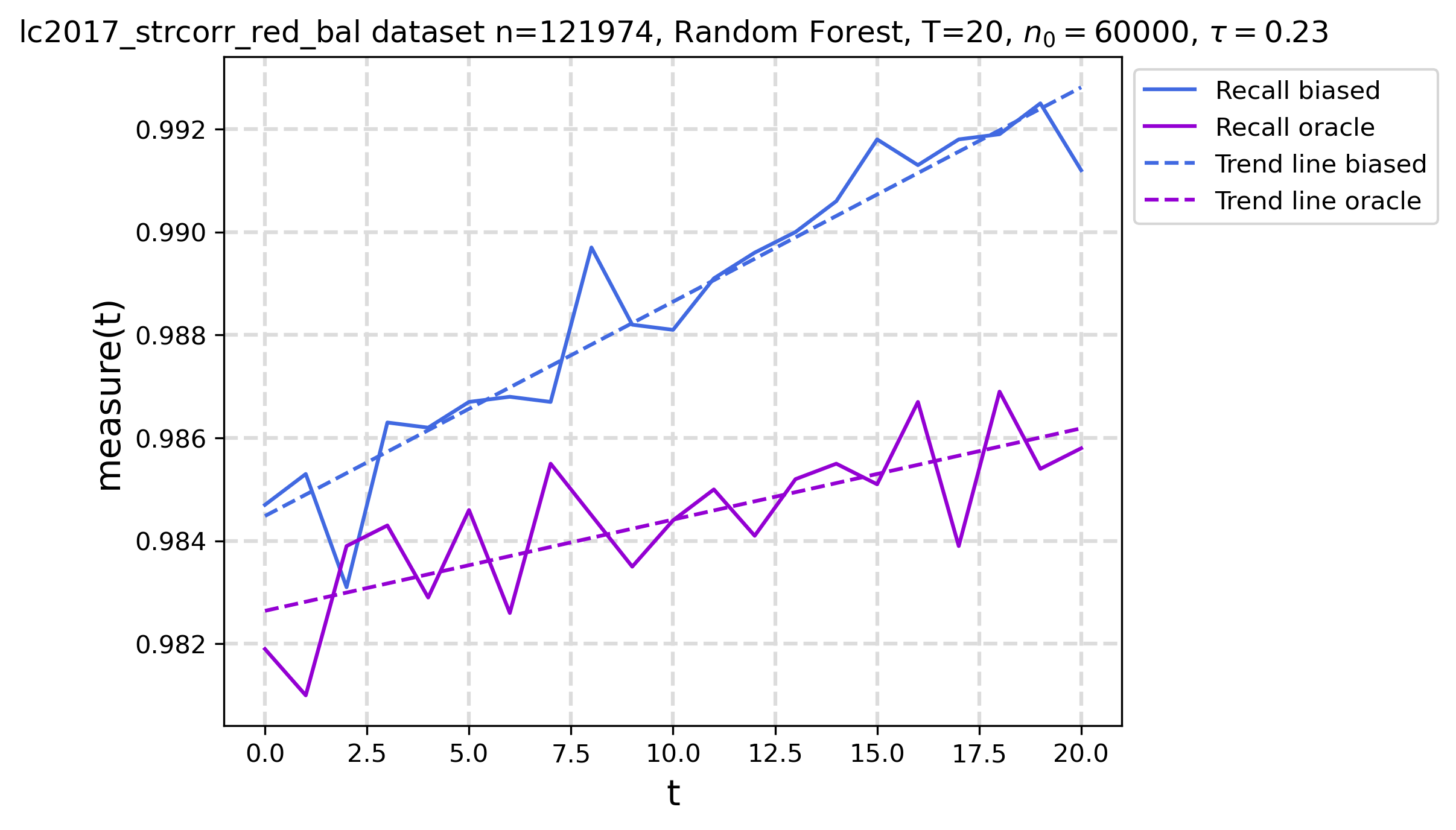}\\
    \end{tabular}
    \caption{Temporal evolution - Random Forest - Decision thresholds for $c \in \{1, 3, 5\}$ - Recall.}
    \label{fig:rf_tempevol_recall}
\end{figure*}

\begin{figure*}[t]
    \centering
    \begin{tabular}{ccc}
    Default & ppdai\_bal\_imbr77 & lc17\_bal\_imbr50\\
    \includegraphics[scale=.25]{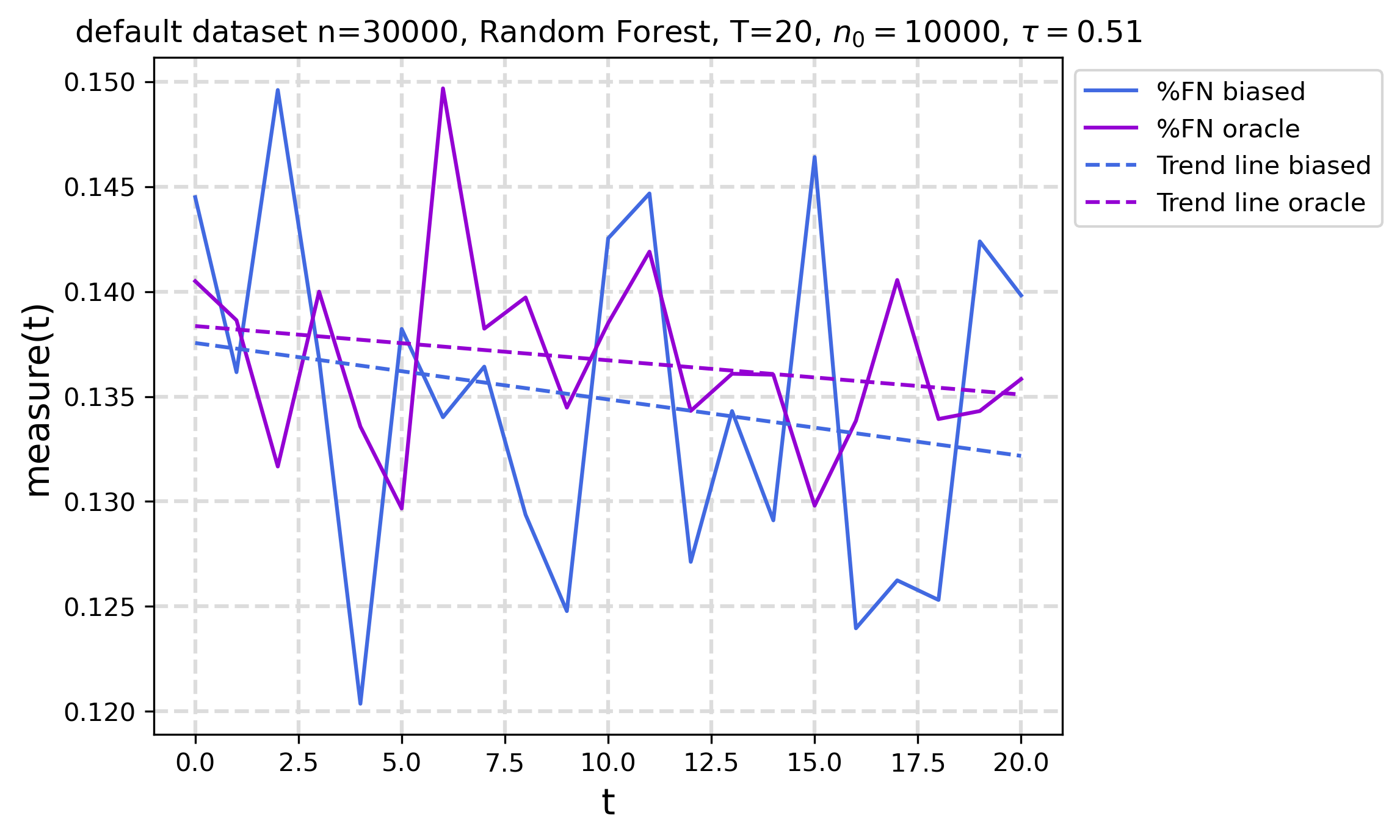} & \includegraphics[scale=.25]{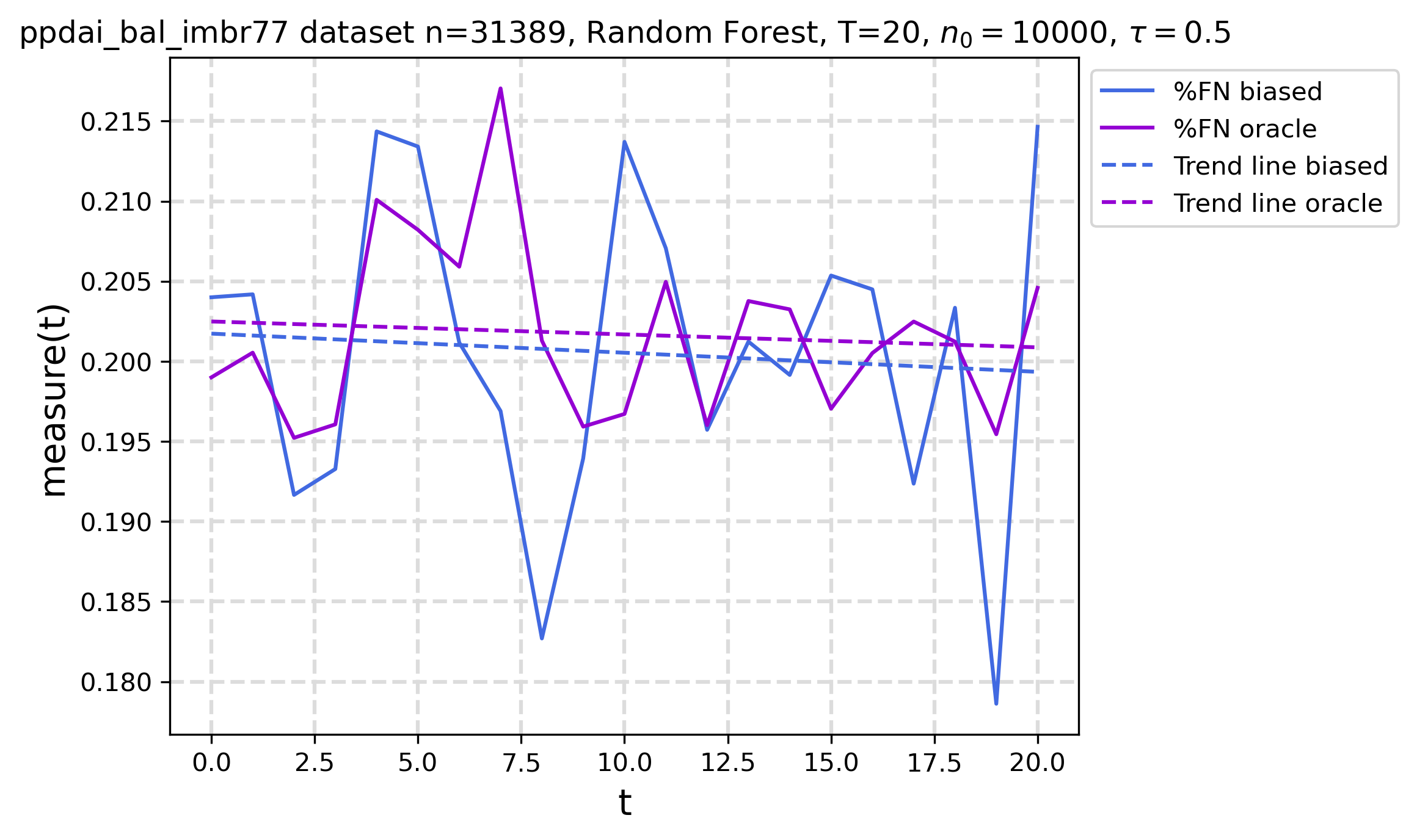} & \includegraphics[scale=.25]{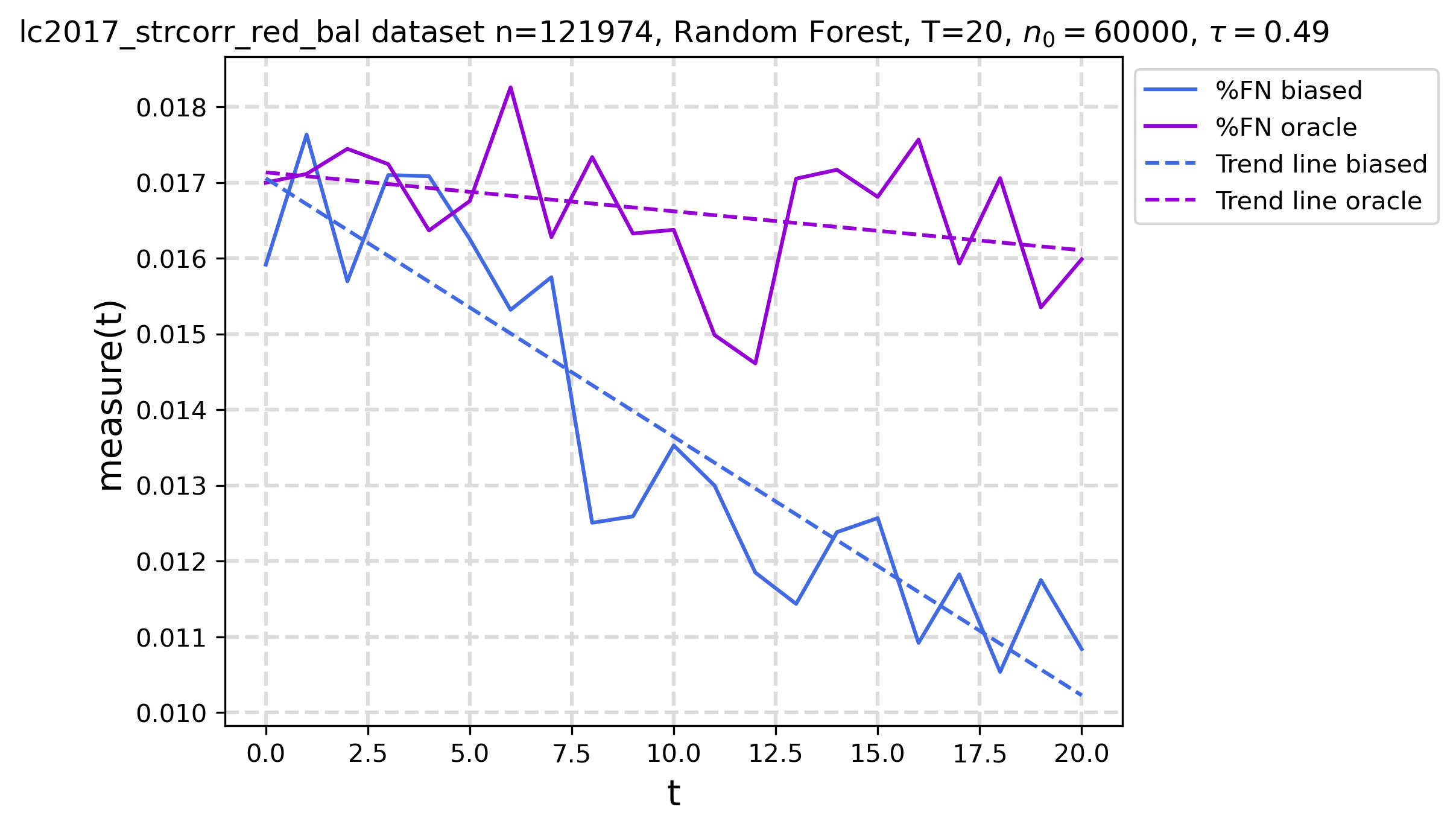}\\
   \includegraphics[scale=.25]{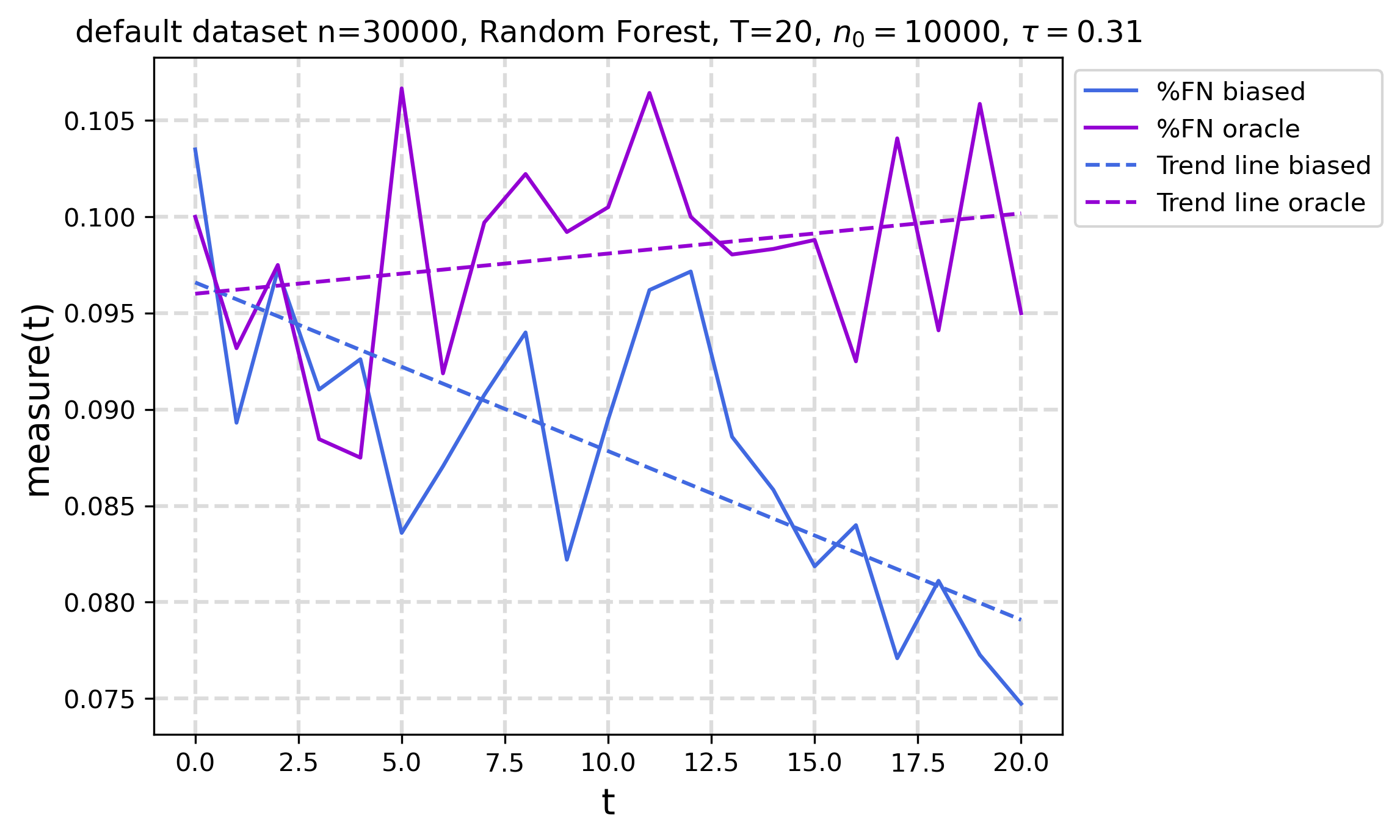} & \includegraphics[scale=.25]{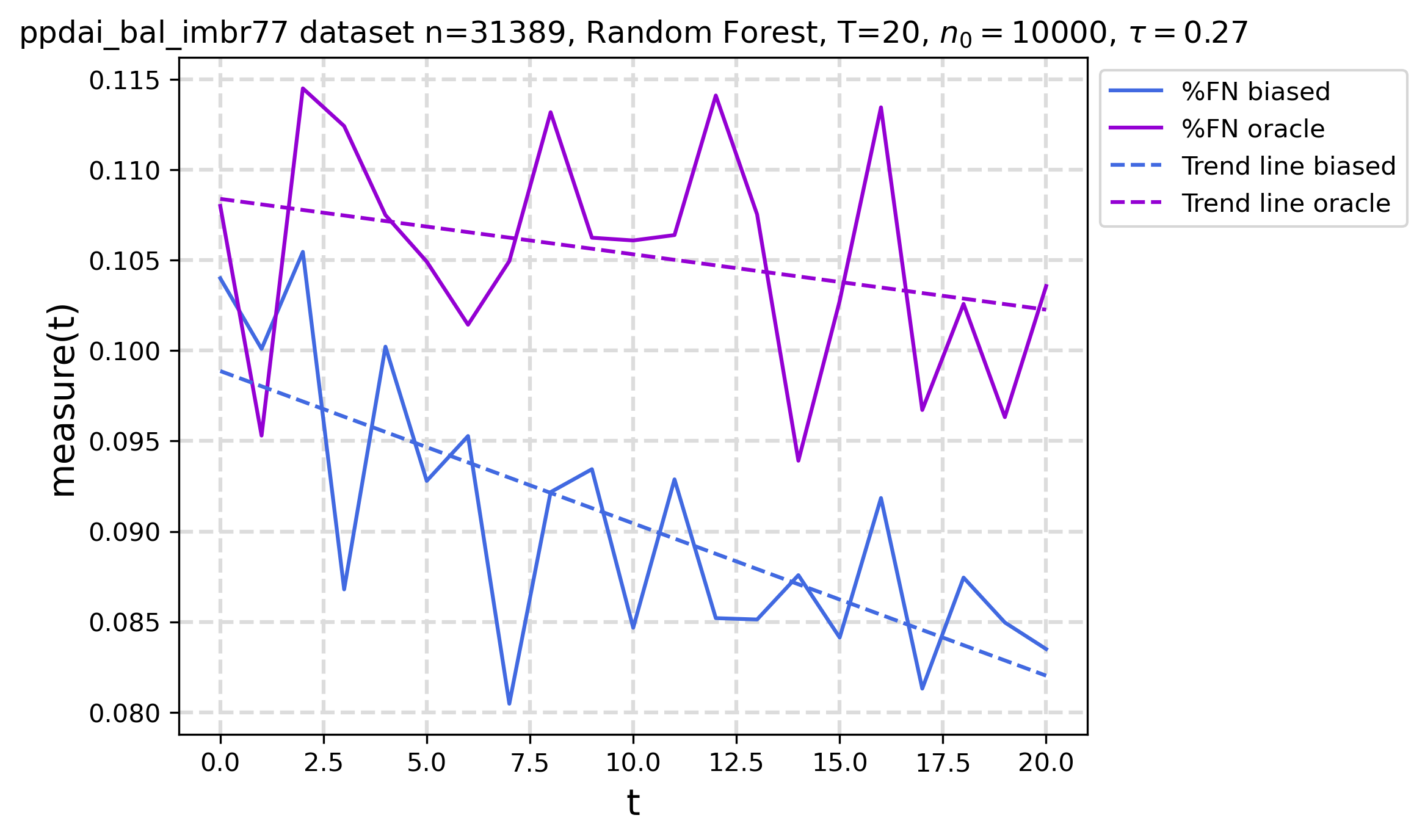} & \includegraphics[scale=.25]{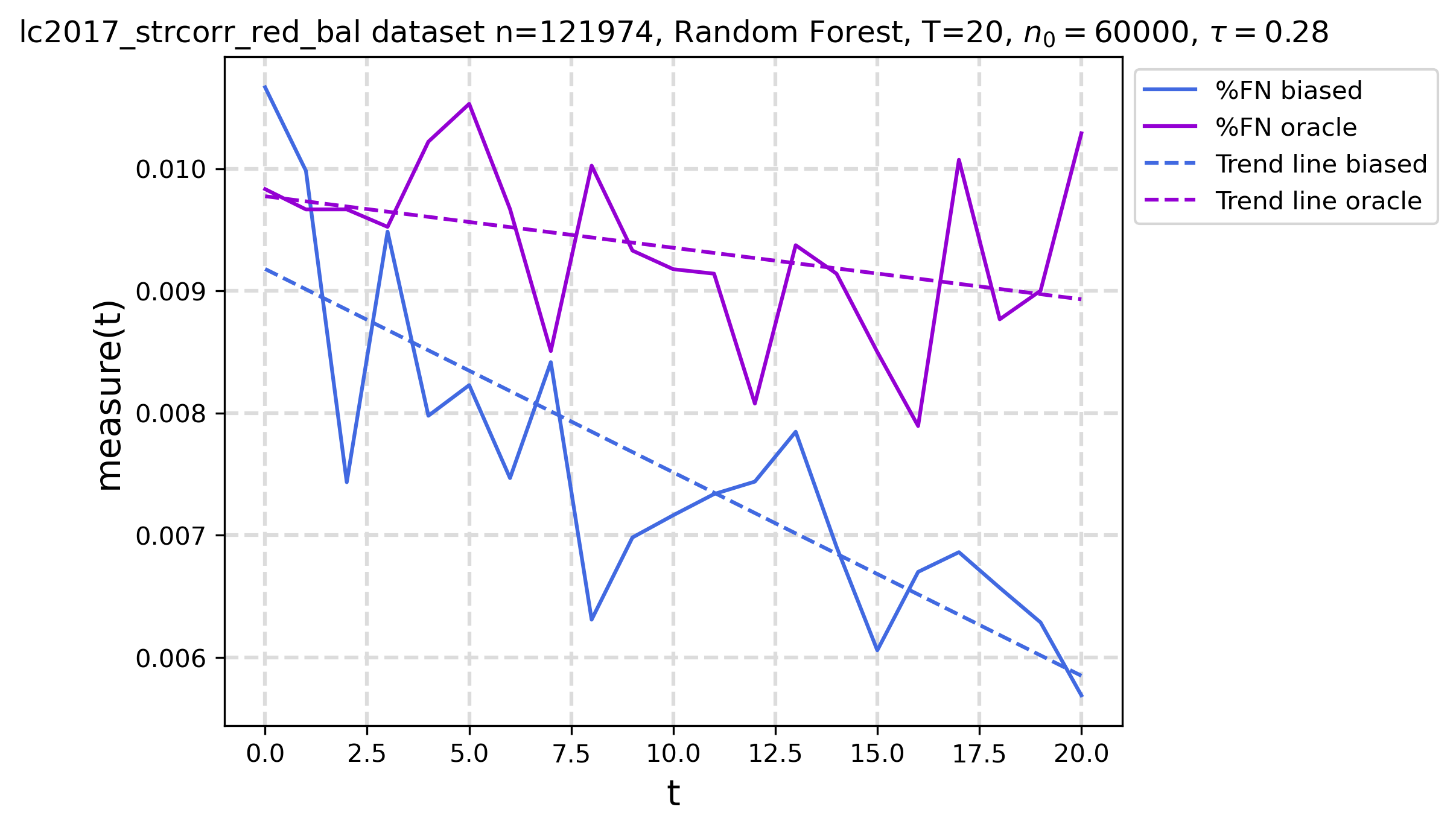}\\
    \includegraphics[scale=.25]{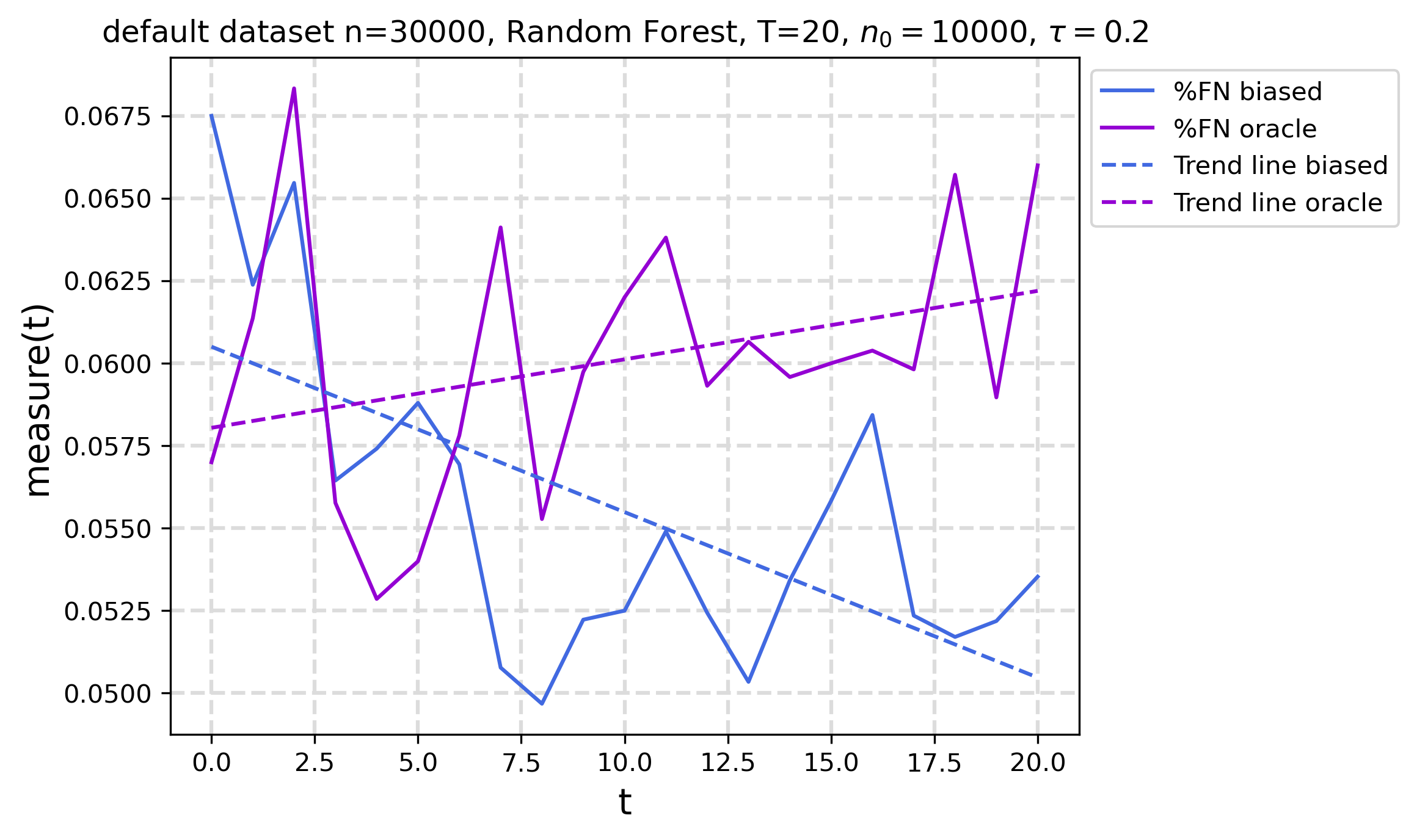} & \includegraphics[scale=.25]{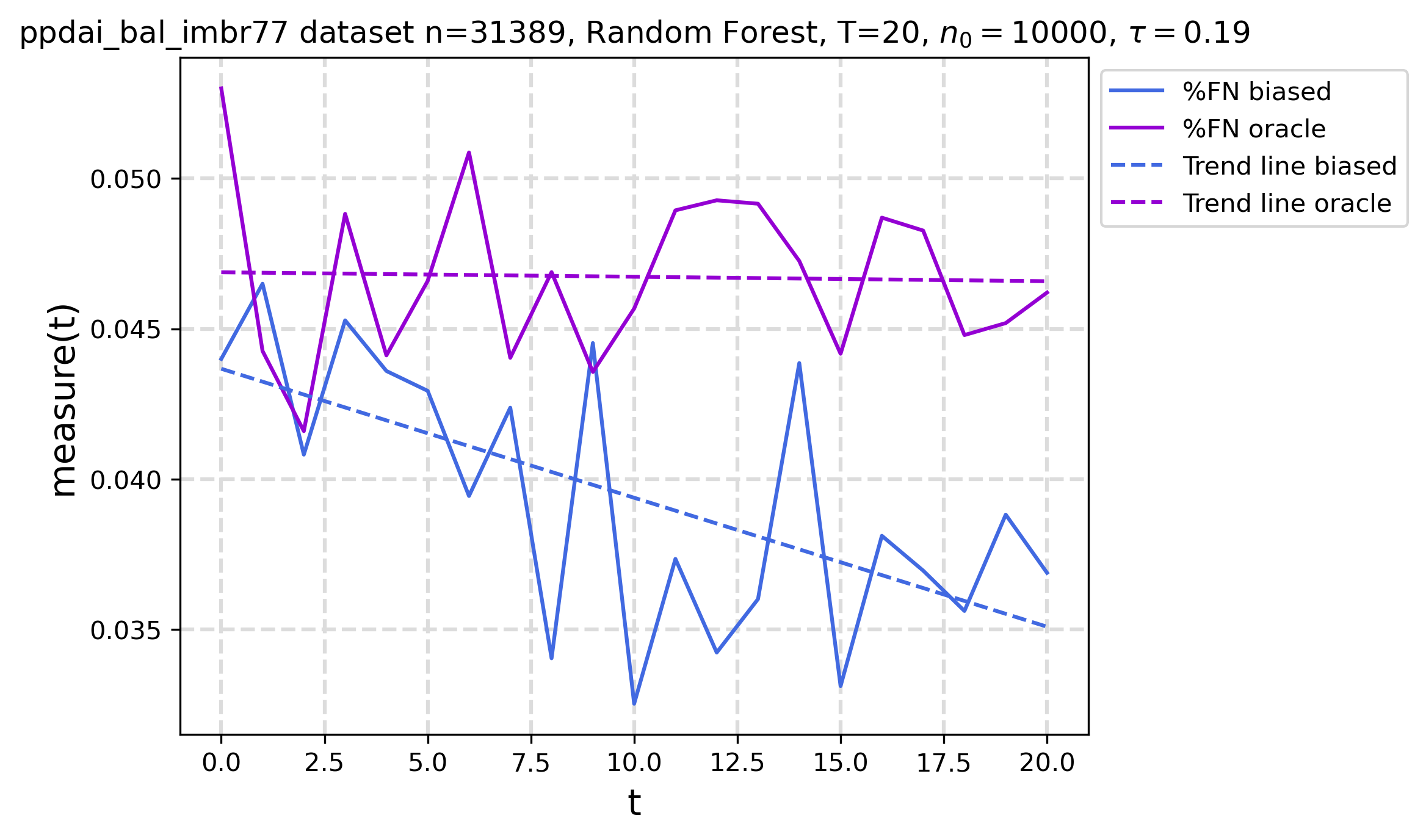} & \includegraphics[scale=.25]{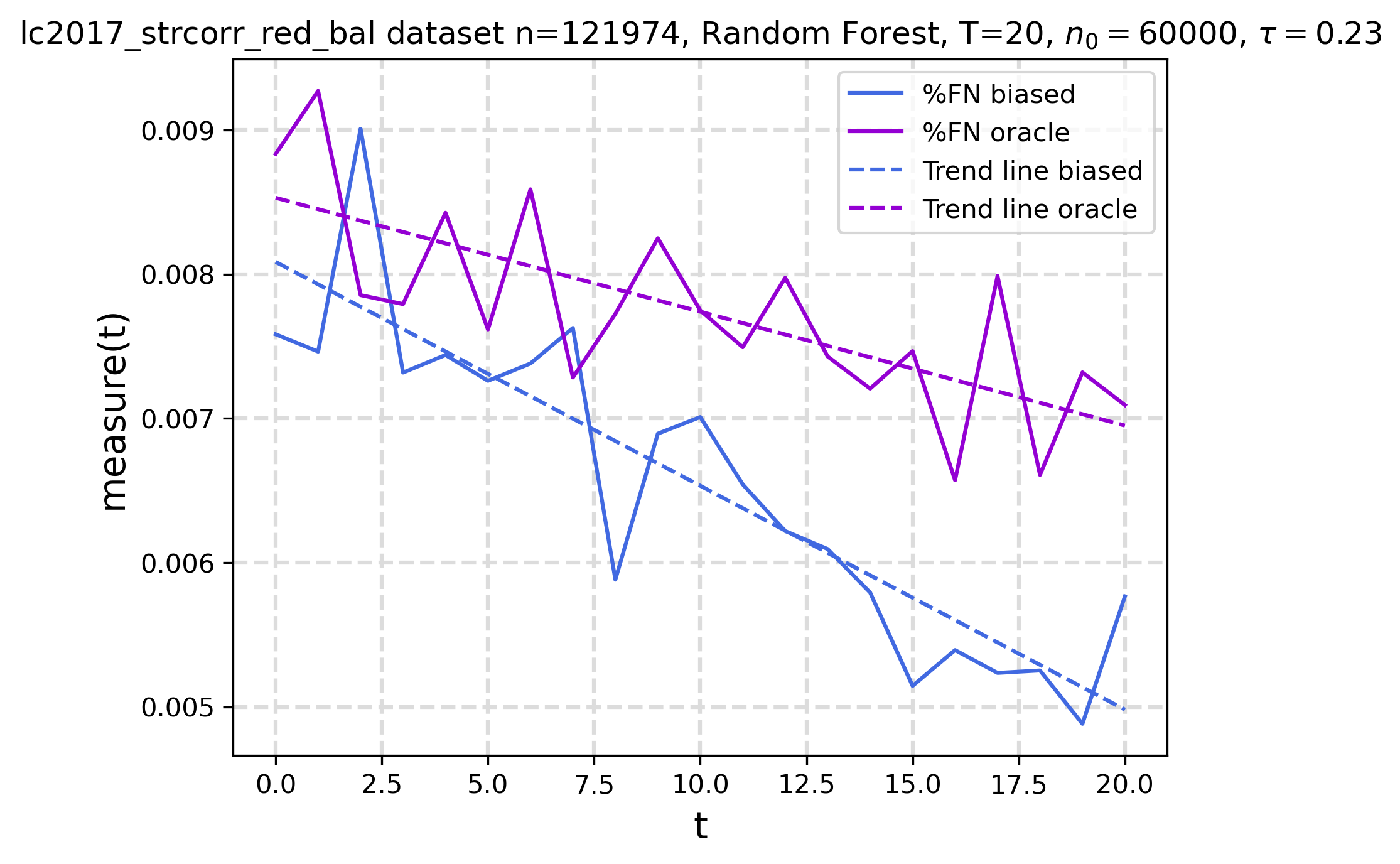}\\
    \end{tabular}
    \caption{Temporal evolution - Random Forest - Decision thresholds for $c \in \{1, 3, 5\}$ - Percentage of False Negatives.}
    \label{fig:rf_tempevol_perFN}
\end{figure*}

\clearpage

\begin{figure*}[t]
    \centering
    \begin{tabular}{ccc}
    Default & ppdai\_bal\_imbr77 & lc17\_bal\_imbr50\\
    \includegraphics[scale=.25]{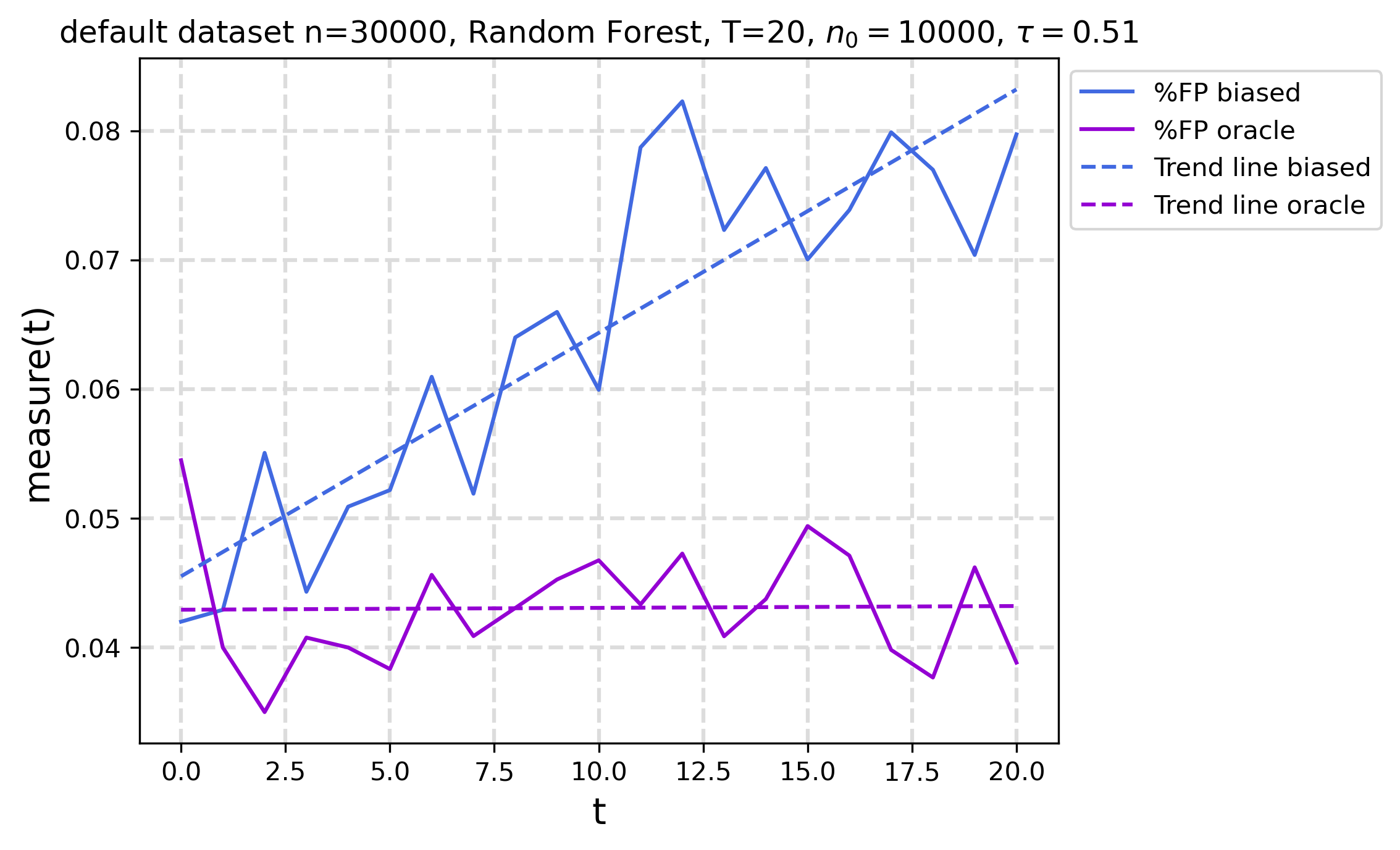} & \includegraphics[scale=.25]{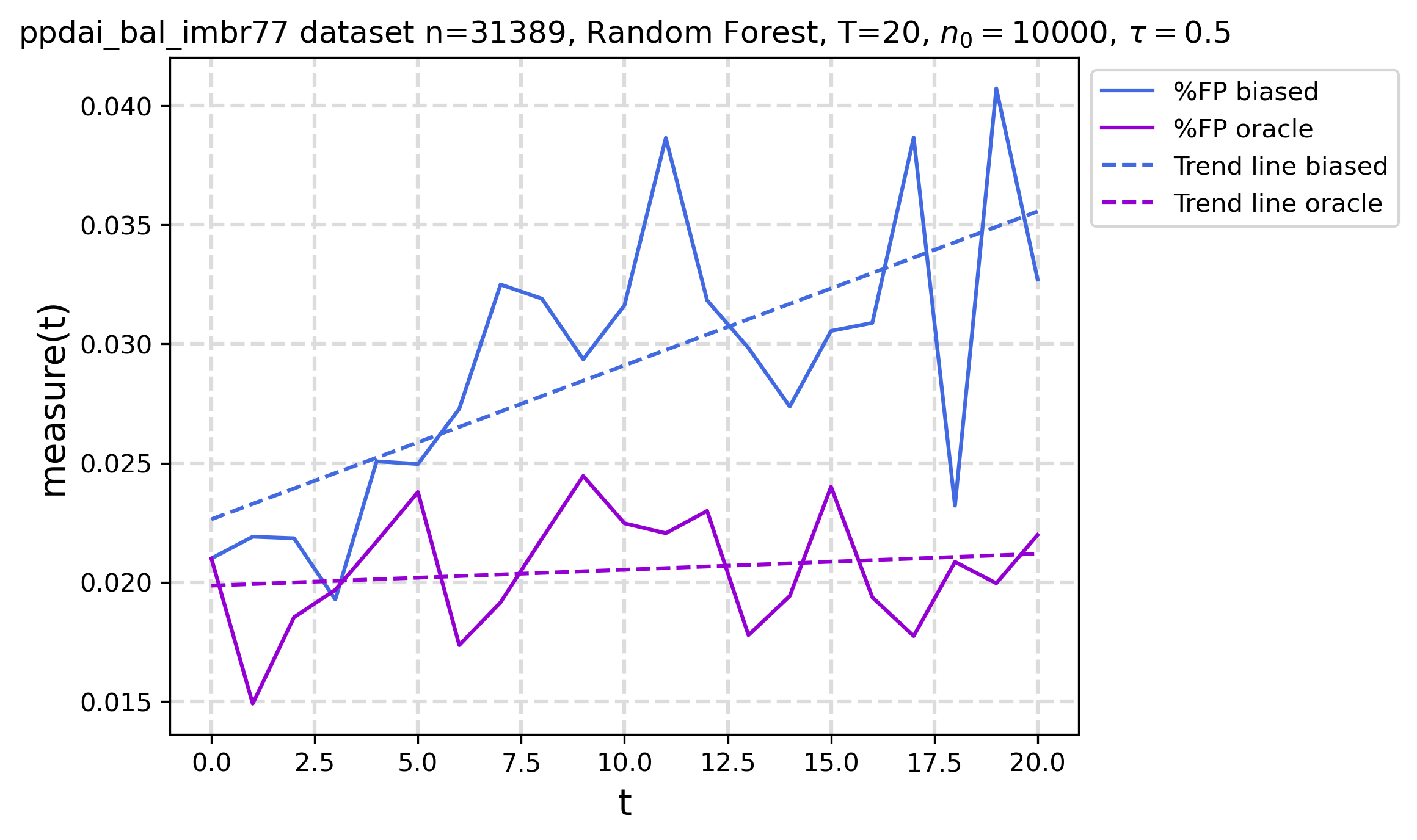} & \includegraphics[scale=.25]{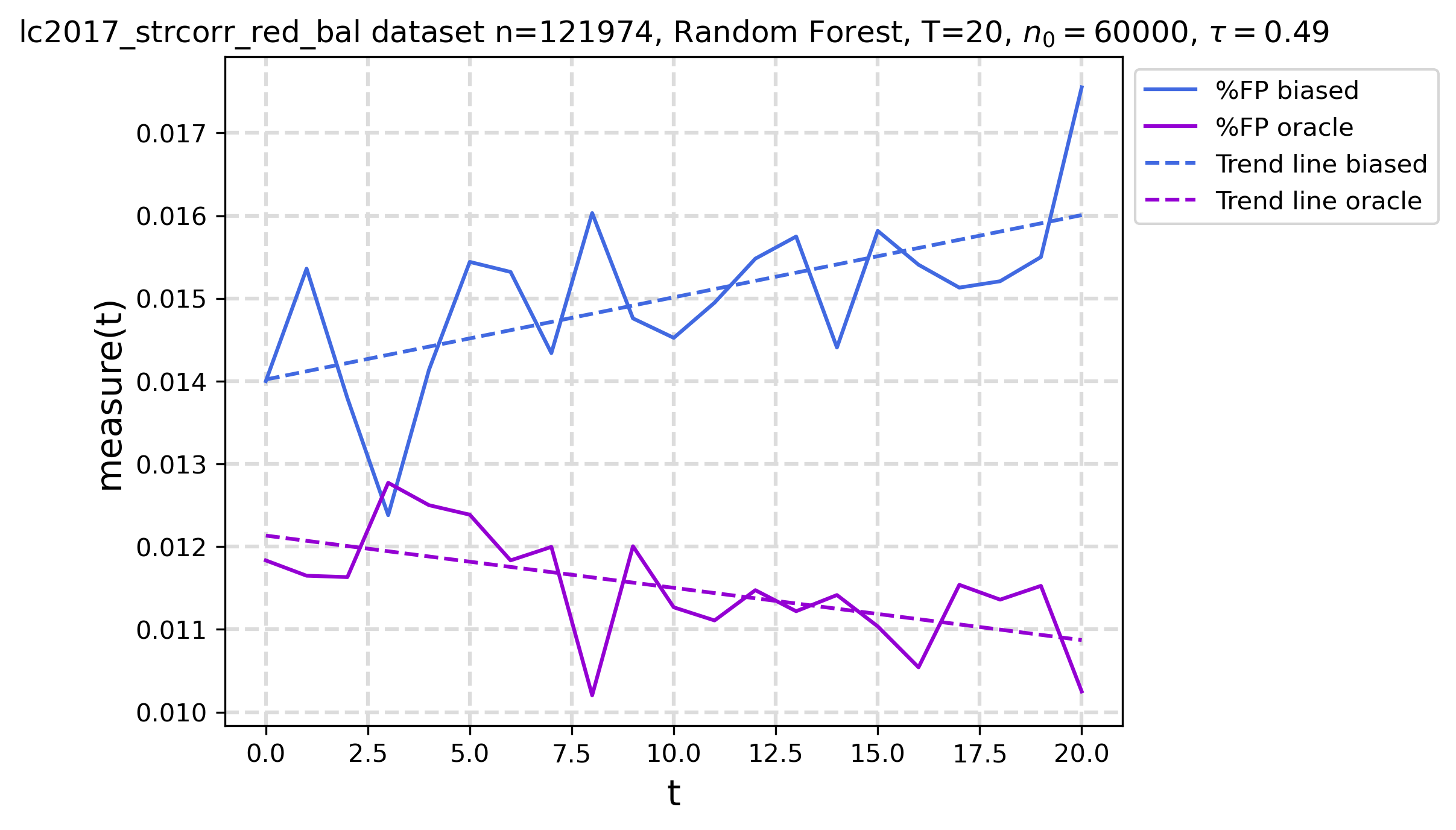}\\
  \includegraphics[scale=.25]{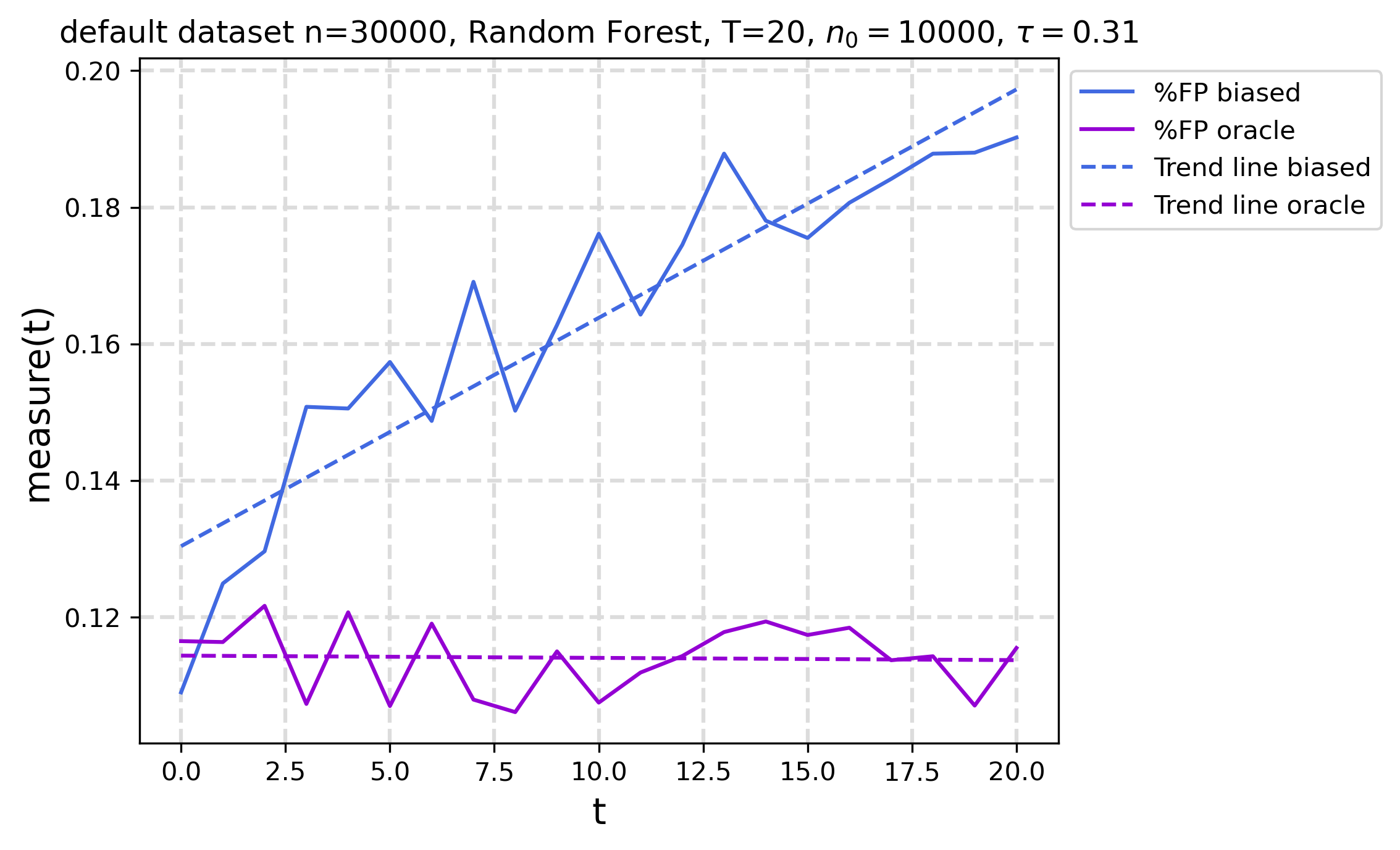} & \includegraphics[scale=.25]{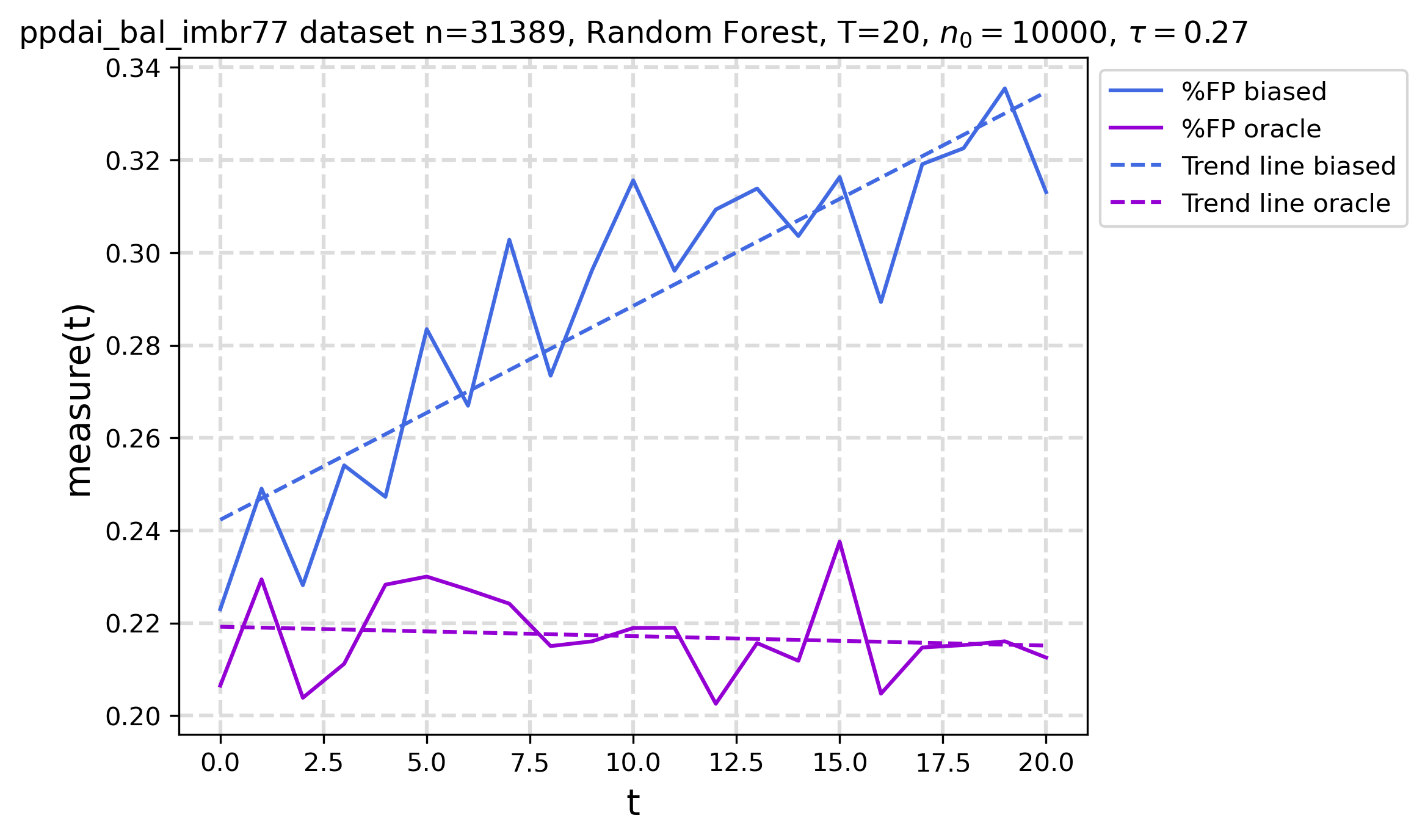} & \includegraphics[scale=.25]{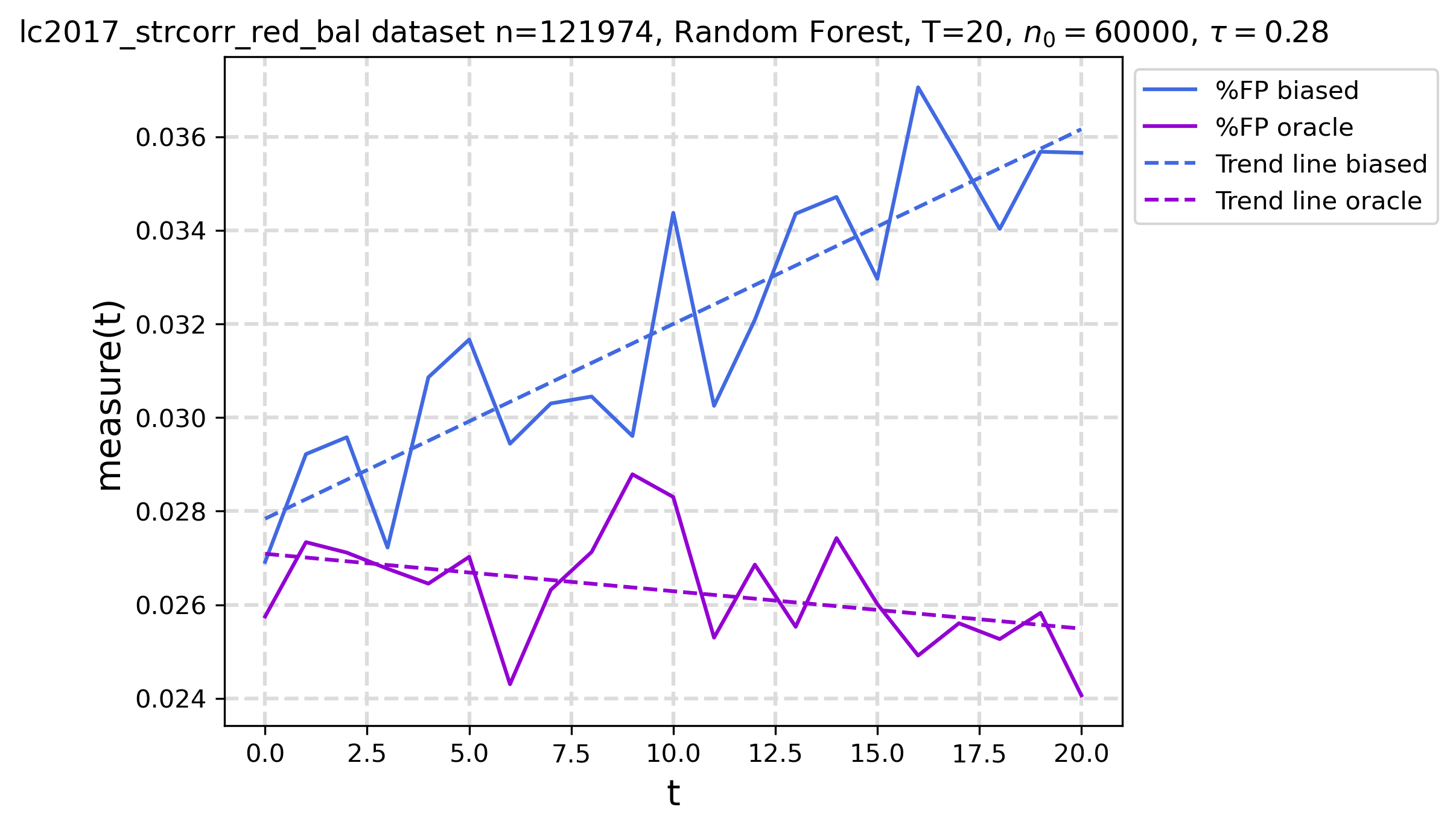}\\
    \includegraphics[scale=.25]{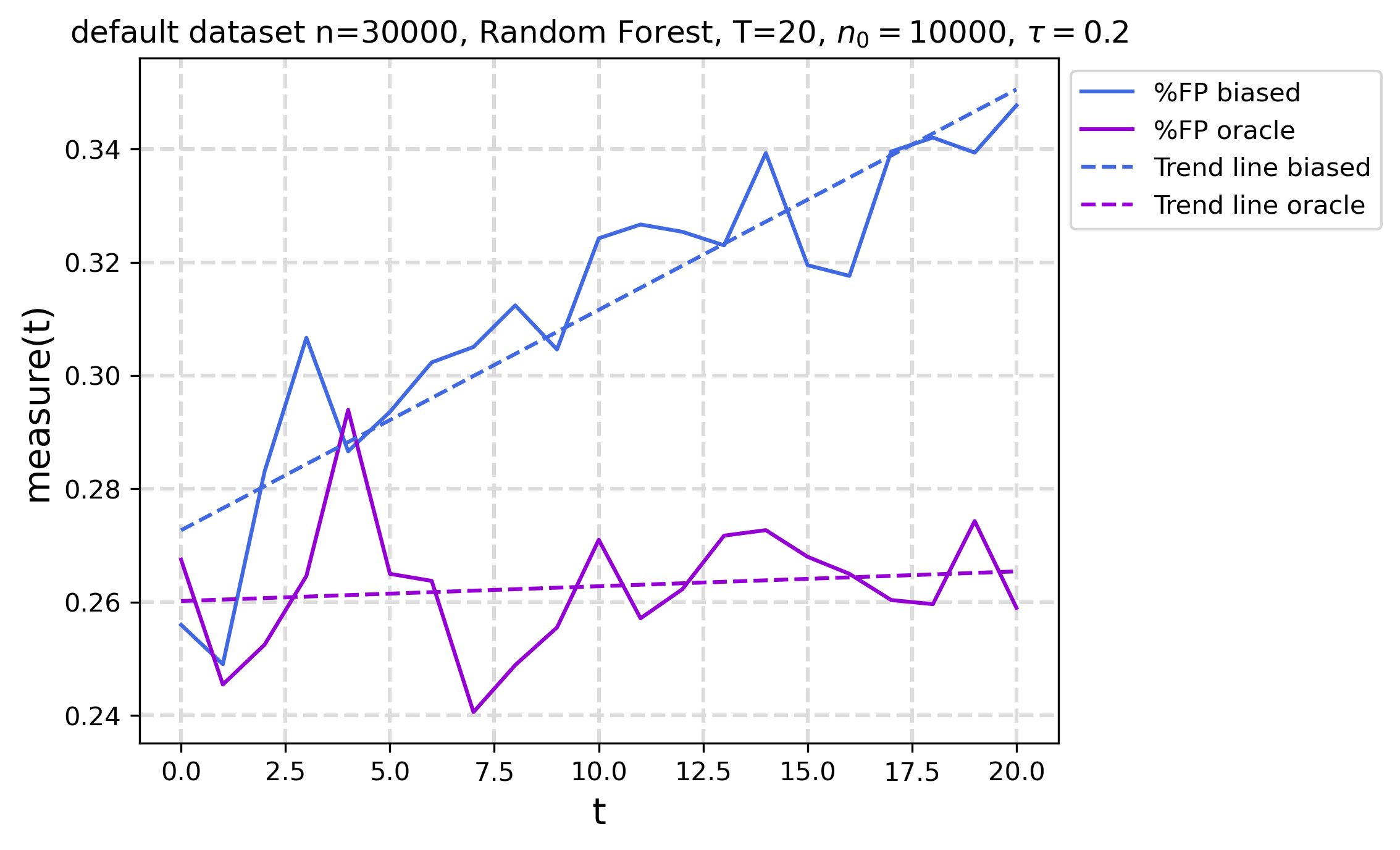} & \includegraphics[scale=.25]{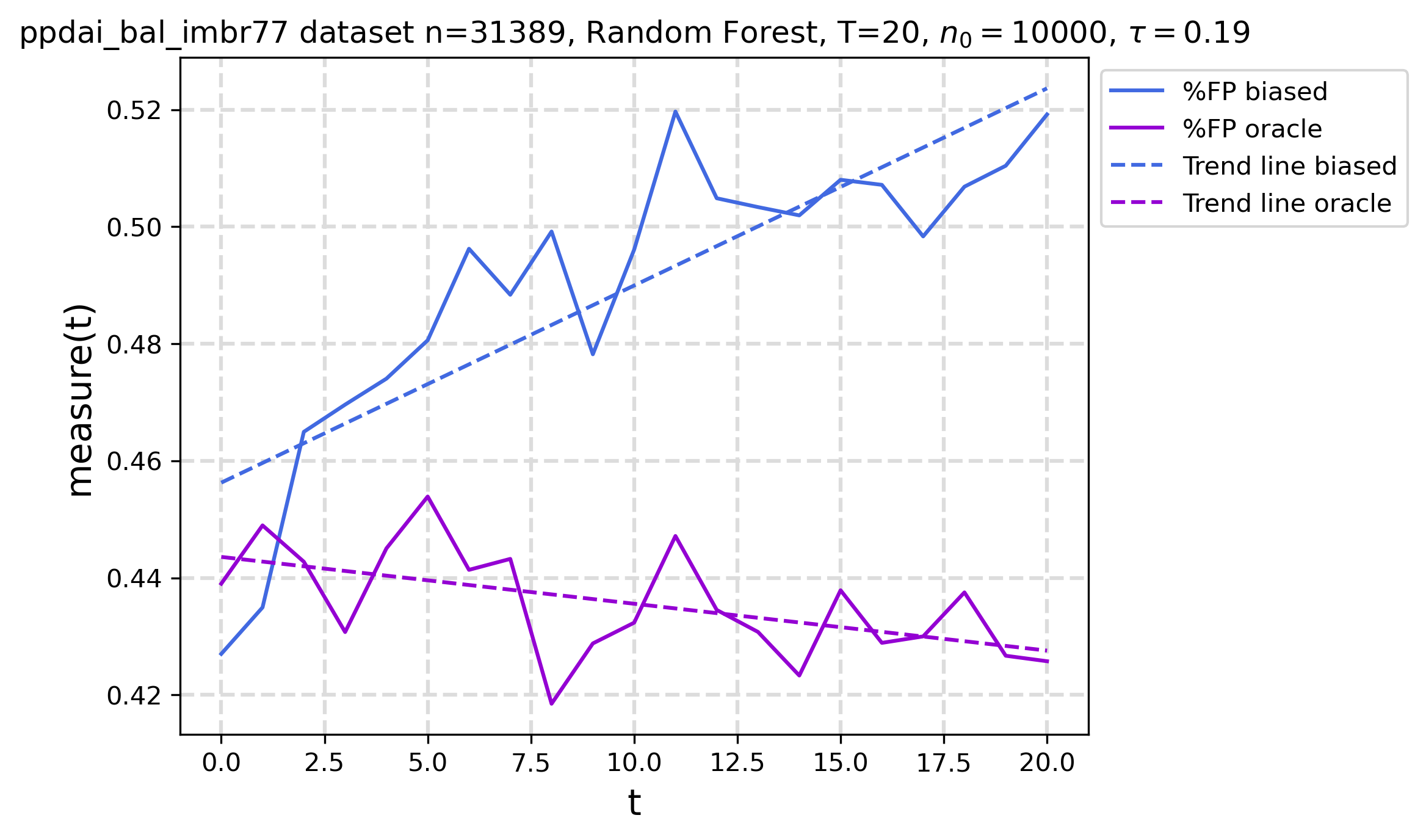} & \includegraphics[scale=.25]{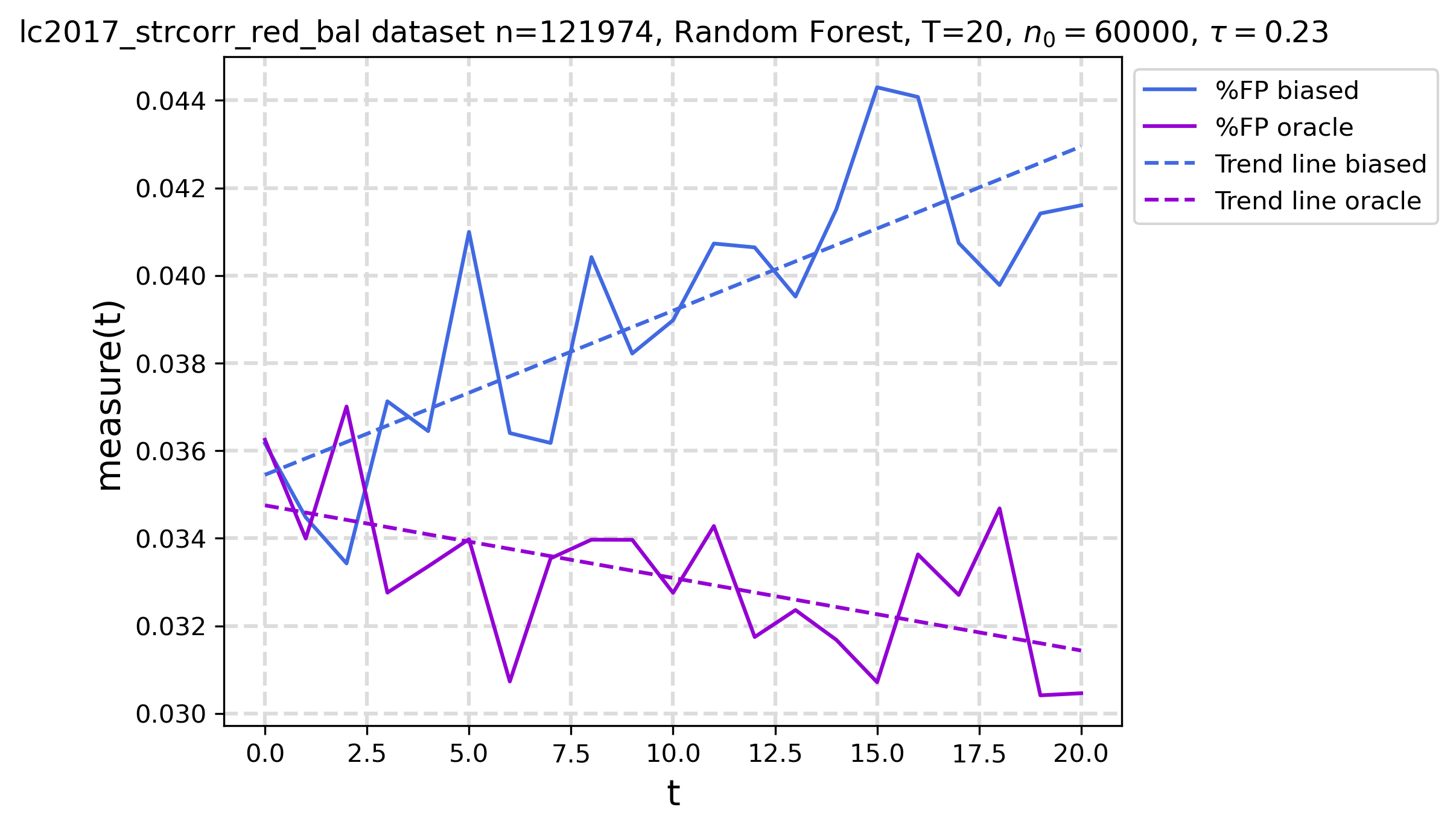}\\
    \end{tabular}
    \caption{Temporal evolution - Random Forest - Decision thresholds for $c \in \{1, 3, 5\}$ - Percentage of False Positives.}
    \label{fig:rf_tempevol_perFPs}
\end{figure*}

\begin{figure*}[t]
    \centering
    \begin{subfigure}{.81\textwidth}
    \begin{minipage}{.5\textwidth}
      \centering
      \includegraphics[scale=.28]{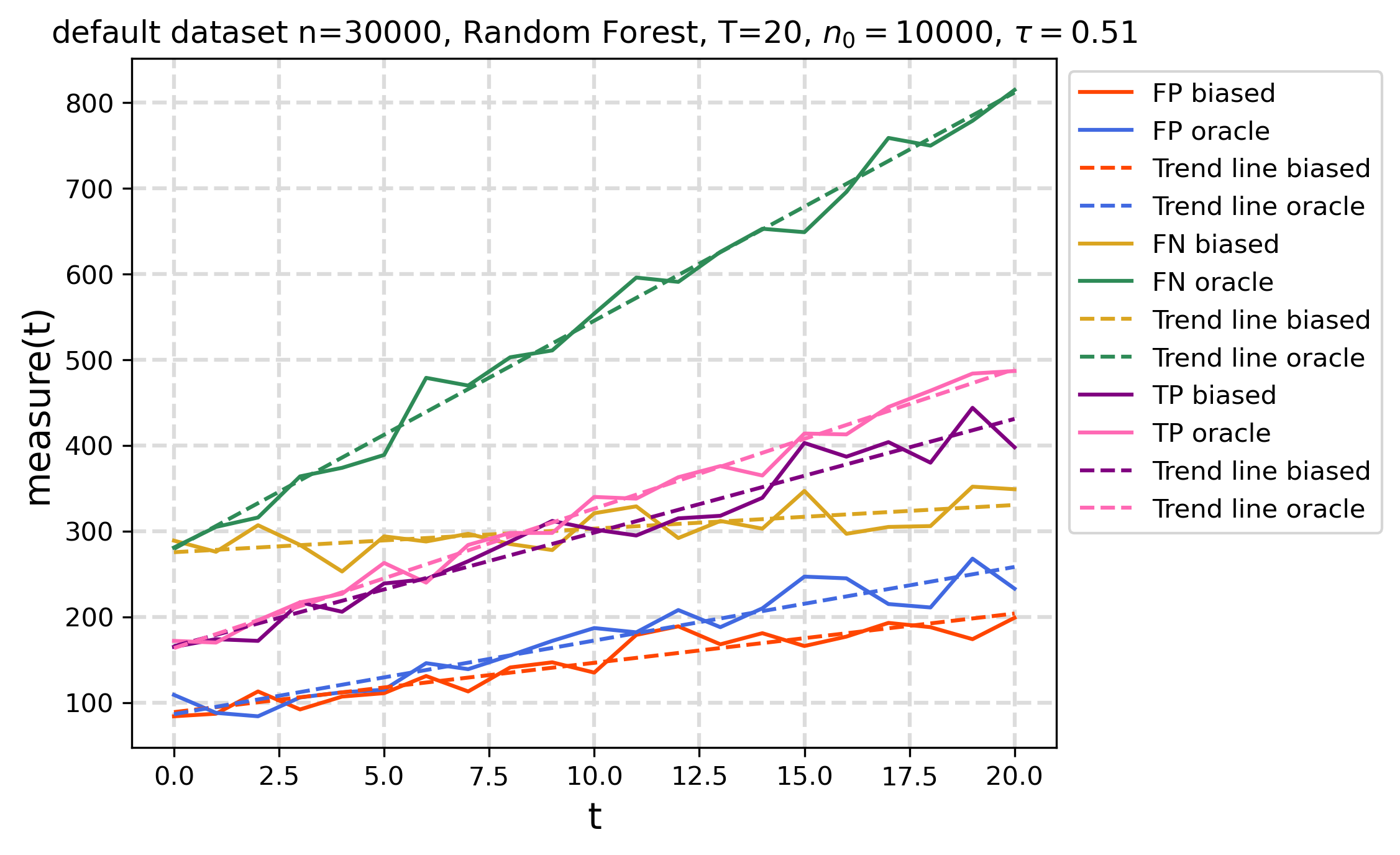}
      \caption{$\tau_{acc}=0.510$ - TPs, FNs, FPs}
    \end{minipage}%
    \begin{minipage}{.5\textwidth}
      \centering
      \includegraphics[scale=.28]{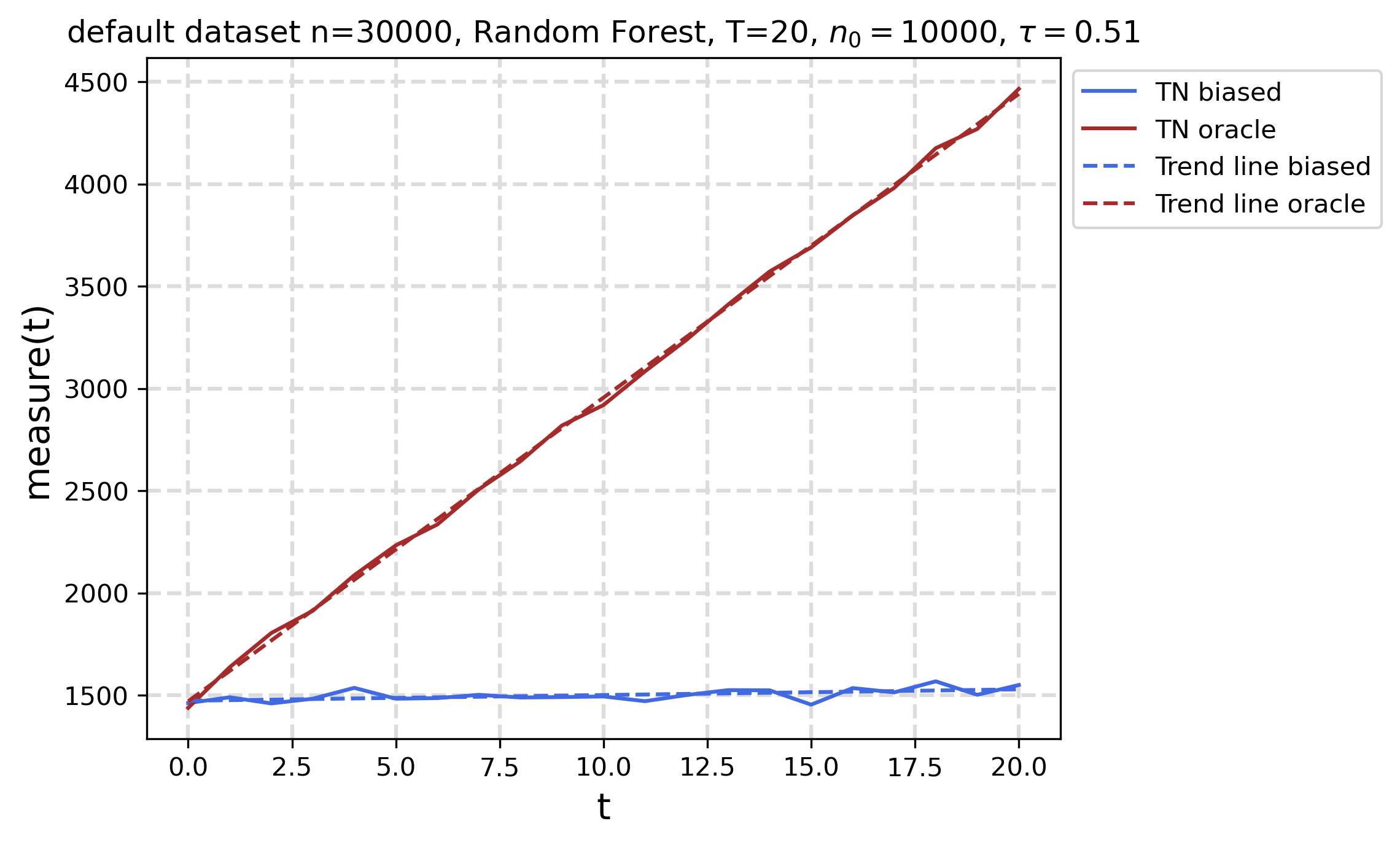}
      \caption{$\tau_{acc}=0.510$ - TNs}
    \end{minipage}
    \end{subfigure}\\
    \begin{subfigure}{.8\textwidth}
    \begin{minipage}{.5\textwidth}
      \centering
      \includegraphics[scale=.28]{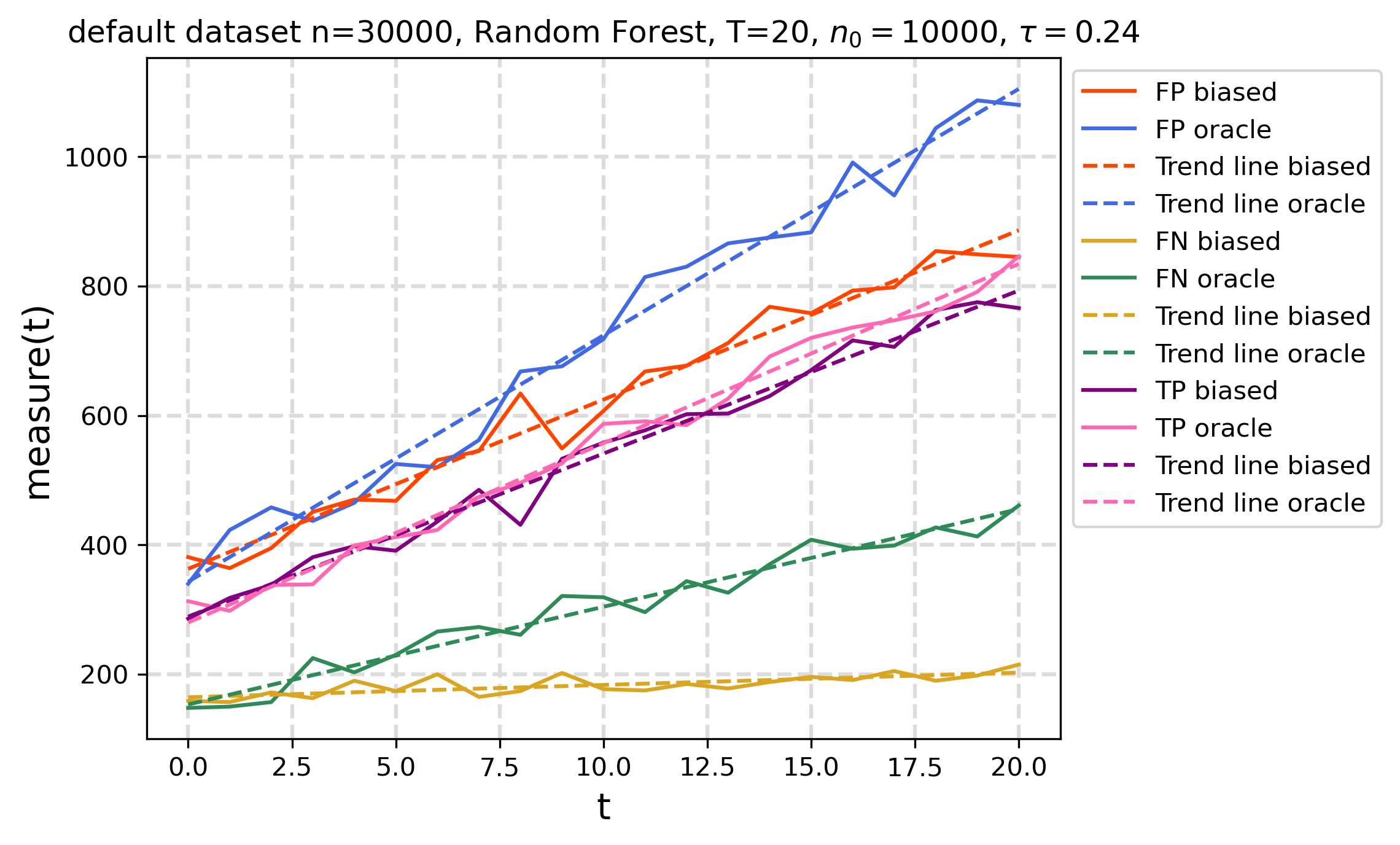}
      \caption{$\tau_{sc,c=4}=0.240$ - TPs, FNs, FPs}
    \end{minipage}%
    \begin{minipage}{.5\textwidth}
      \centering
      \includegraphics[scale=.28]{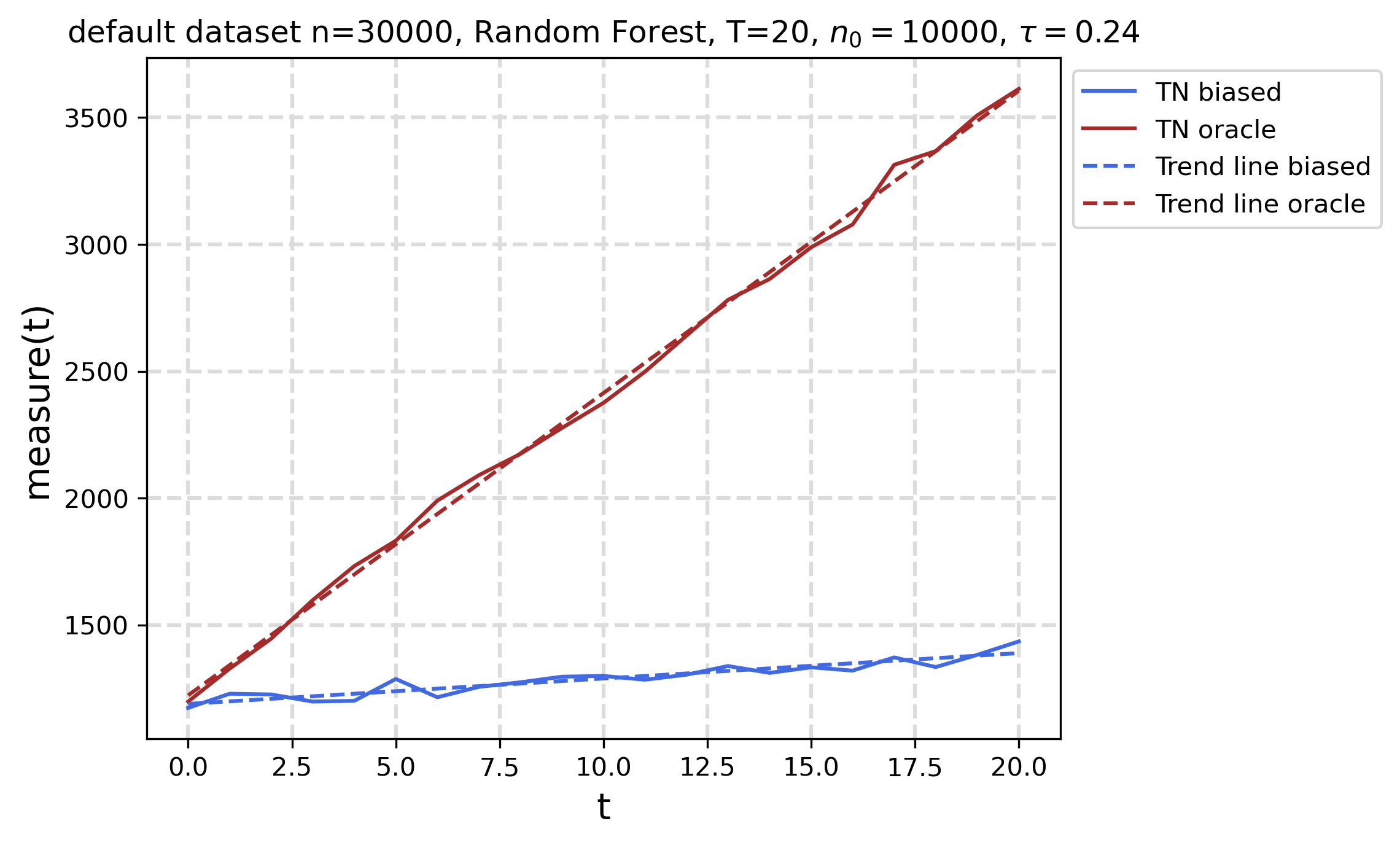} %
      \caption{$\tau_{sc,c=4}=0.240$ - TNs}
    \end{minipage}
    \end{subfigure}%
    \caption{Temporal evolution - Oracle vs Biased -  Random Forest \& Default datasets.}
    \label{fig:oraclevsfilt_tpfptnfn}
\end{figure*}

Then, the TNs of the Oracle models increase faster than the number of FNs, while the ones of the Biased models remain constant, which explains the accuracy gap. Precision is not directly affected by the filtering, and thus both types of model exhibit very similar behavior. Regarding recall, the number of FNs remains constant for the Biased model, but increases for the Oracle, which again explains the observed gap. In the Biased models, since no FNs are being added but tuples are, this results in a decrease in the percentage of FNs with time. Regarding the percentage of FPs for the Biased models, since no FNs nor TNs are being added in each iteration, given that FPs and TPs grow with similar rates, the model will proportionally have more FPs with time.

Perhaps the most interesting observation is that if an external observer analyzes the performance measures of the Biased models in production (\ie in a setting where the Oracle model is not observable), 
one would incorrectly conclude that the system is improving: precision either slightly increases or remains the same and recall is steady increasing over time and the degradation in accuracy could be attributed to alternative valid reasons (\eg behavioral or regulatory shifts in the market, more noise in the data, etc.), when in fact the quality of the training data (and thus of the trained models) is getting worse due to survival bias.

\subsection{Trade-off between Most Accurate and Socially Optimal models}\label{sec:acc_vs_sc}

Now we experimentally measure the accuracy loss when using the socially optimal models. We vary $\tau\in [0,1]$ using .0001 increments. For each $\tau$ we plot both the accuracy and the social cost for $c_{FN}\in\{1,\dots ,10\}$. The results for the RF model are shown in Table \ref{fig:rf_acc_sc_cFN_1_10}, and the ones for GBDT can be found in Appendix \ref{app:acc_vs_sc} (Table \ref{fig:gbdt_acc_sc_cFN_1_10}). The red curve in the charts connects the points having the minimal social cost and the corresponding accuracy for each value of $c_{FN}$. We include more values of $c_{FN}$ to have a more complete representation of the curve. The analytical results for the RF model and the different values of $c_{FN}$ are given in Table \ref{tab:tauacc_tausc_values_rf_lc}. The results for the other two datasets are included in the Additional Material.  %
We measure the increment in accuracy when using $\tau_{acc}$ relative to $\tau_{sc}$ as $\frac{acc(\tau_{acc})-acc(\tau_{sc})}{acc(\tau_{sc})}\cdot 100$. What is remarkably interesting is how short the red curve is for the case of the balanced LC dataset, meaning that even for high numbers of $c_{FN}$, the socially optimal model's accuracy loss is particularly small. Even for the case where an FN has five times greater impact than a FP, the accuracy loss is less than $1.3\%$ relative to the most accurate model. This encourages the use of this optimization criterion in model training.

\begin{figure*}[t]
    \begin{minipage}{.32\textwidth}
        \includegraphics[width=\linewidth]{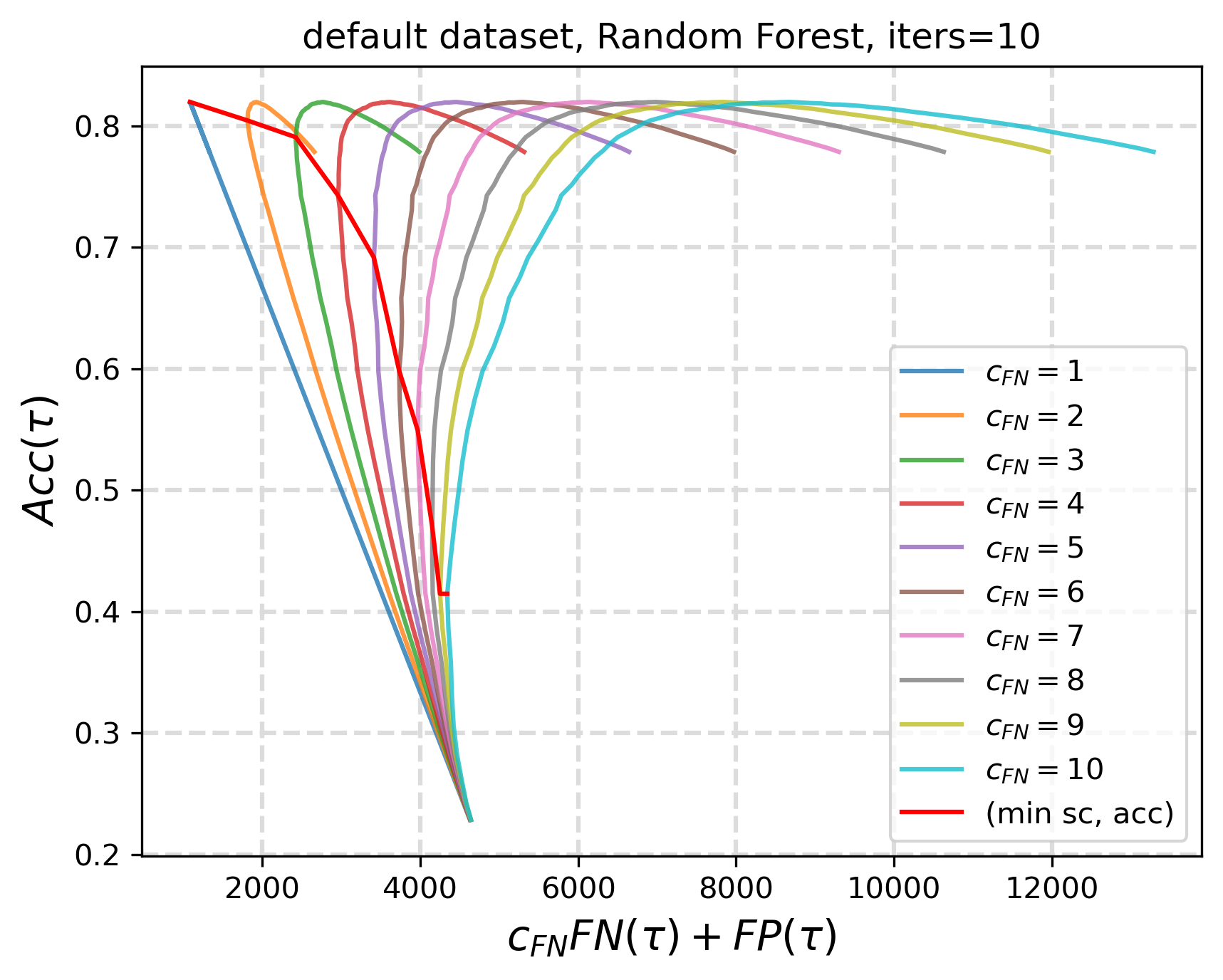}
    \end{minipage}
    \begin{minipage}{.32\textwidth}
        \includegraphics[width=\linewidth]{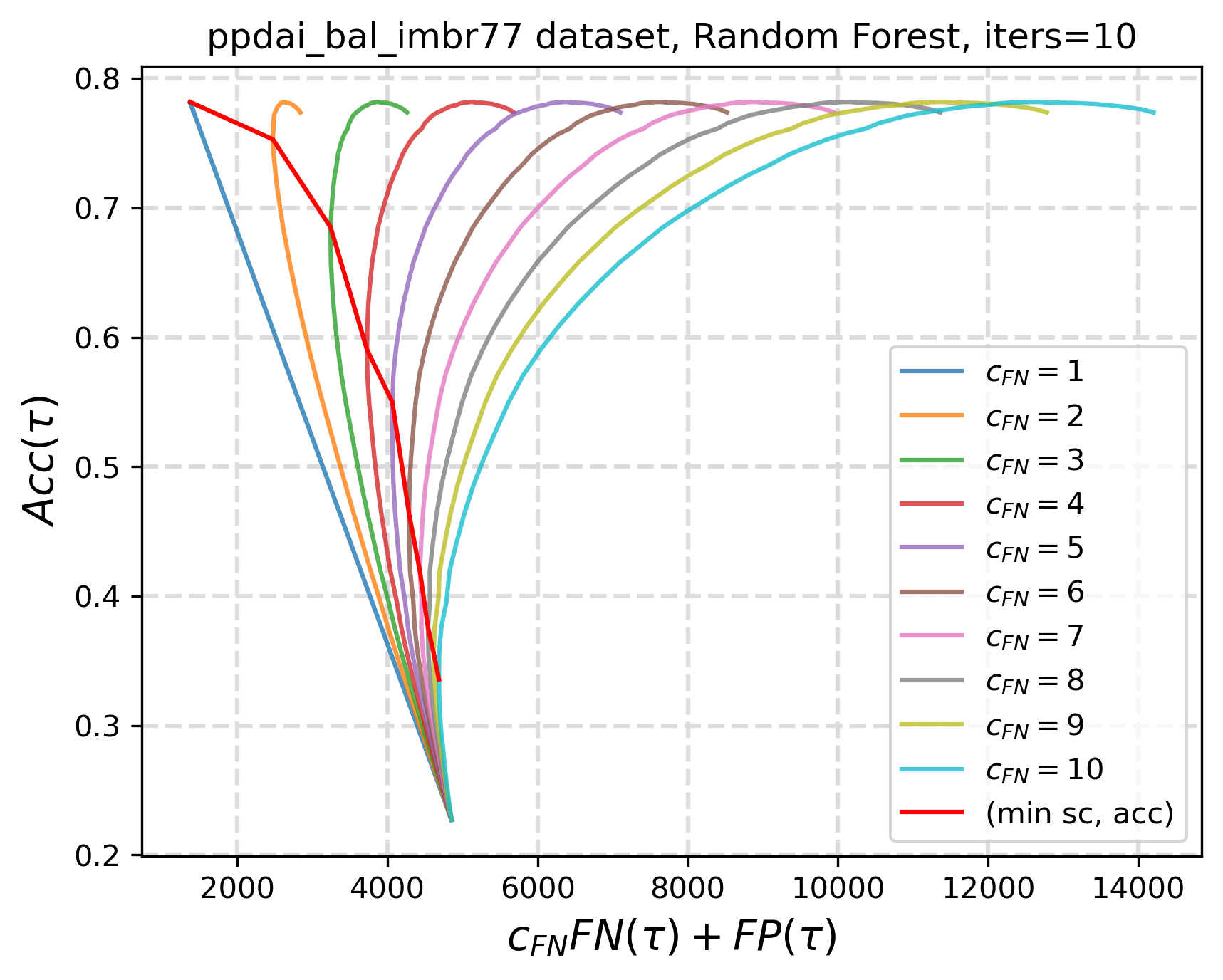}
    \end{minipage}
    \begin{minipage}{.32\textwidth}
        \includegraphics[width=\linewidth]{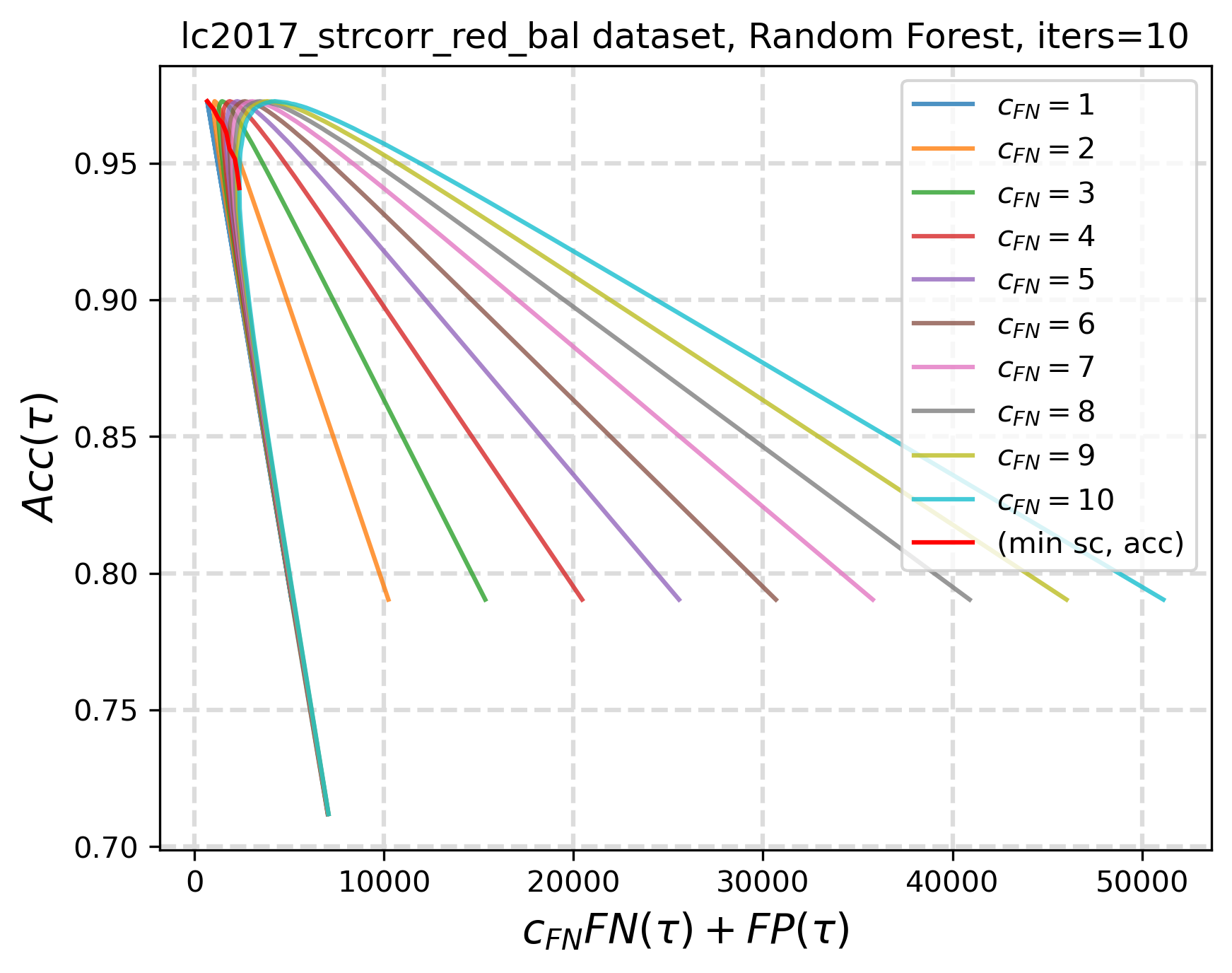}
    \end{minipage}
    \centering
    \caption{Accuracy vs social cost for $\tau\in [0,1]$ for the Random Forest model and all datasets.}
    \label{fig:rf_acc_sc_cFN_1_10}
\end{figure*}

\begin{table*}[t]
    \centering
    \begin{tabular}{|c|c|c|c|c|}
        \hline
        $c$ & Instance & $\tau$ & Acc($\tau$) & Relative diff. (\%)\\
        \hline
        $1,\dots,5$ & Max. acc. = Min. s.c. for $c=1$ & 0.490 & .9730 & 0\\
        \hline
        2 & Min. s.c. & 0.340 & 0.970 & 0.309\\
        \hline
        3 & Min. s.c. & 0.280 & 0.966 & 0.725\\
        \hline
        4 & Min. s.c. & 0.260 & 0.965 & 0.829\\
        \hline
        5 & Min. s.c. & 0.230 & 0.961 & 1.249\\
        \hline
    \end{tabular}
    \caption{Accuracy loss for socially optimal models, for the Random Forest model and the balanced LC dataset. Values are rounded to three decimal places. }
    \label{tab:tauacc_tausc_values_rf_lc}
\end{table*}

\section{Conclusions}
\label{sec:conclusions}

 Our work shows the importance of {\em legitimacy and competence}, the first and new instrumental principle of the ACM Principles for Responsible Algorithmic Systems \cite{ACM22}, in the context of predictive ML applications related to human behavior. In fact, current predictive models are really just heuristics that do not work for all people.  
 
 One main issue is pseudoscience \cite{pseudo2024} that affects the design and evaluation of ML models, triggering ethical, legal and technical issues that may have harmful societal impact. In our case the pseudoscience is reflected in using data from existing clients to assess the risk of new clients. 
 
 A second issue is technical mistakes, like not considering survival bias or not using the best decision threshold to optimize the target measure (accuracy in our case). Third, evaluating wrongly by considering a simple average, like accuracy, to take a decision. Indeed, in most binary classification applications, all errors do not have the same impact, so we really need to weight carefully the cost of false positives and false negatives, as we have done here. 
  
As a side product of our work, we obtain competitive models for risk prediction for different bank lending datasets. We plan to improve our feature selection even further in the near future and also find analytic expressions for the different types of predictions and thus be able to characterize the quantities of interest for a more in-depth study. We will also mitigate survival bias to verify how our results extend to that more complex case.

Finally, the simulation methodology used in this paper can be applied to many cases where enough data is available. In the future we want to use the same idea to study well-known AI incidents from the point of view of pseudo-science and/or bias.

\bibliography{bibliography}

\clearpage

\section{Additional Experimental Results}

In this appendix, we include complementary experimental results of our work. 

\subsection{Decision thresholds for different costs}\label{subsec:thresholds_acc_sc}

Table \ref{tab:taus_cin1_5} lists the complete set of decision thresholds for all models and datasets.

\subsection{Temporal Evolution}
\label{app:tempevol}

In this section, we include the temporal evolution results for the five measures considered for the case of the GBDT models: Accuracy (Figure \ref{fig:gbdt_tempevol_accuracy}), Precision (Figure \ref{fig:gbdt_tempevol_precision}) and Recall (Figure  \ref{fig:gbdt_tempevol_recall}), percentage of FNs (Figure \ref{fig:gbdt_tempevol_perFN}) and percentage of FPs (Figure \ref{fig:gbdt_tempevol_perFP}).

\subsection{Trade-off between Most Accurate and Socially Optimal models}\label{app:acc_vs_sc}

Figure \ref{fig:gbdt_acc_sc_cFN_1_10} shows the curves for the accuracy and social cost values for different thresholds for the GBDT algorithm. The analytical values for the RF model are shown in Table \ref{tab:tauacc_tausc_values_rf_default} for the Default dataset and in Table \ref{tab:tauacc_tausc_values_rf_ppdai} for the PPDai dataset. 

\vspace*{1cm}

\begin{table}[h]
    \begin{tabular}{|c|c|c|c|c|c|c|}
        \hline
        Model & Default &  & ppdai\_bal\_imbr77 &  & lc17\_bal\_imbr50 & -\\
        \hline
        Random Forest & $\tau_{acc}=\tau_{sc,c=1}$ & .510 & $\tau_{acc}$ & .500 & $\tau_{acc}=\tau_{sc,c=1}$ & .490\\
        \hline
        Random Forest & $\tau_{sc,c=2}$ & .350 & $\tau_{sc,c=2}$ & .350 & $\tau_{sc,c=2}$ & .340\\
        \hline
        Random Forest & $\tau_{sc,c=3}$ & .310 & $\tau_{sc,c=3}$ & .270 & $\tau_{sc,c=3}$ & .280\\
        \hline
        Random Forest & $\tau_{sc,c=4}$ & .240 & $\tau_{sc,c=4}$ & .210 & $\tau_{sc,c=4}$ & .260\\
        \hline
        Random Forest & $\tau_{sc,c=5}$ & .200 & $\tau_{sc,c=5}$ & .190 & $\tau_{sc,c=5}$ & .230\\
        \hline
        GBDT & $\tau_{acc}=\tau_{sc,c=1}$ & .512 & $\tau_{acc}$ & .499 & $\tau_{acc}=\tau_{sc,c=1}$ & .534\\
        \hline
        GBDT & $\tau_{sc,c=2}$ & .313 & $\tau_{sc,c=2}$ & .289 & $\tau_{sc,c=2}$ & .303\\
        \hline
        GBDT & $\tau_{sc,c=3}$ & .220 & $\tau_{sc,c=3}$ & .241 & $\tau_{sc,c=3}$ & .211\\
        \hline
        GBDT & $\tau_{sc,c=4}$ & .213 & $\tau_{sc,c=4}$ & .222 & $\tau_{sc,c=4}$ & .156\\
        \hline
        GBDT & $\tau_{sc,c=5}$ & .176 & $\tau_{sc,c=5}$ & .185 & $\tau_{sc,c=5}$ & .141\\
        \hline
    \end{tabular}
    \caption{Decision thresholds for $c \in \{1,\dots,5\}$.}
    \label{tab:taus_cin1_5}
\end{table}

\begin{table}[h]
    \centering
    \begin{tabular}{|c|c|c|c|c|}
        \hline
        $c$ & Instance & $\tau$ & Acc($\tau$) & Relative diff. (\%)\\
        \hline
        $1,\dots,5$ & Max. acc. = Min. s.c. for $c=1$ & 0.510 & 0.820 & 0\\
        \hline
        2 & Min. s.c. & 0.350 & 0.804 & 1.990\\
        \hline
        3 & Min. s.c. & 0.310 & 0.791 & 3.666\\
        \hline
        4 & Min. s.c. & 0.240 & 0.743 & 10.363\\
        \hline
        5 & Min. s.c. & 0.200 & 0.691 & 18.669\\
        \hline
    \end{tabular}
    \caption{Accuracy loss for socially optimal models, for the Random Forest model and the  Default dataset. Values are rounded to three decimal places. }
    \label{tab:tauacc_tausc_values_rf_default}
\end{table}

\clearpage

\begin{figure*}[t]
    \centering
    \begin{tabular}{ccc}
        Default & ppdai\_bal\_imbr77 & lc17\_bal\_imbr50\\
        \includegraphics[scale=.25]{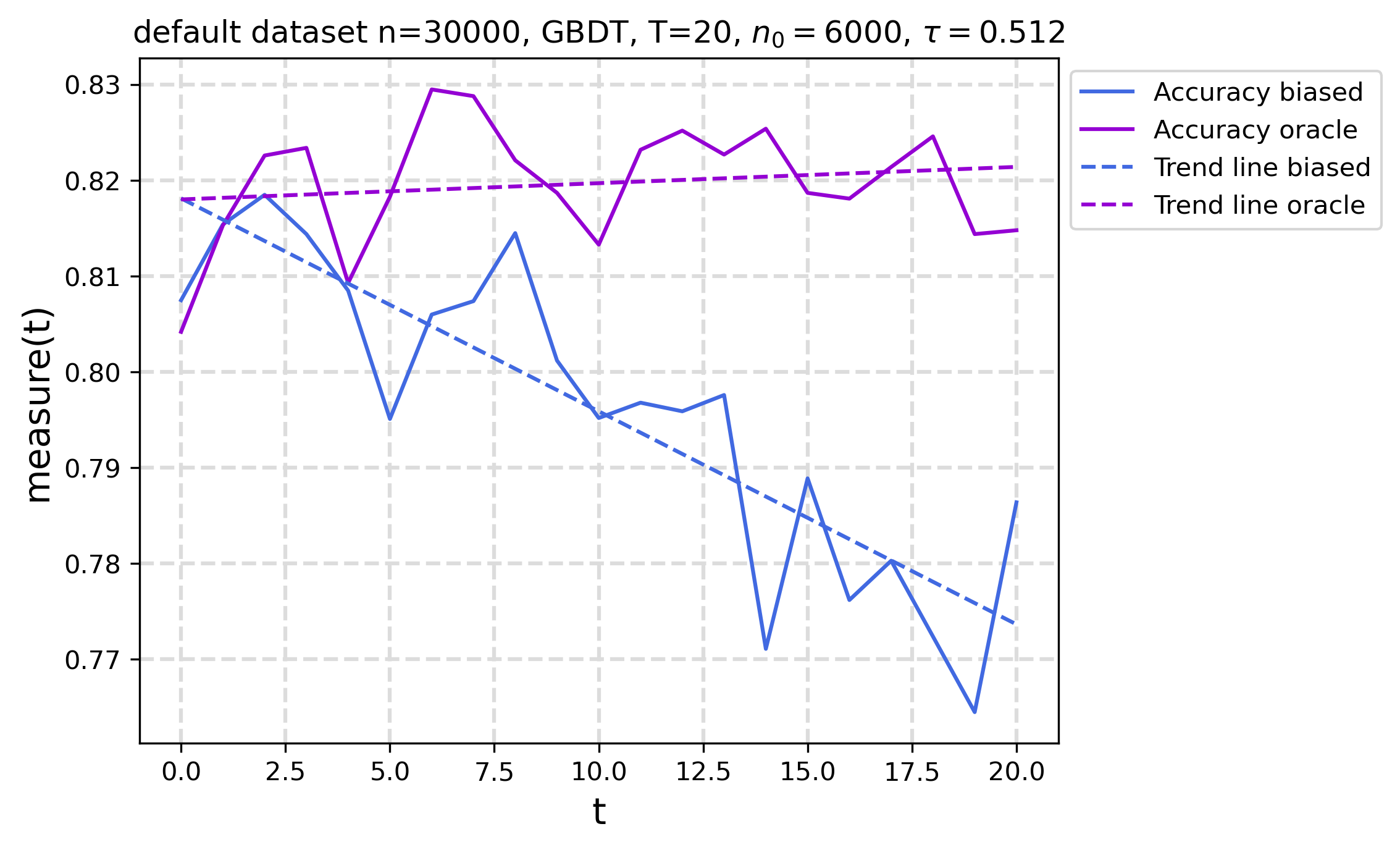} & \includegraphics[scale=.25]{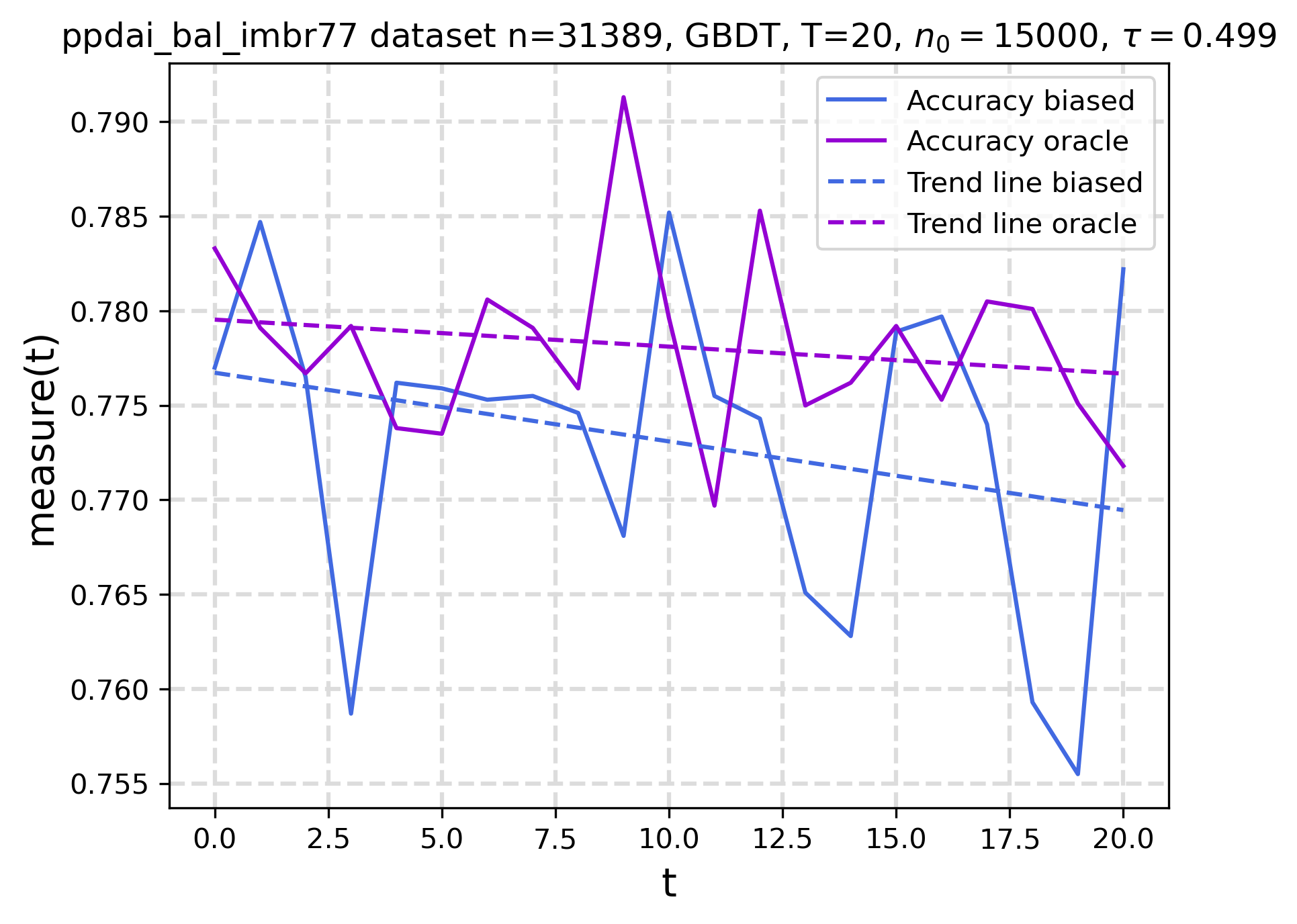} & \includegraphics[scale=.25]{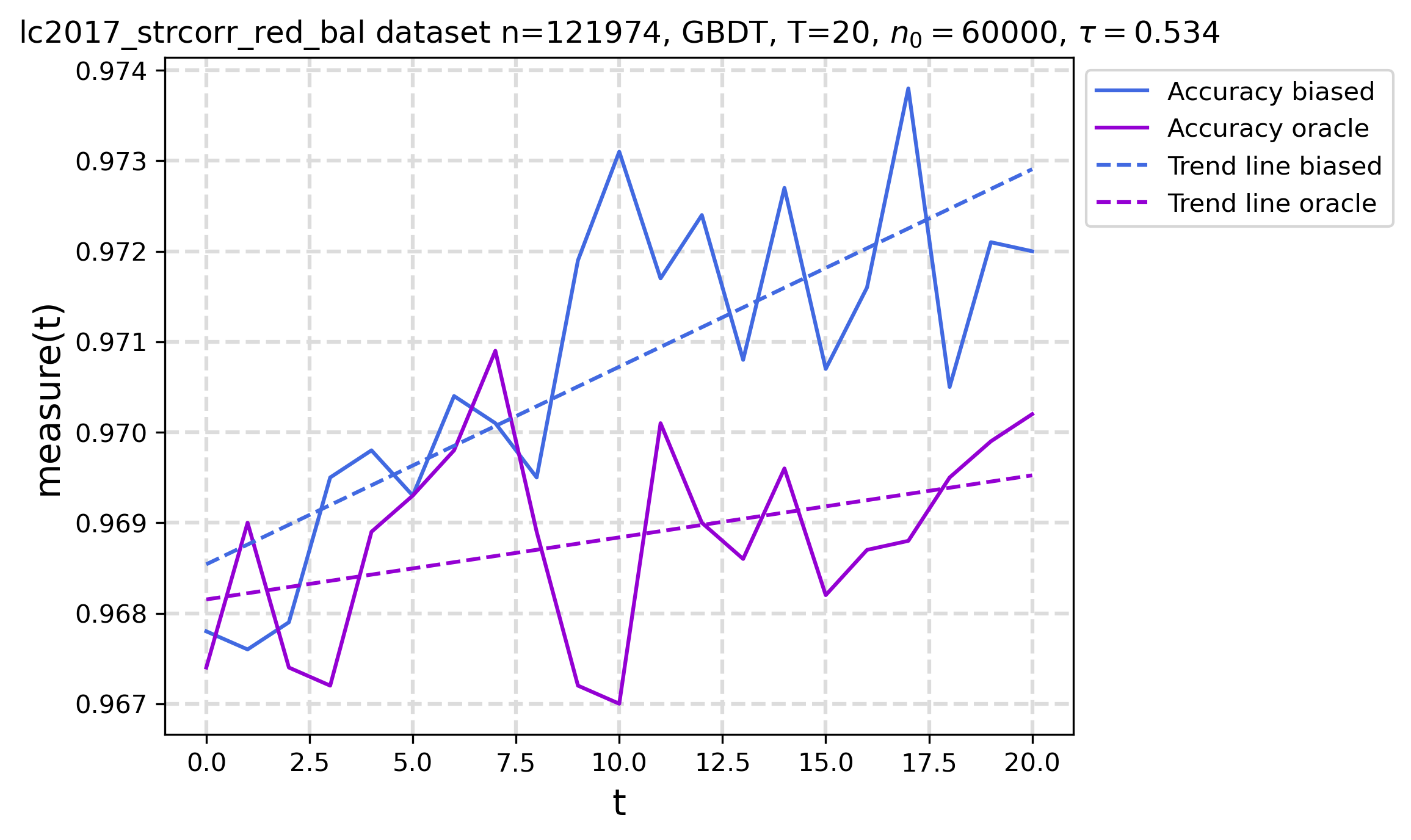}\\
       \includegraphics[scale=.25]{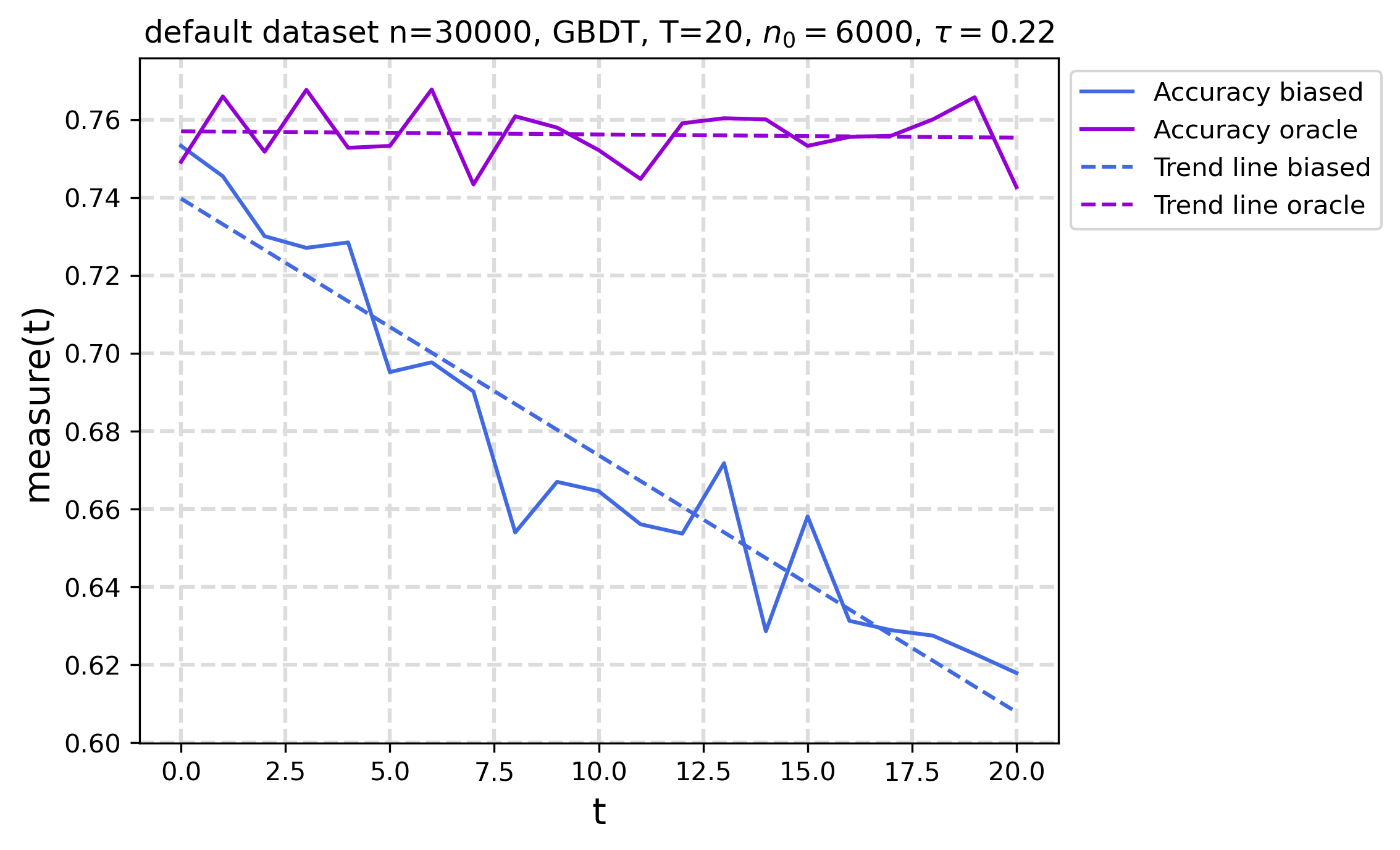} & \includegraphics[scale=.25]{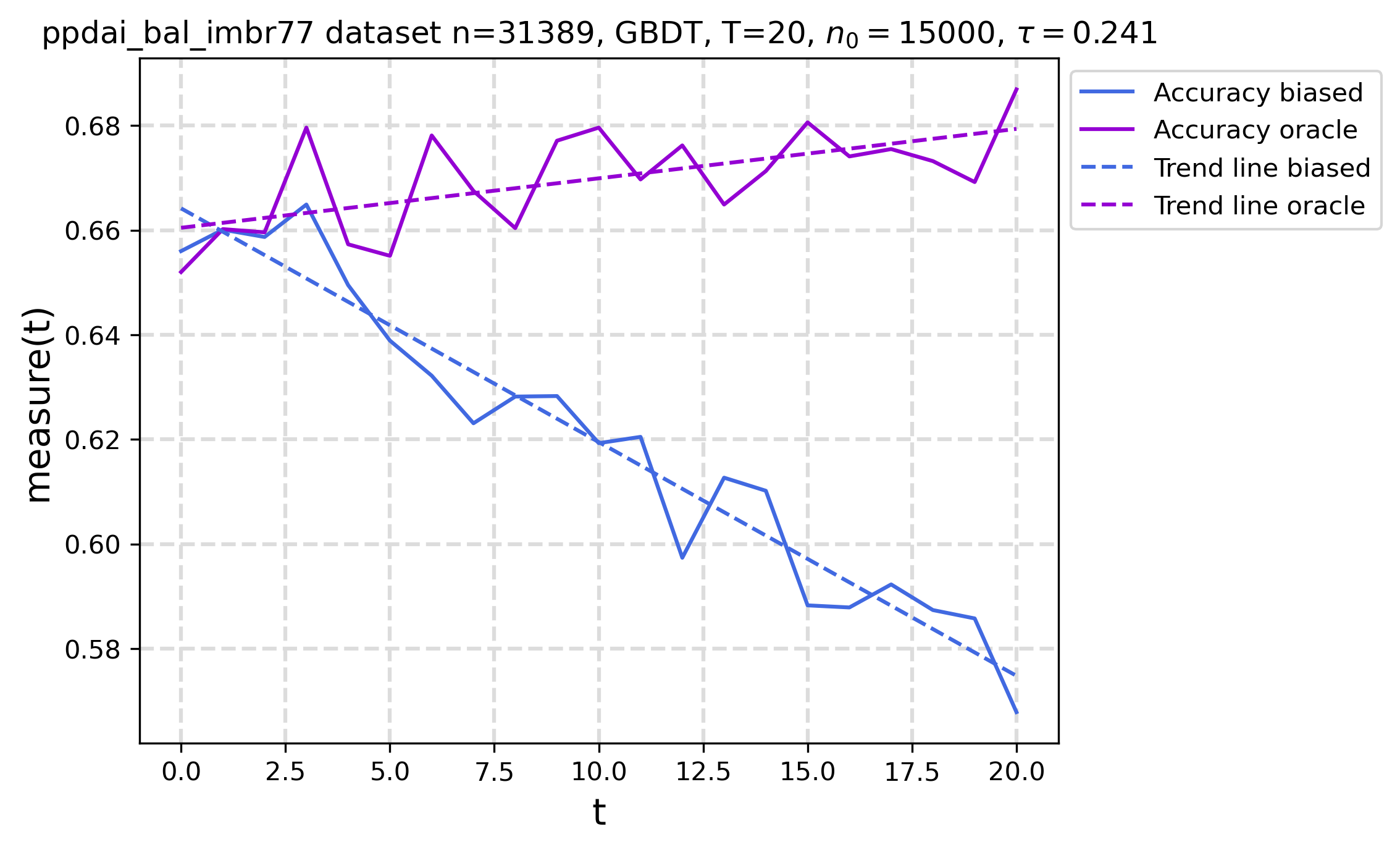} & \includegraphics[scale=.25]{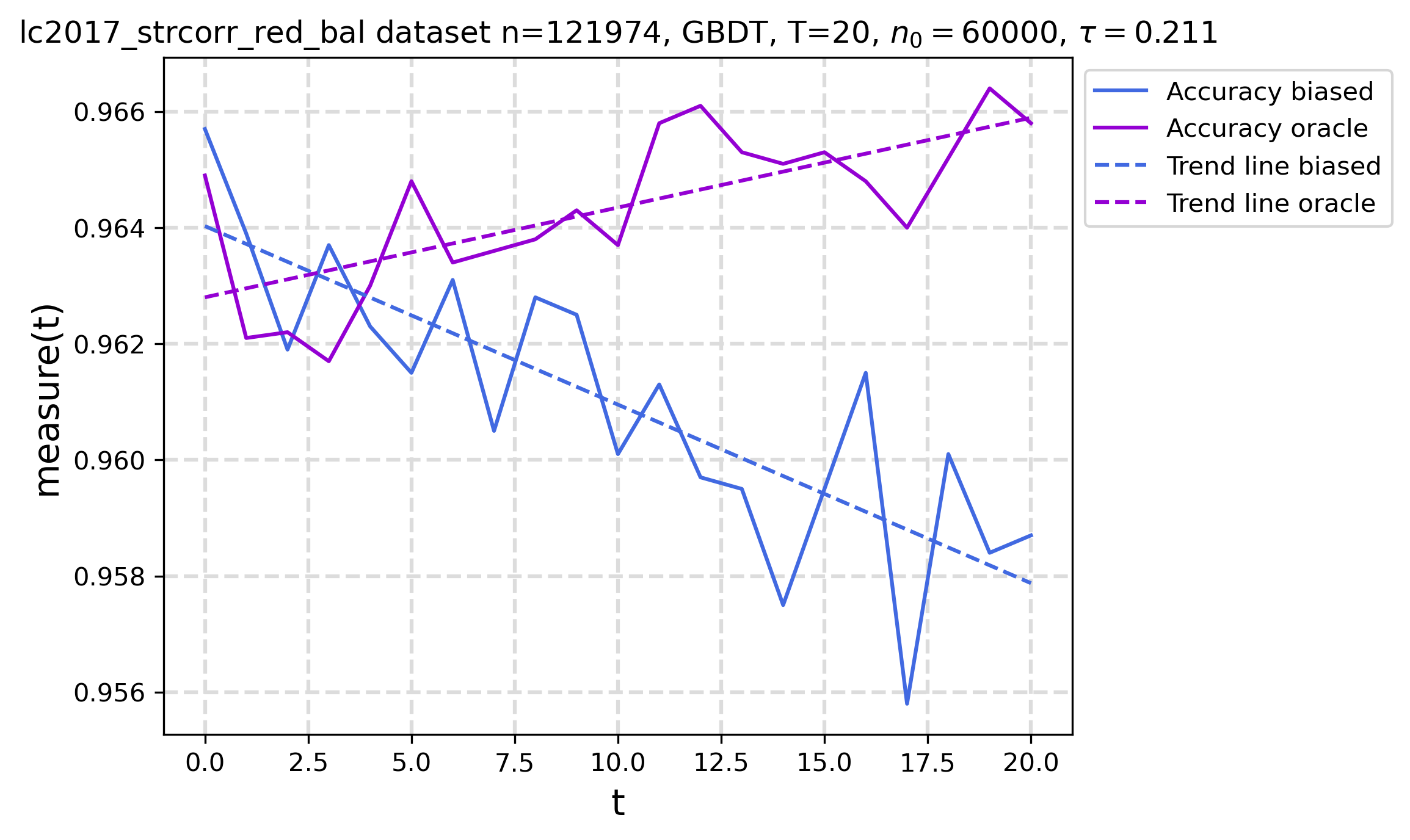}\\
        \includegraphics[scale=.25]{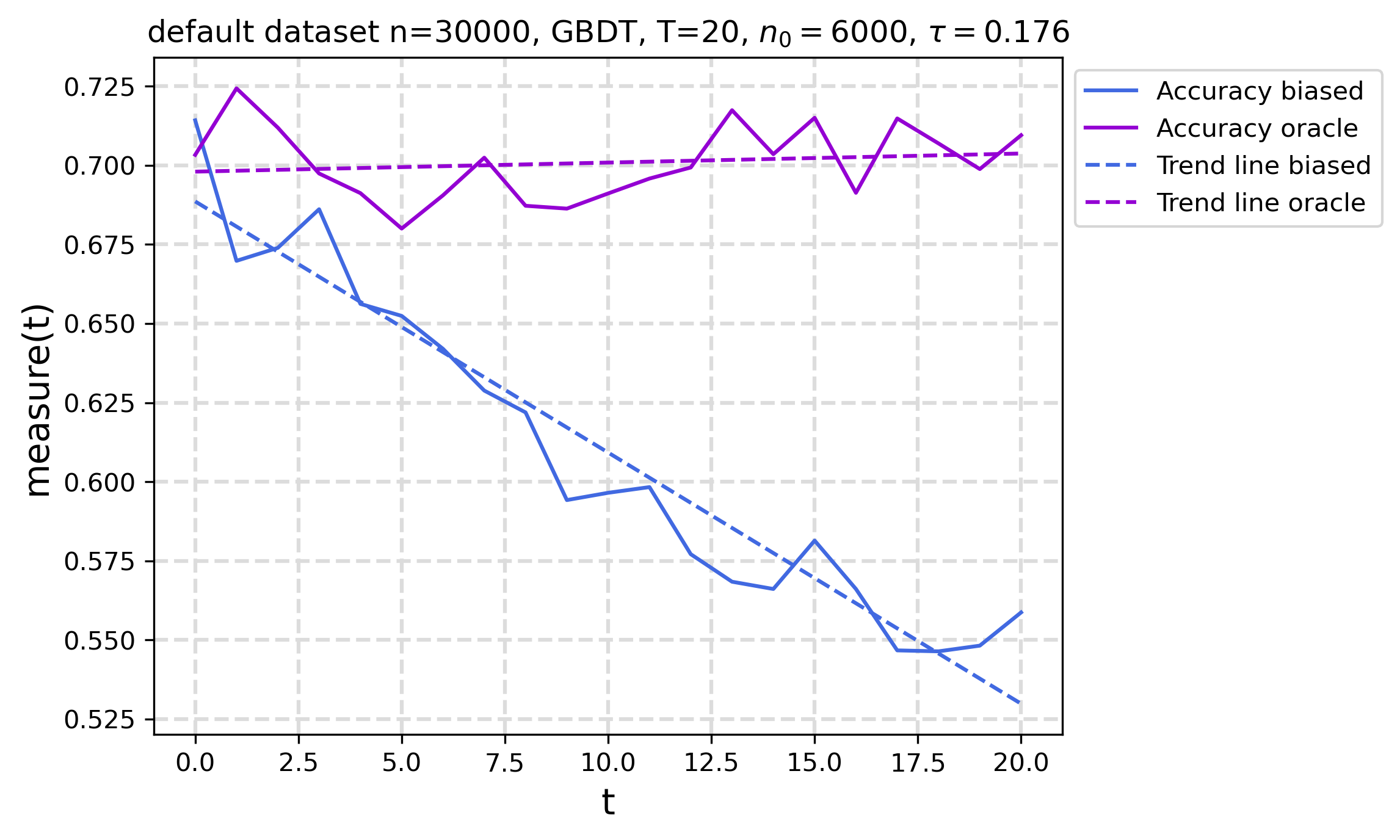} & \includegraphics[scale=.25]{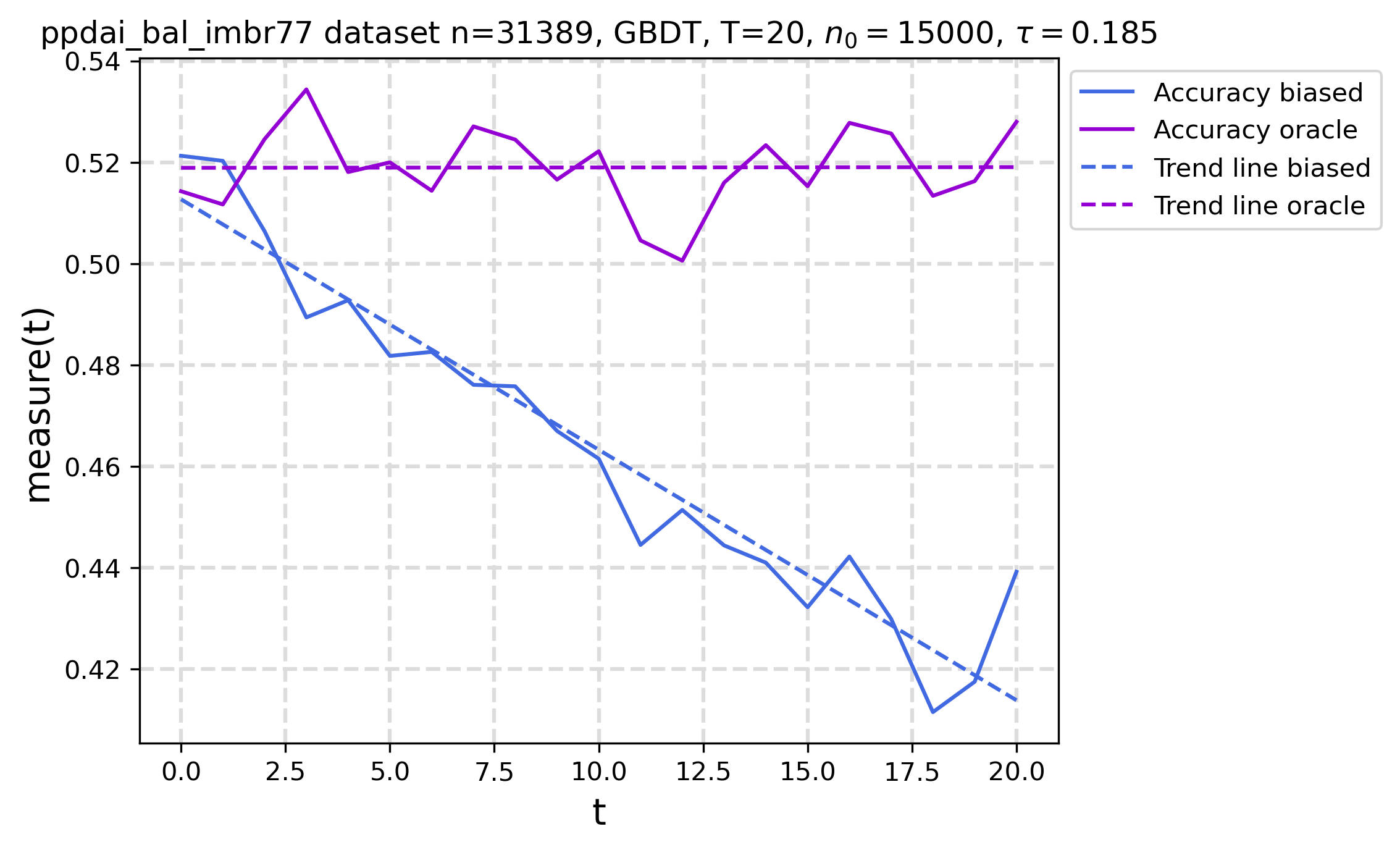} & \includegraphics[scale=.25]{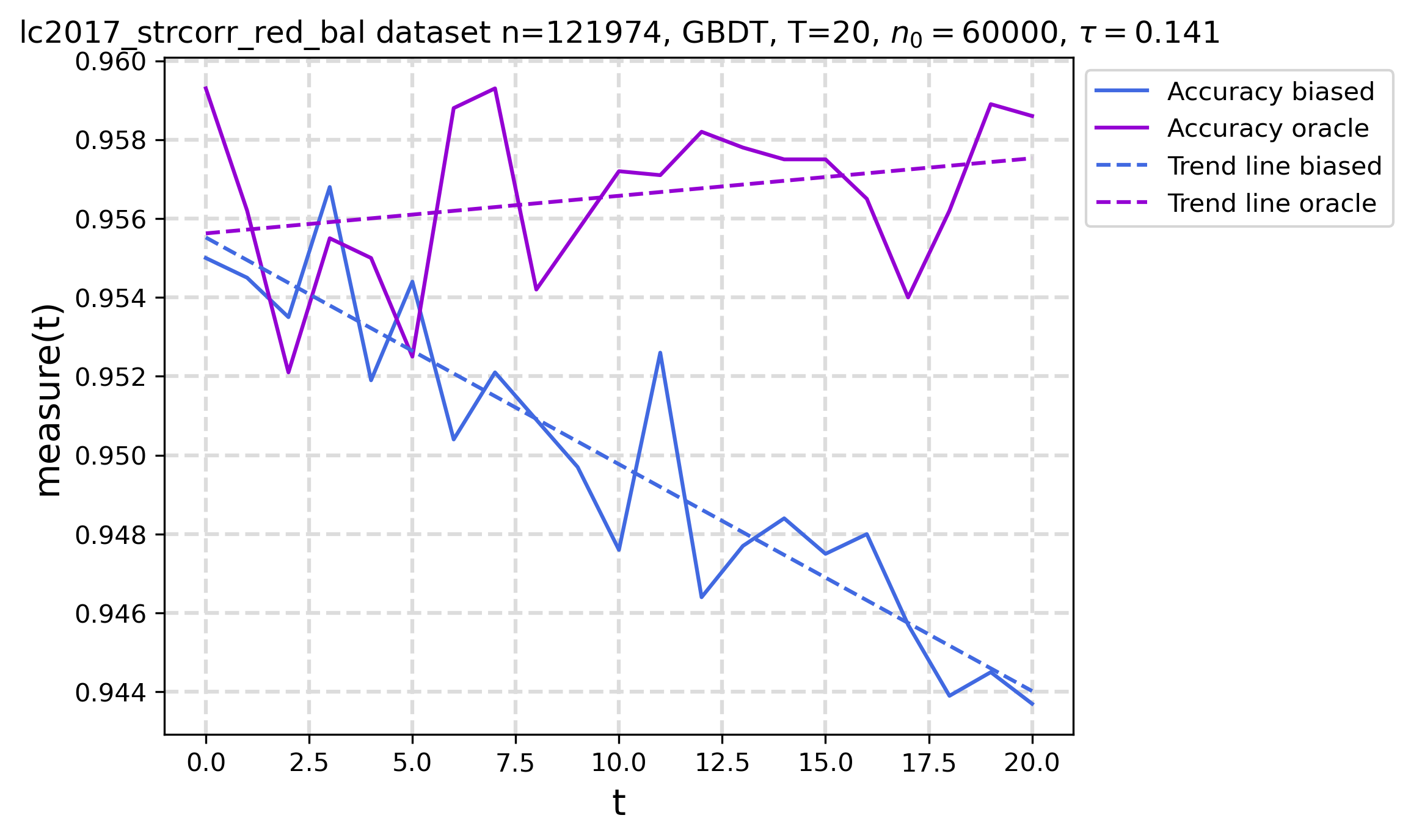}\\
    \end{tabular}
    \caption{Temporal evolution - GBDT - Decision thresholds for $c \in \{1, 3, 5\}$ - Accuracy}
    \label{fig:gbdt_tempevol_accuracy}
\end{figure*}

\begin{figure*}[t]
    \centering
    \begin{tabular}{ccc}
        Default & ppdai\_bal\_imbr77 & lc17\_bal\_imbr50\\
        \includegraphics[scale=.25]{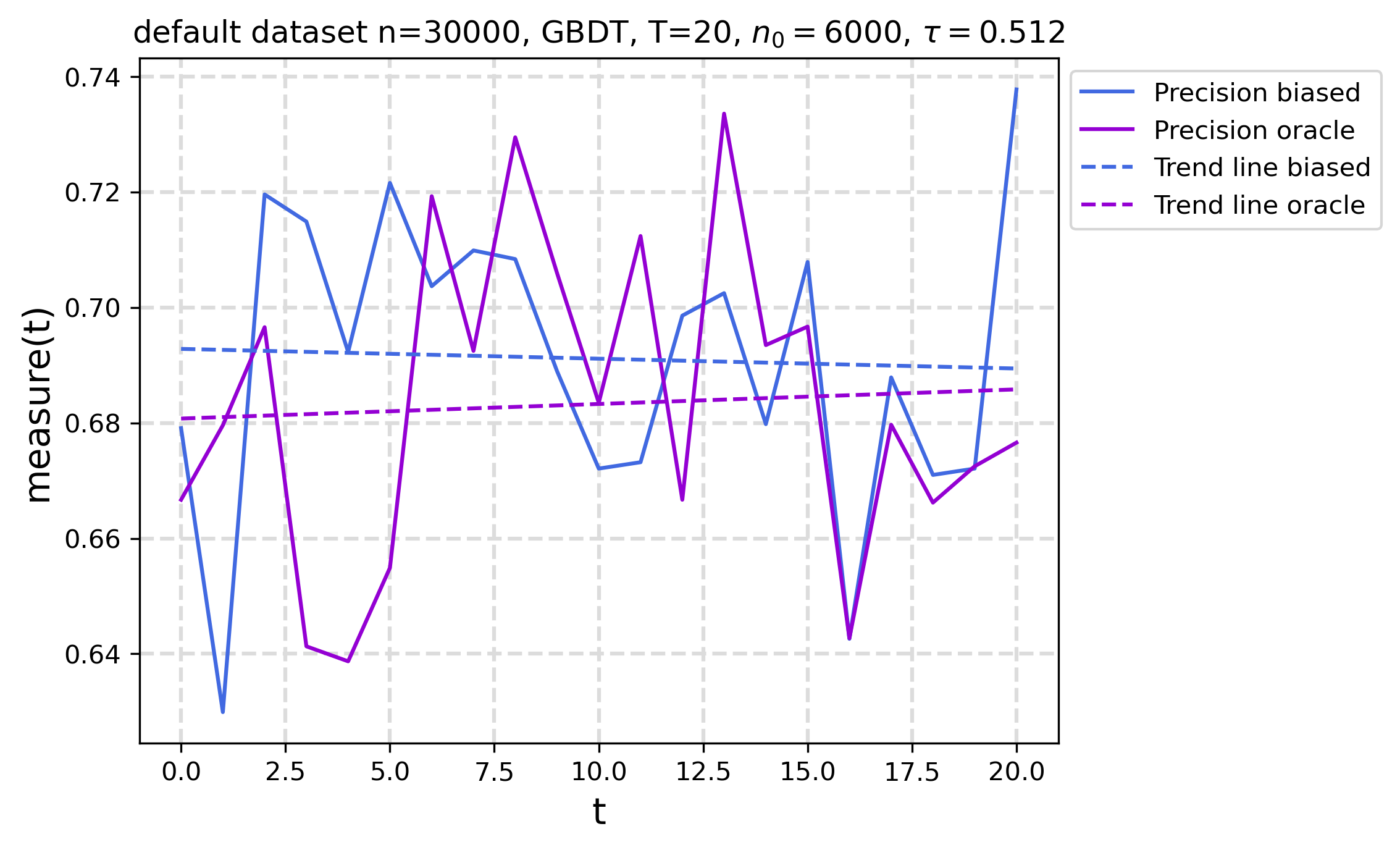} & \includegraphics[scale=.25]{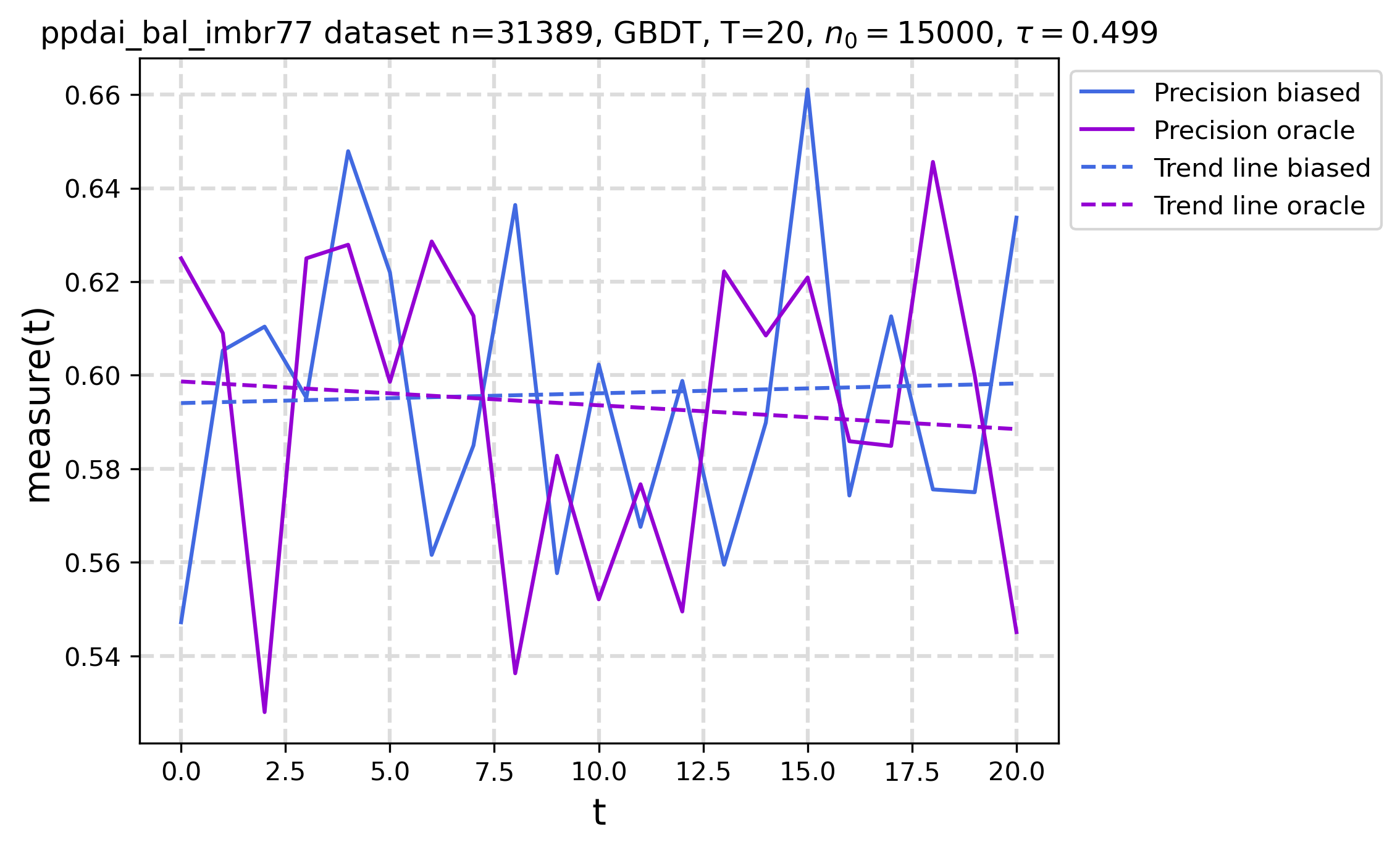} & \includegraphics[scale=.25]{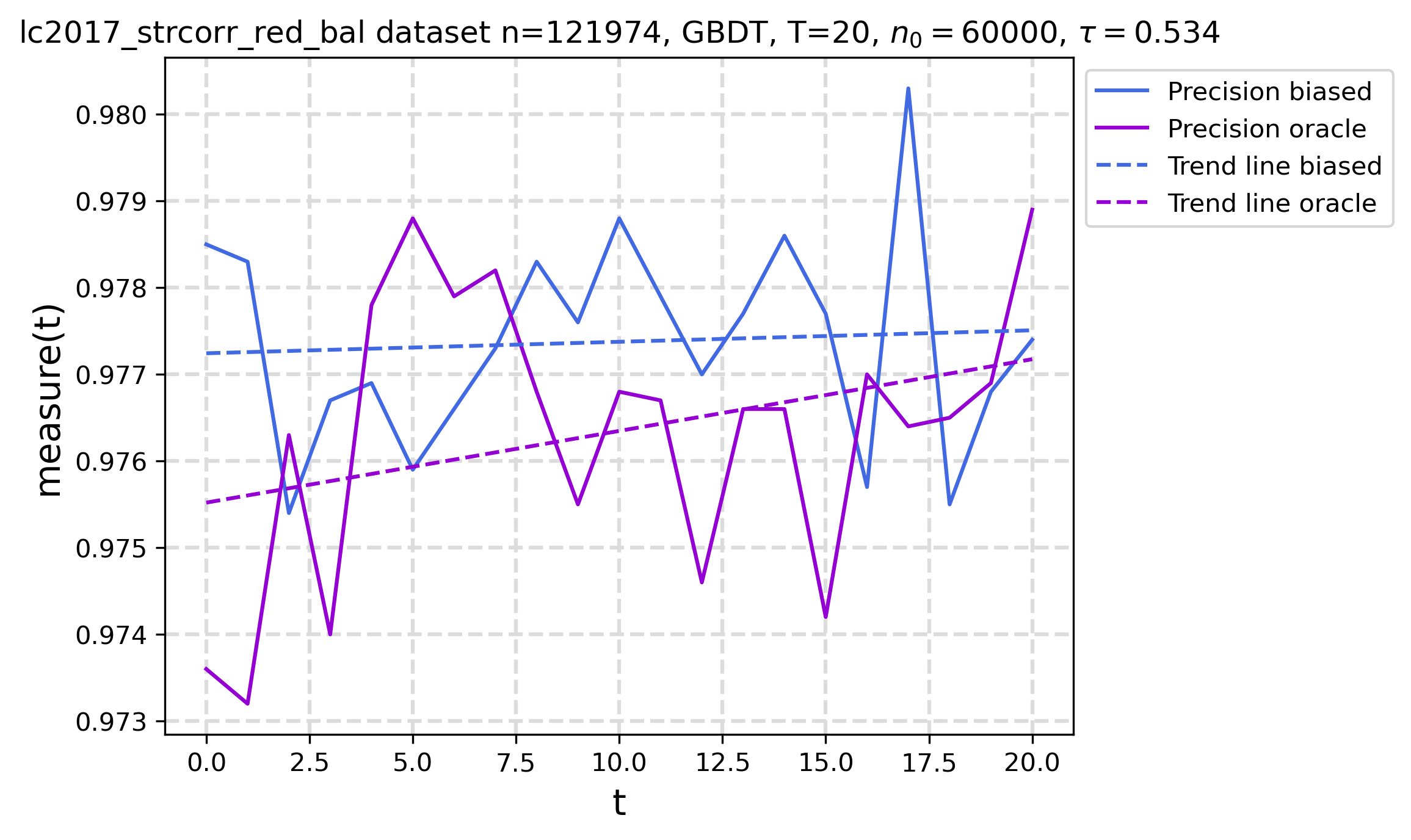}\\
       \includegraphics[scale=.25]{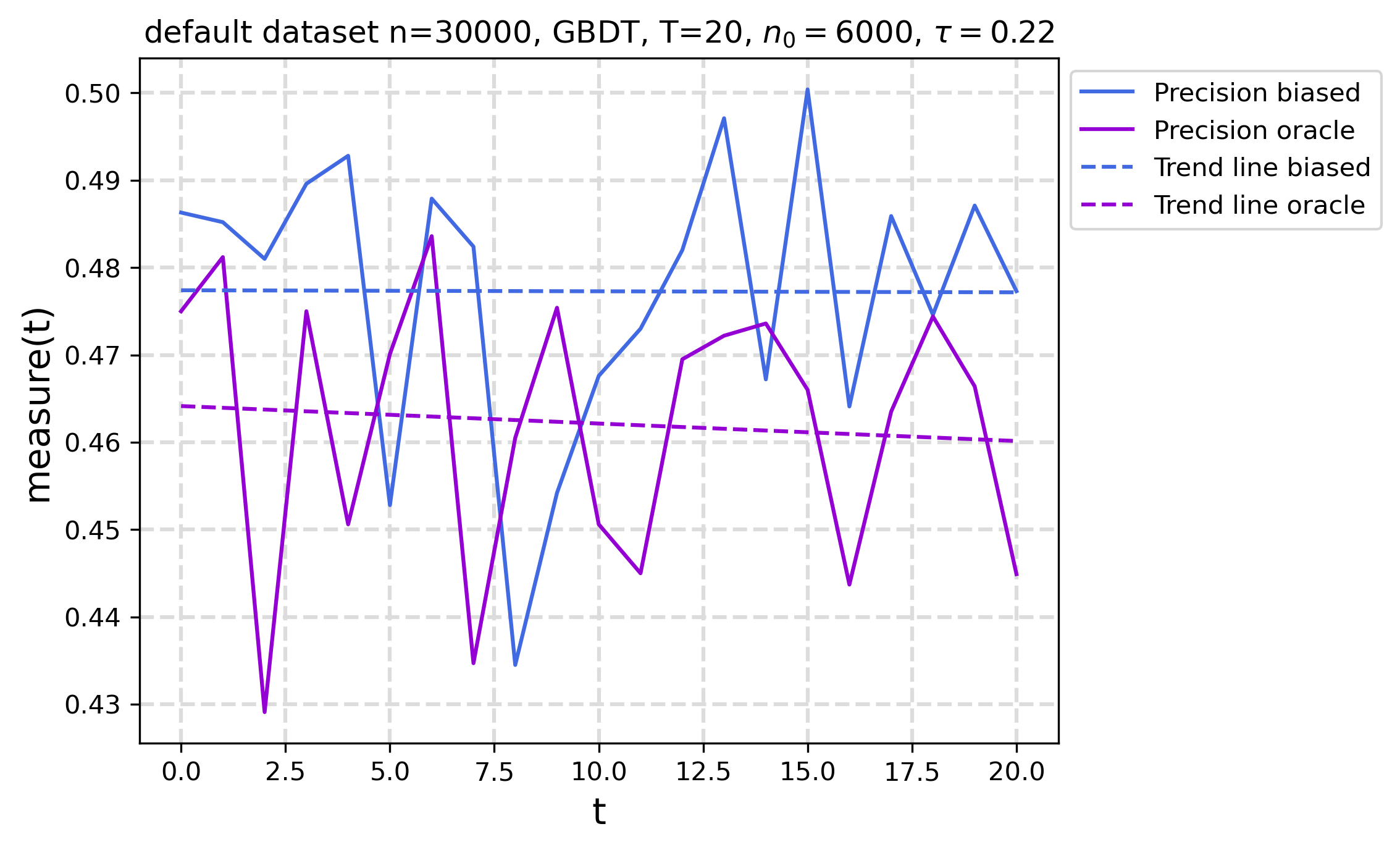} & \includegraphics[scale=.25]{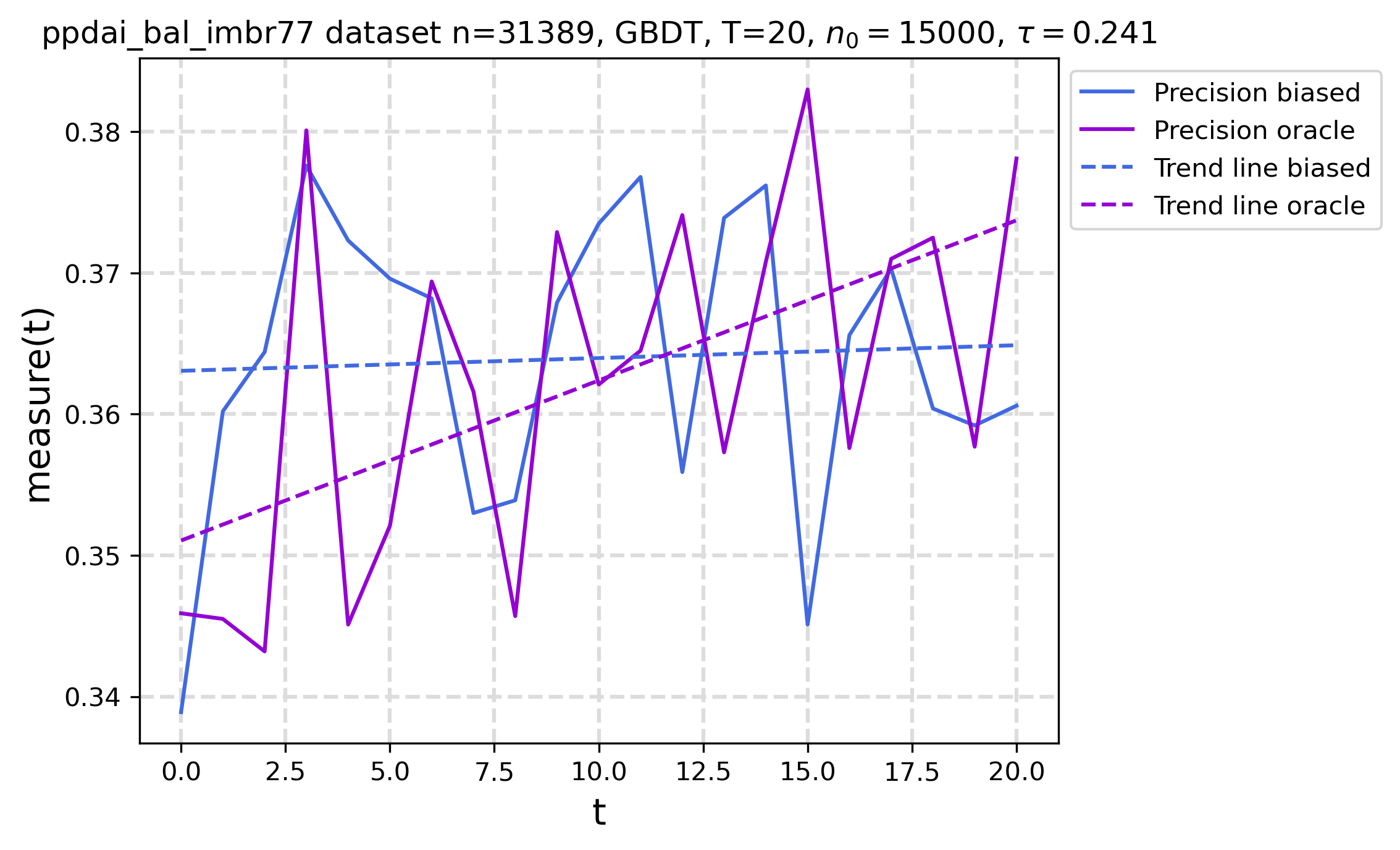} & \includegraphics[scale=.25]{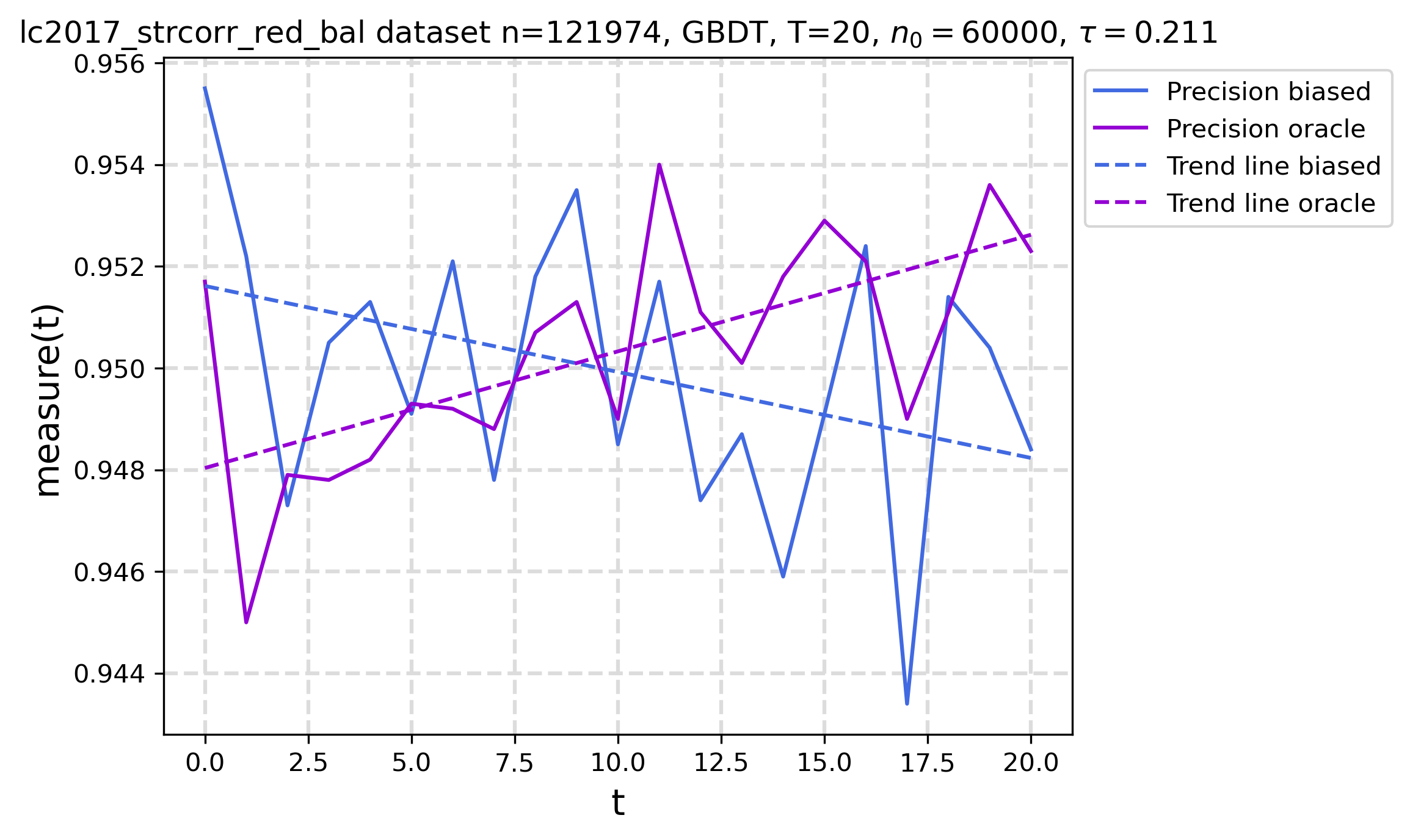}\\
    \end{tabular}
    \caption{Temporal evolution - GBDT - Decision thresholds for $c \in \{1, 3, 5\}$ - Precision}
    \label{fig:gbdt_tempevol_precision}
\end{figure*}

\begin{figure*}[t]
    \centering
    \begin{tabular}{ccc}
        Default & ppdai\_bal\_imbr77 & lc17\_bal\_imbr50\\
        \includegraphics[scale=.25]{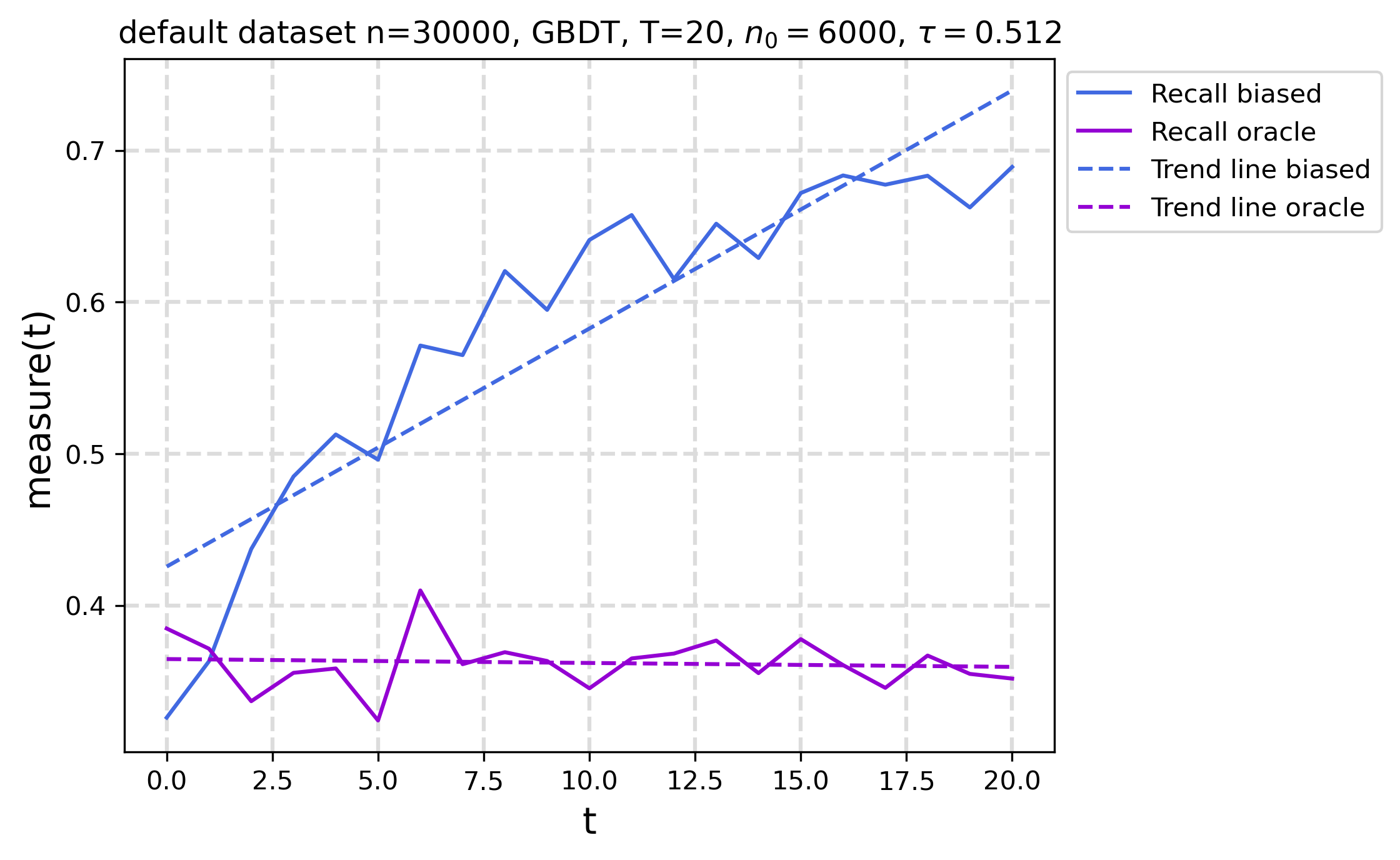} & \includegraphics[scale=.25]{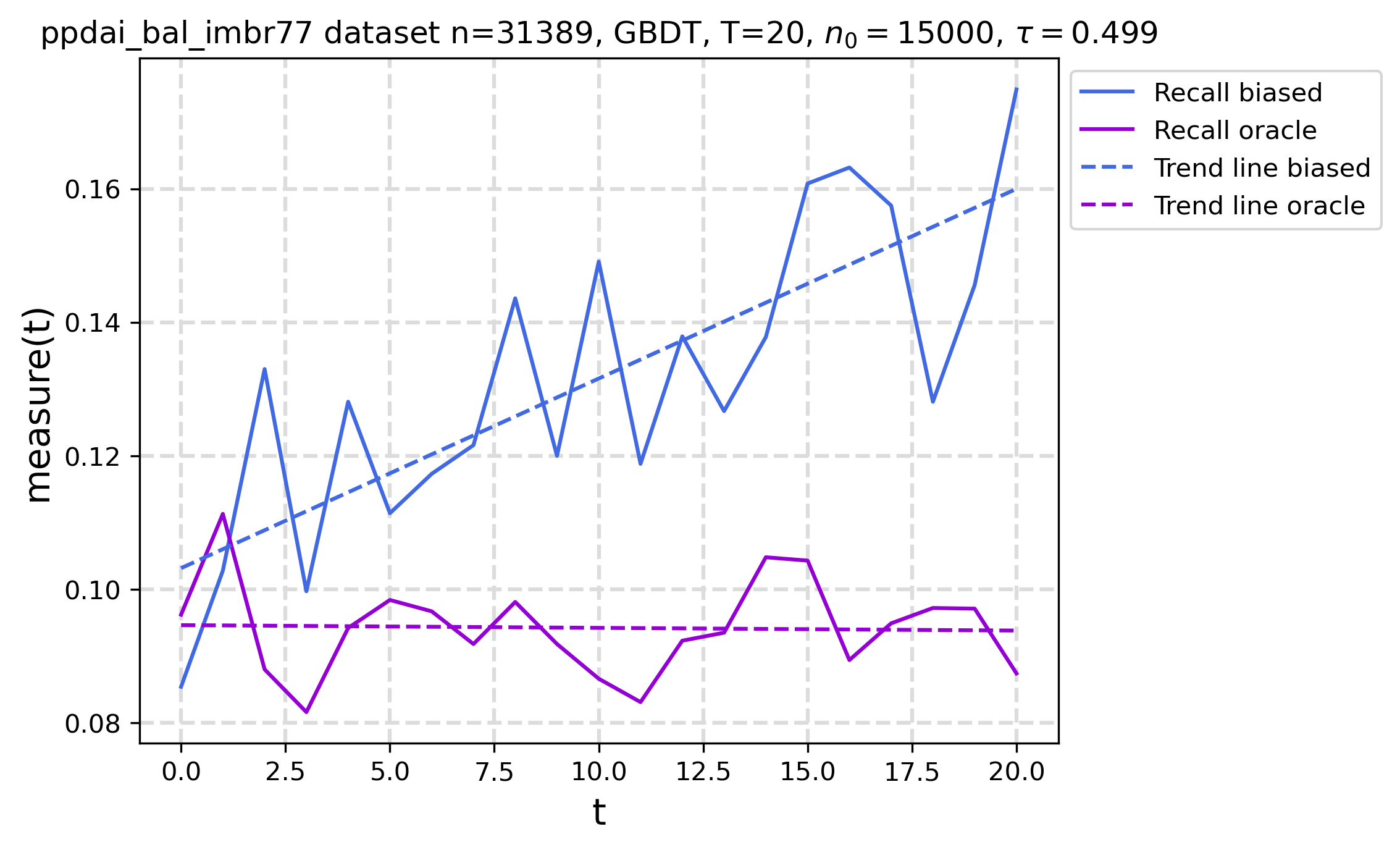} & \includegraphics[scale=.25]{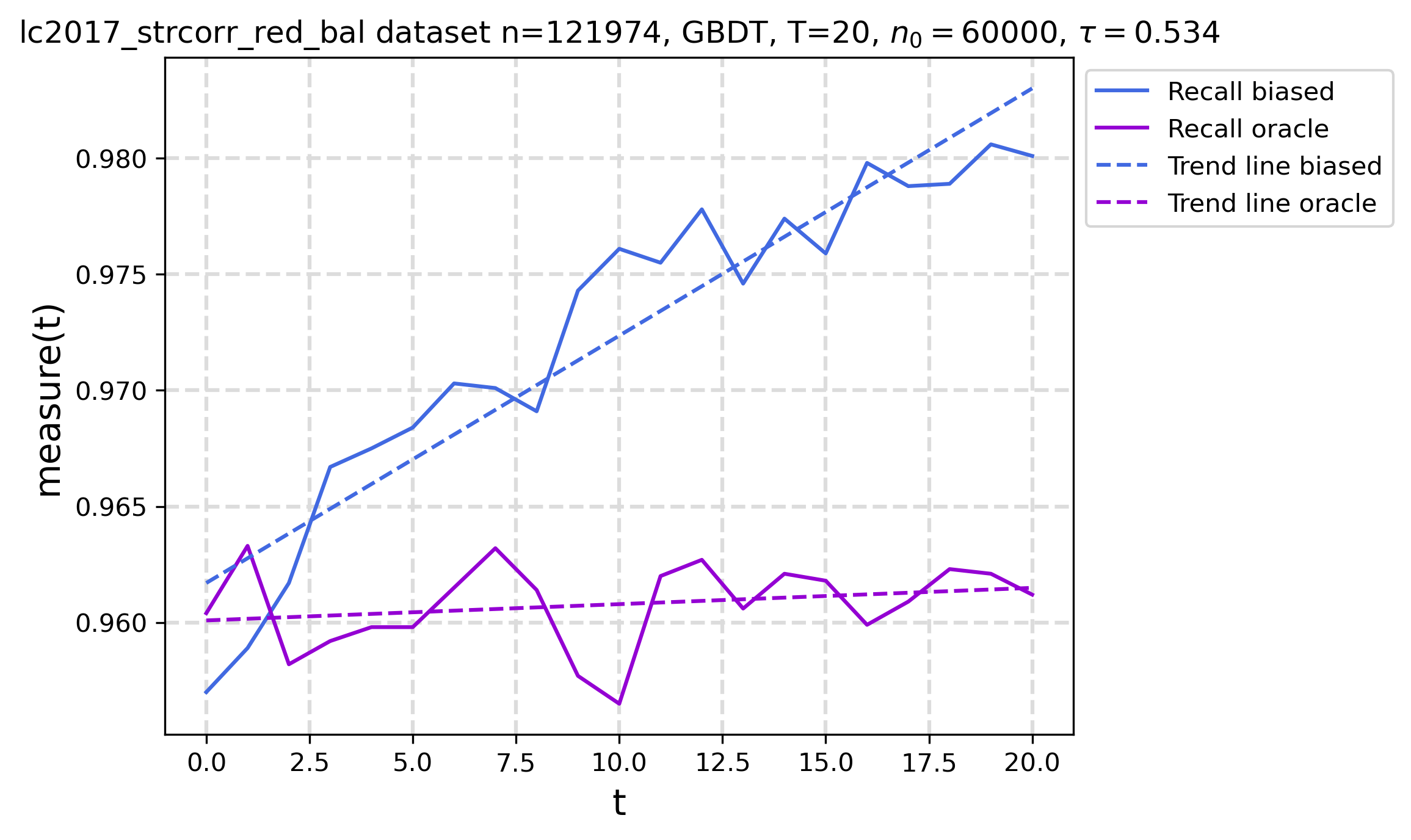}\\
       \includegraphics[scale=.25]{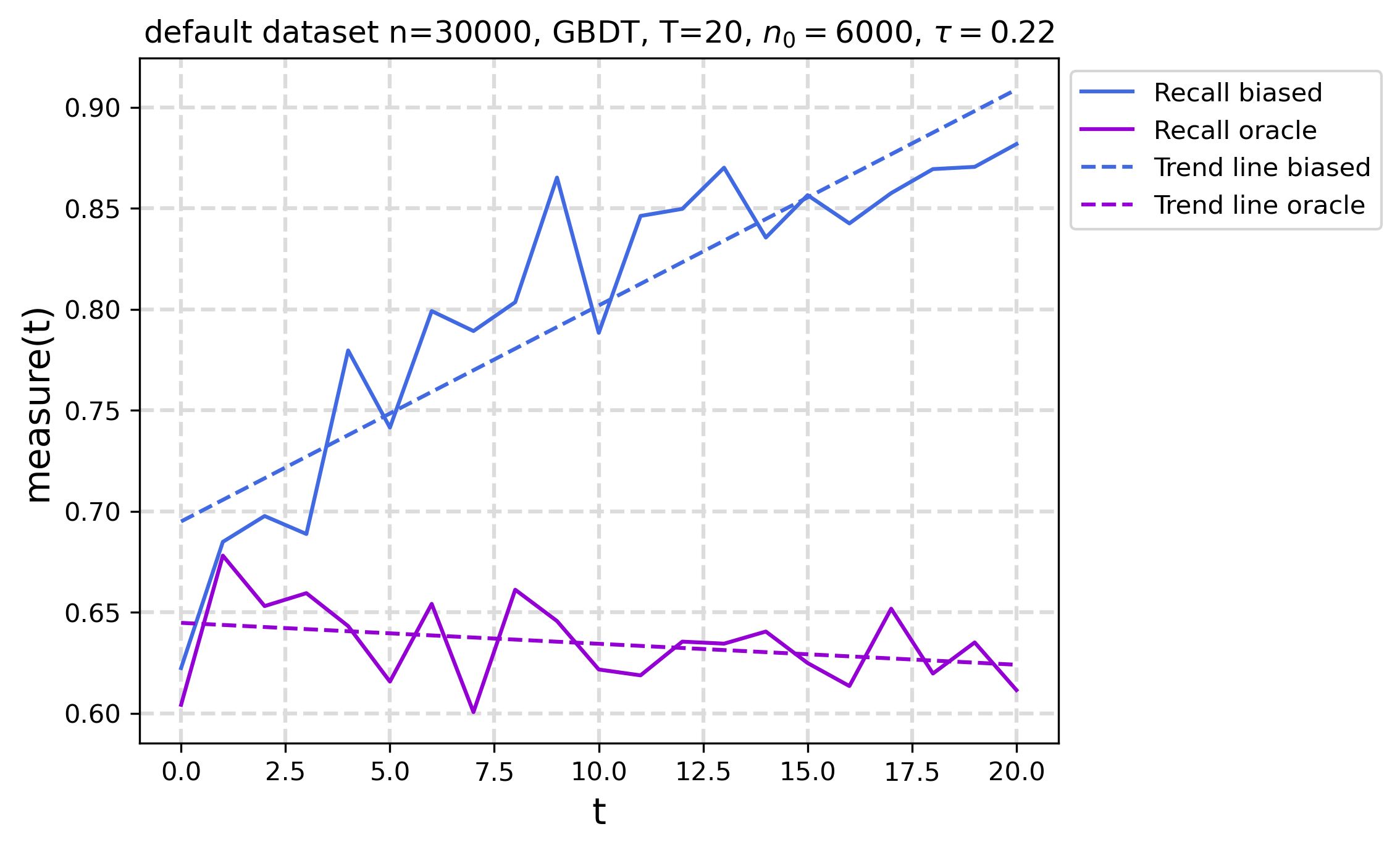} & \includegraphics[scale=.25]{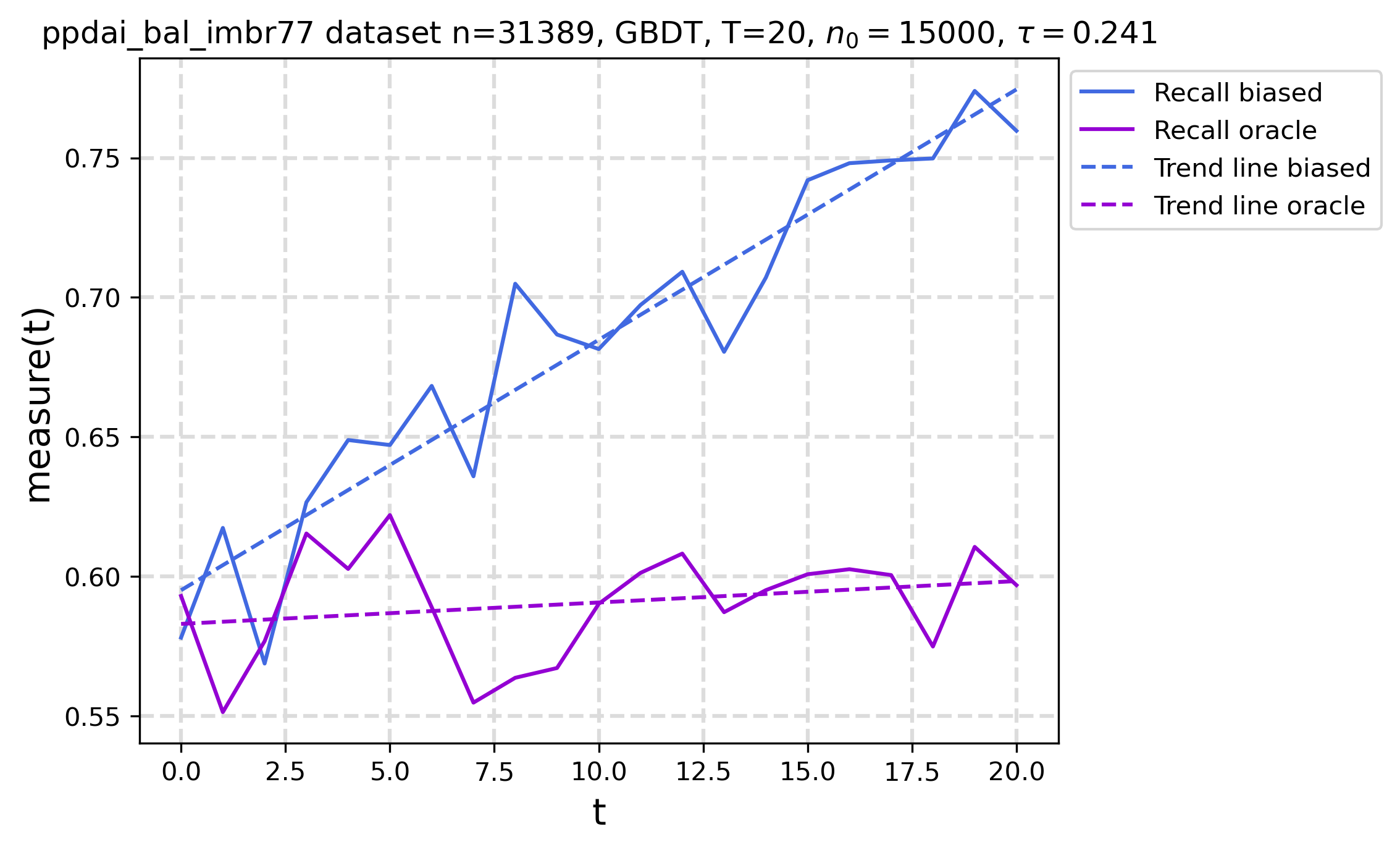} & \includegraphics[scale=.25]{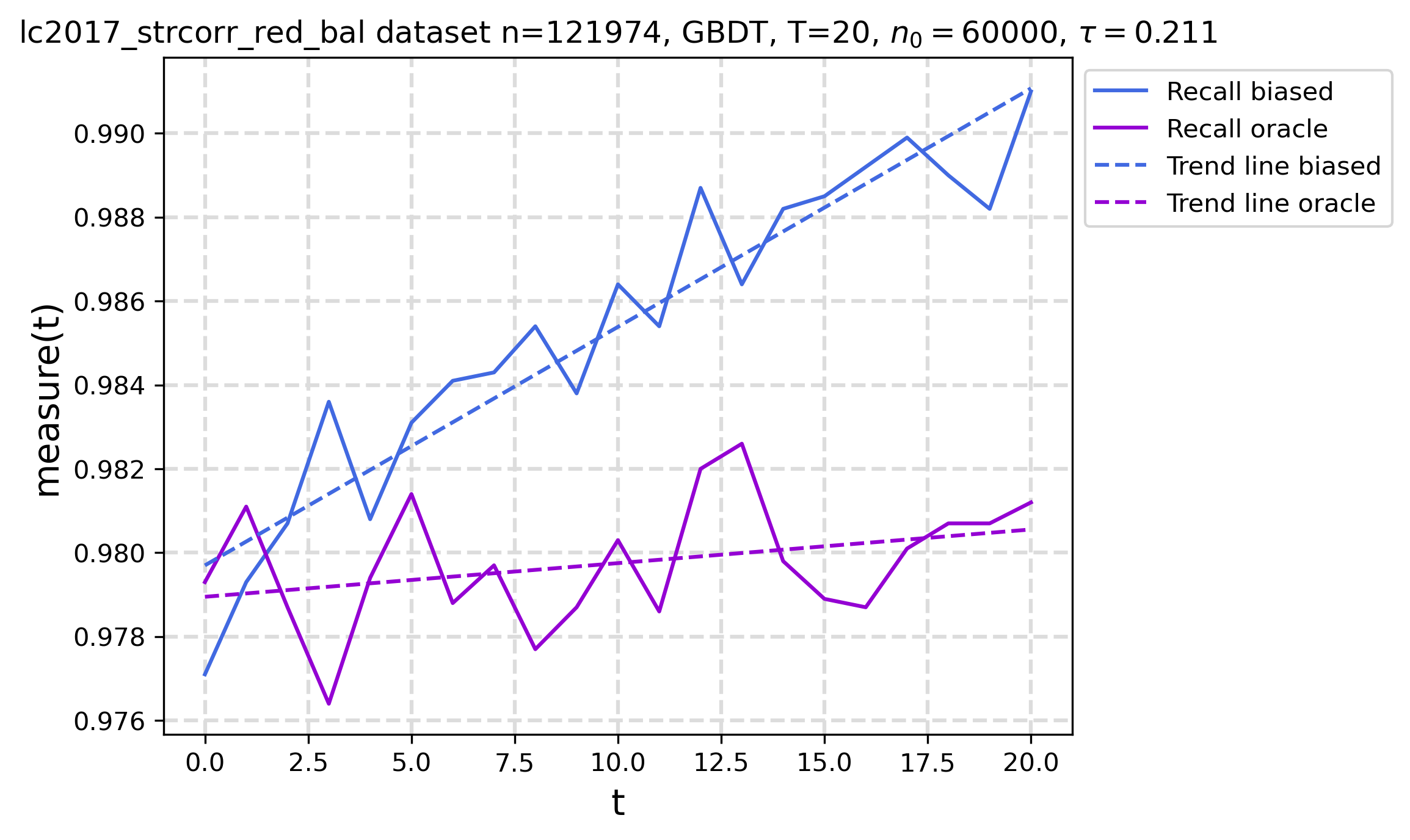}\\
        \includegraphics[scale=.25]{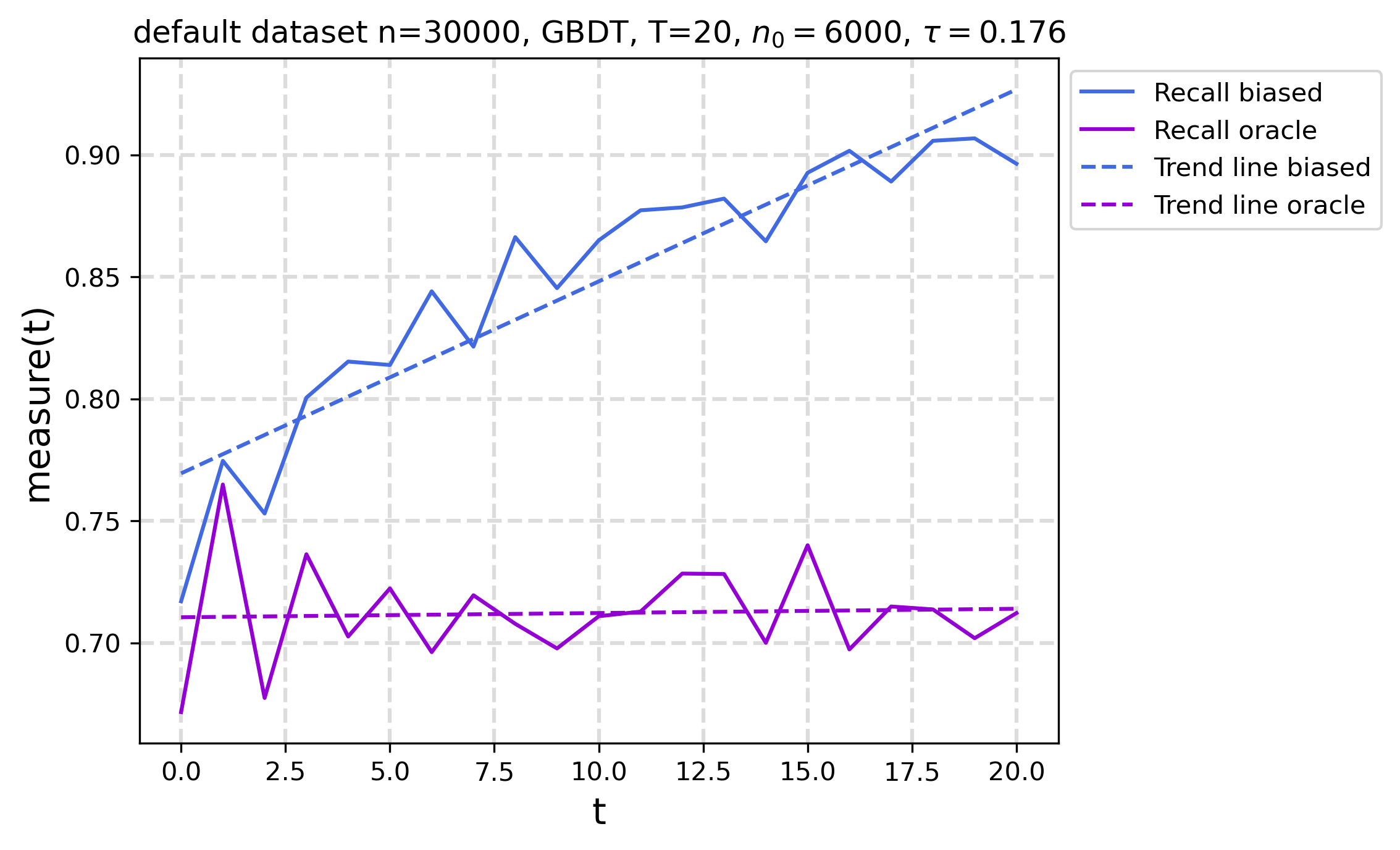} & \includegraphics[scale=.25]{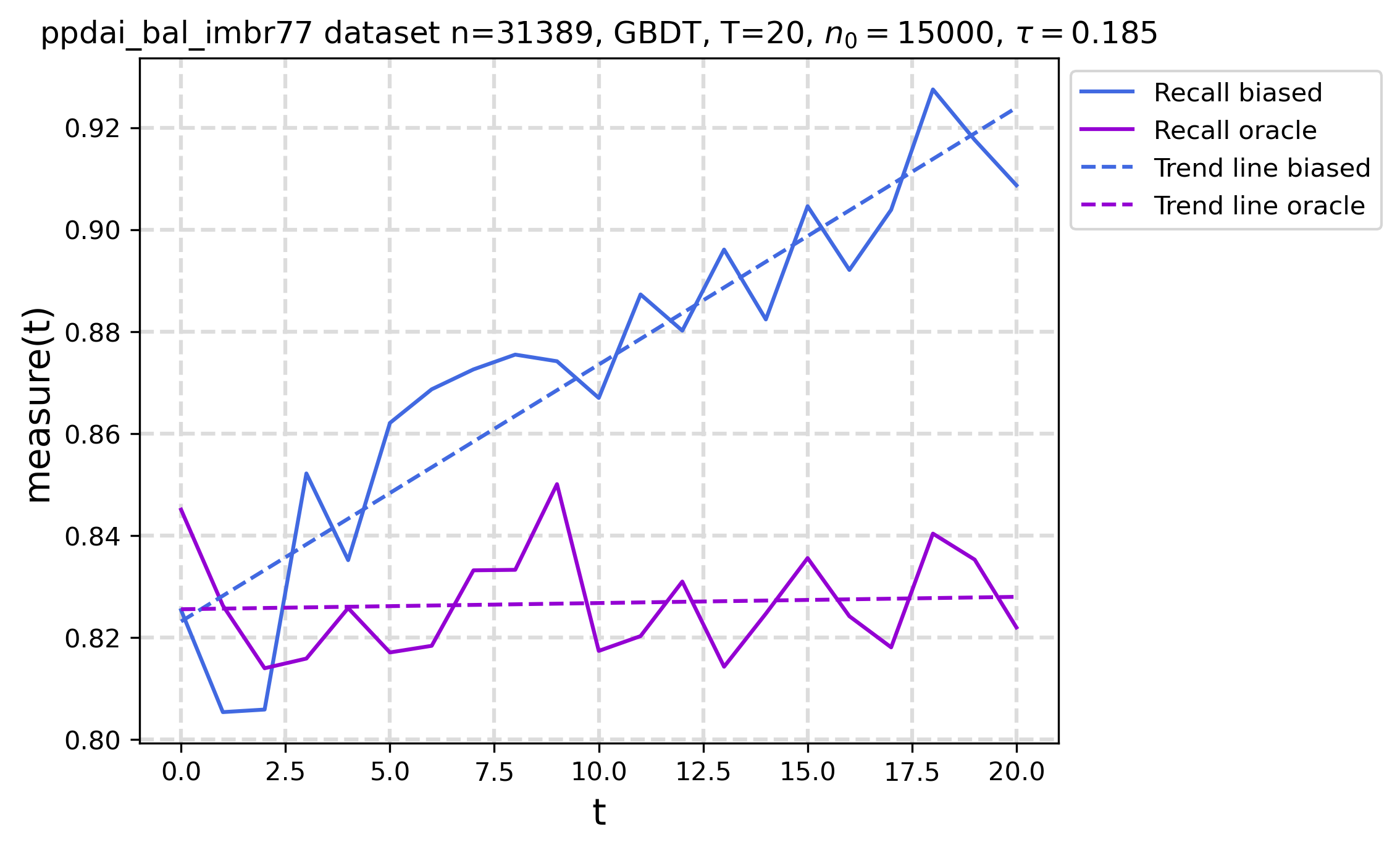} & \includegraphics[scale=.25]{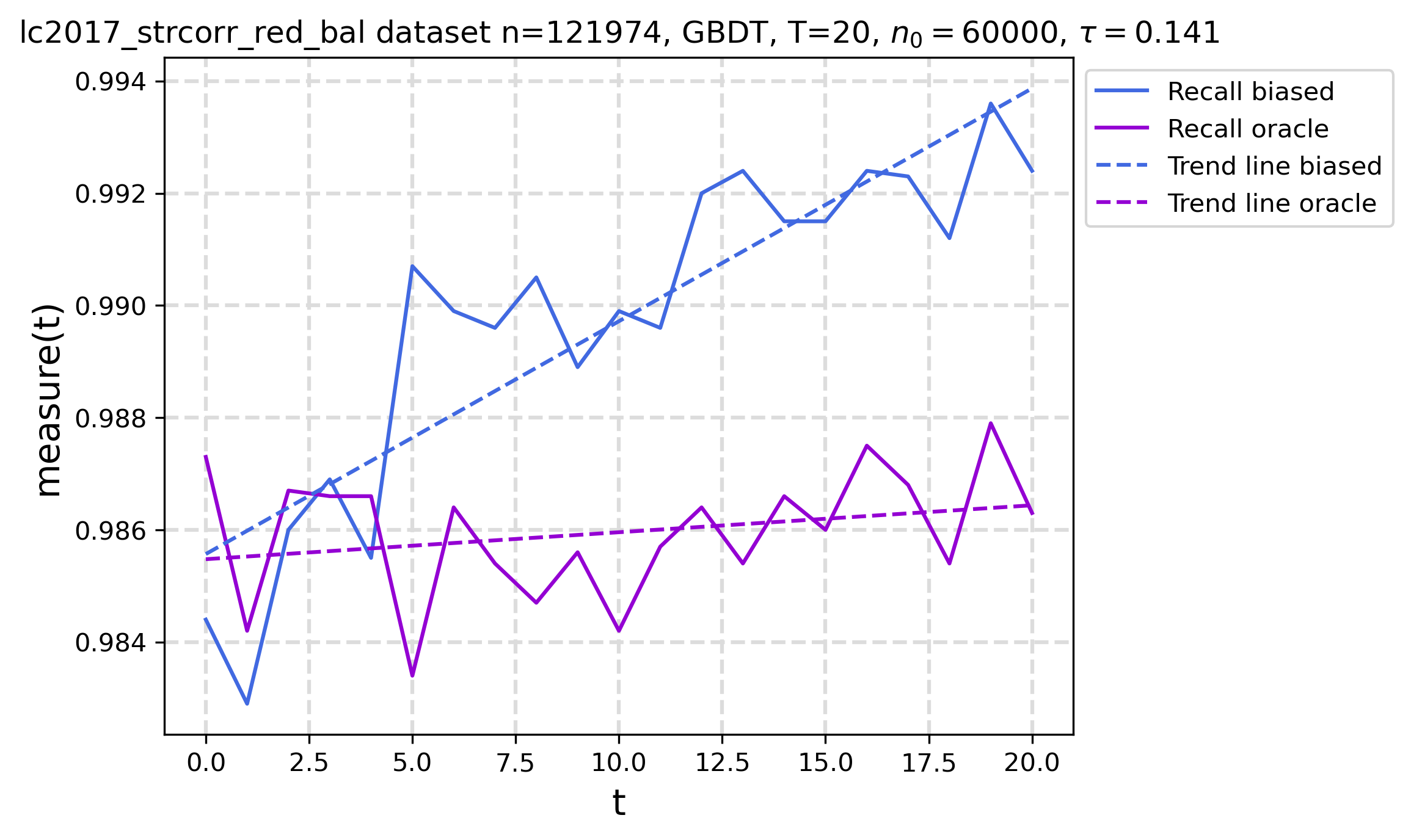}\\
    \end{tabular}
    \caption{Temporal evolution - GBDT - Decision thresholds for $c \in \{1, 3, 5\}$ - Recall}
    \label{fig:gbdt_tempevol_recall}
\end{figure*}

\begin{figure*}[t]
    \centering
    \begin{tabular}{ccc}
        Default & ppdai\_bal\_imbr77 & lc17\_bal\_imbr50\\
        \includegraphics[scale=.25]{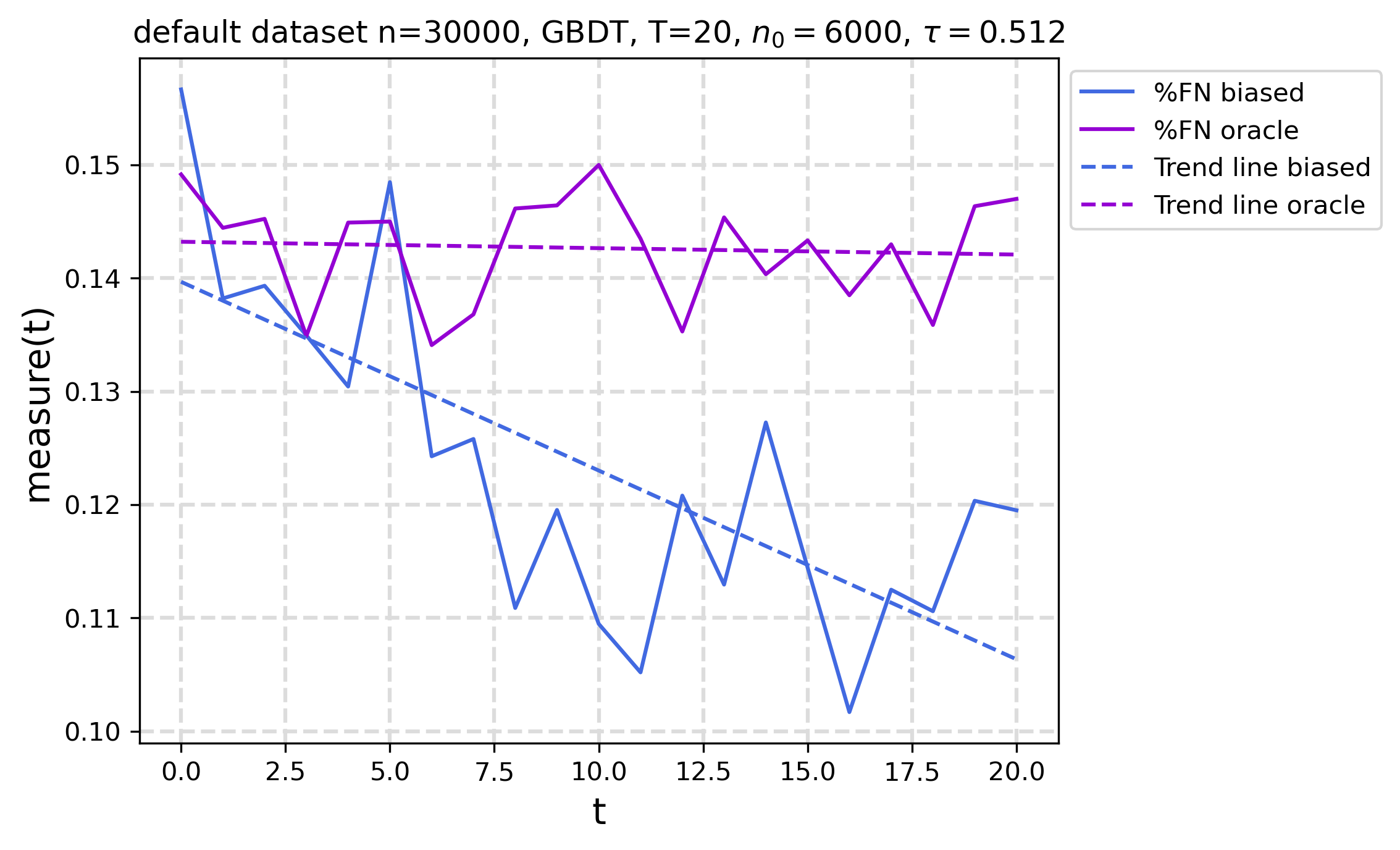} & \includegraphics[scale=.25]{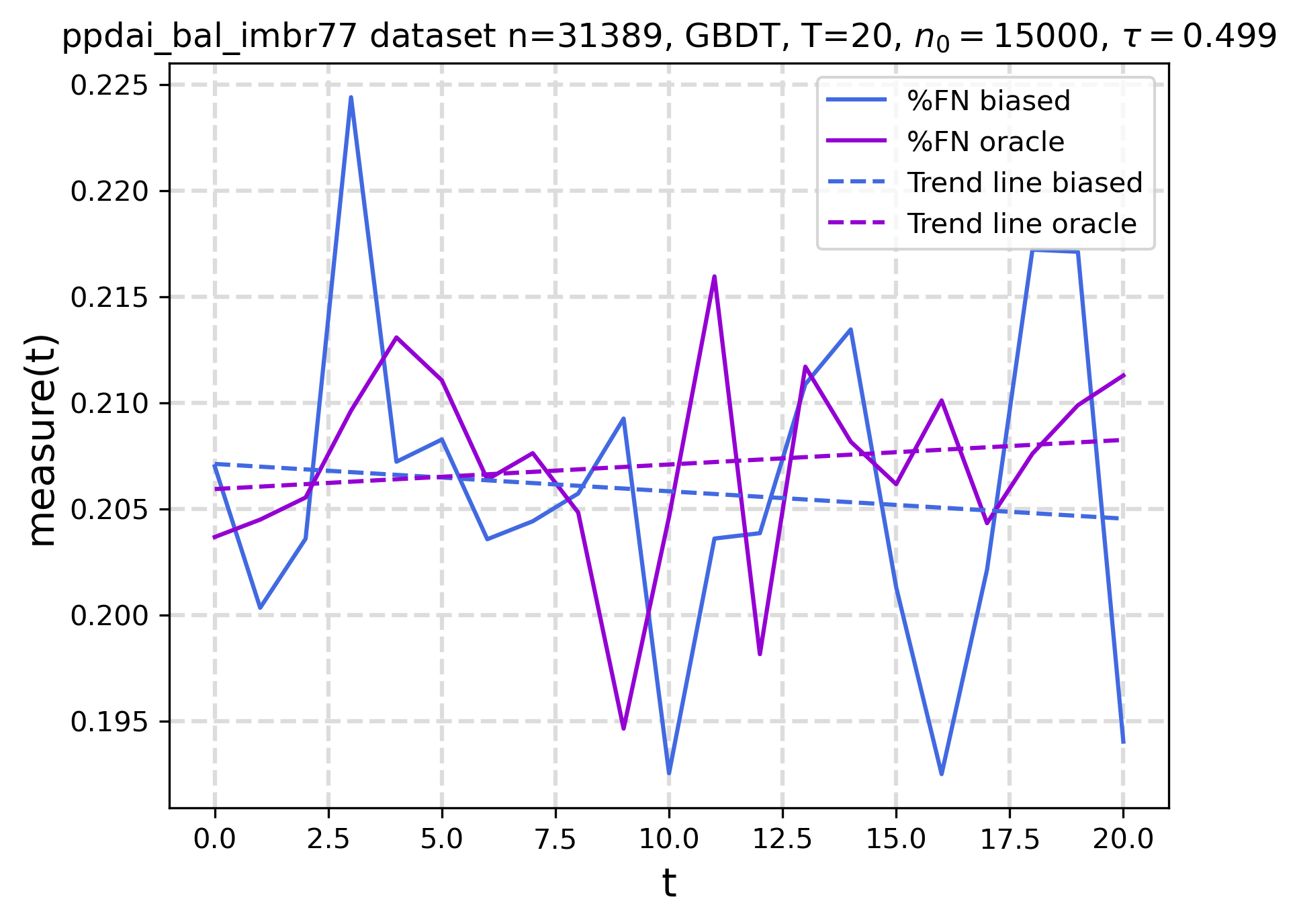} & \includegraphics[scale=.25]{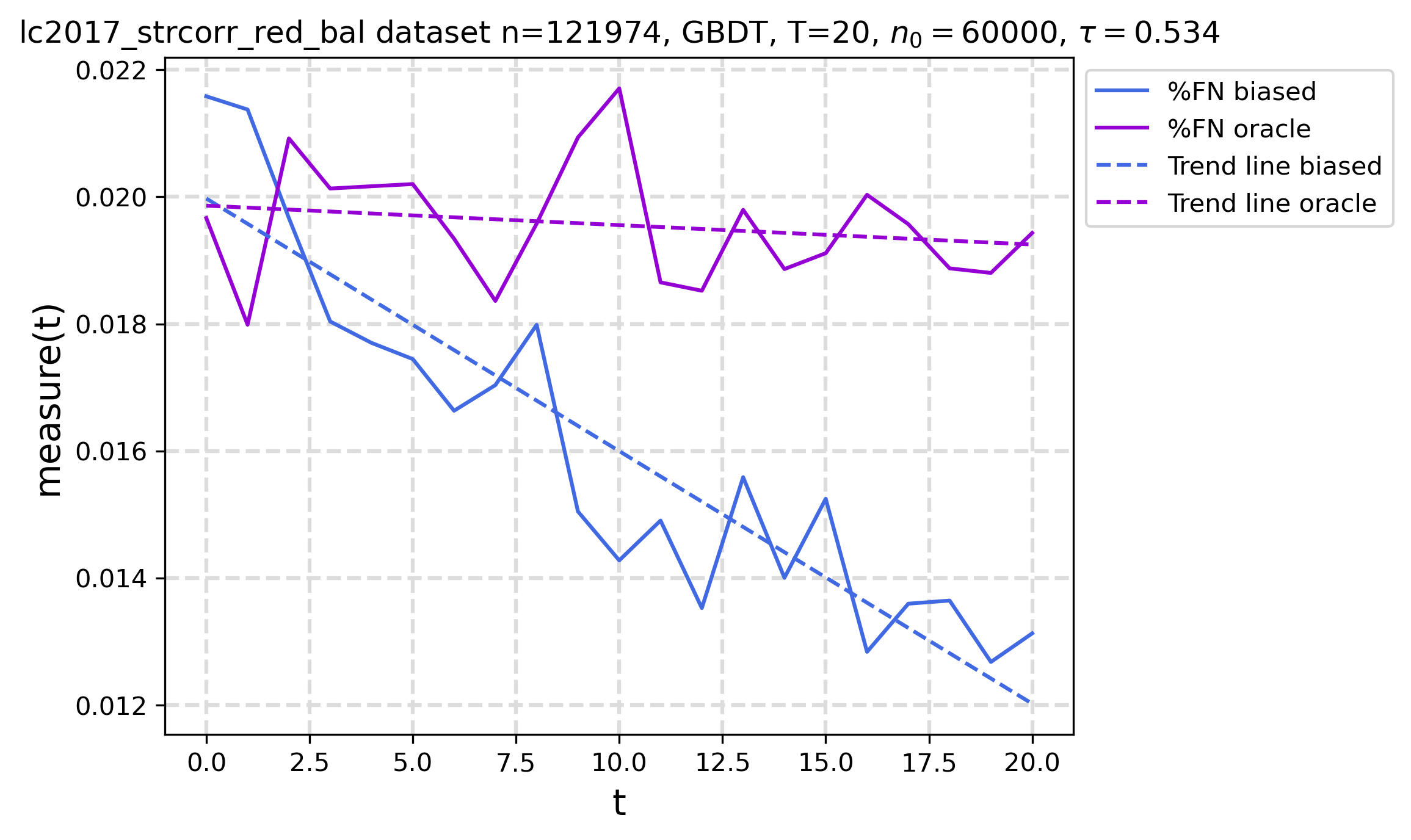}\\
       \includegraphics[scale=.25]{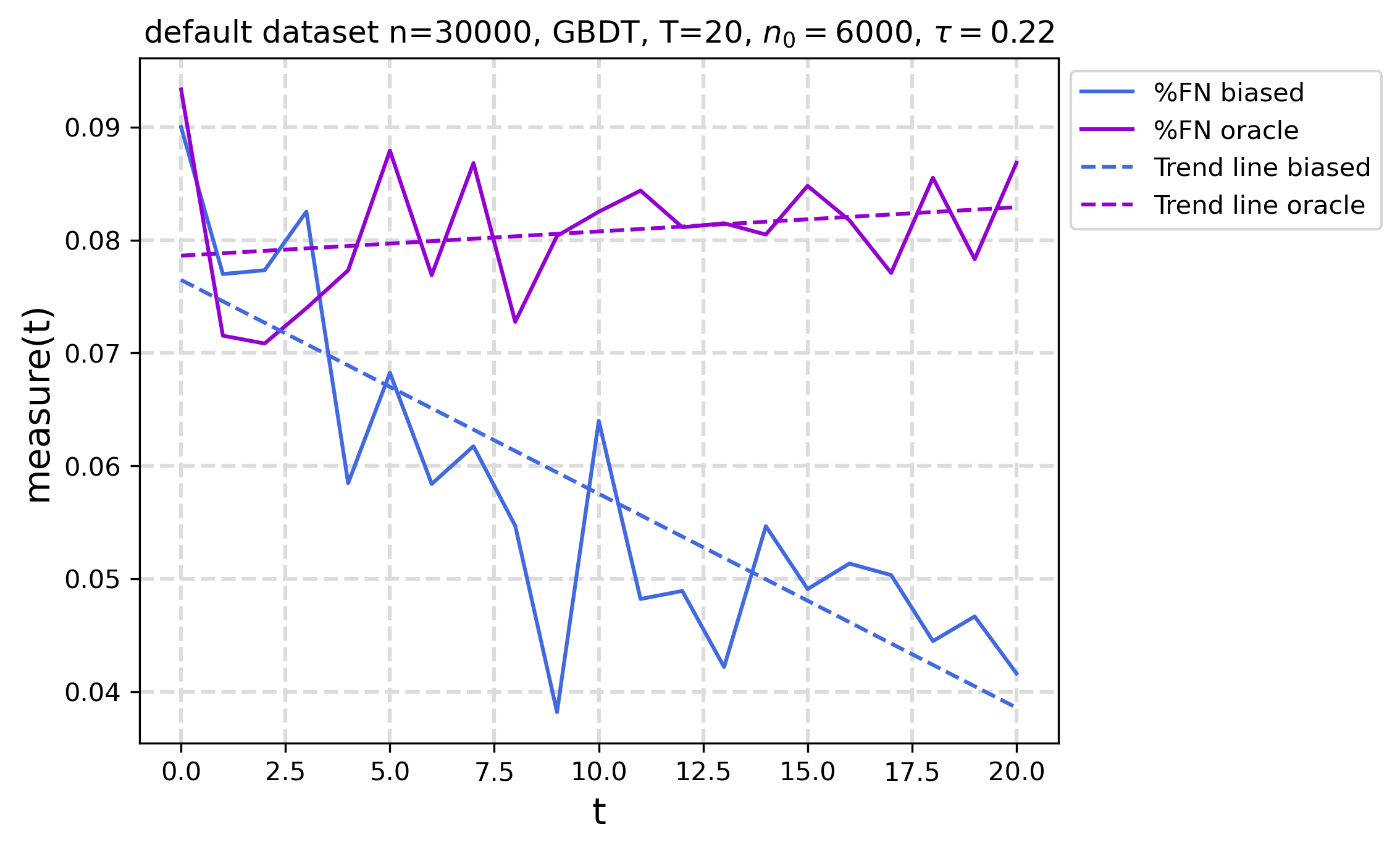} & \includegraphics[scale=.25]{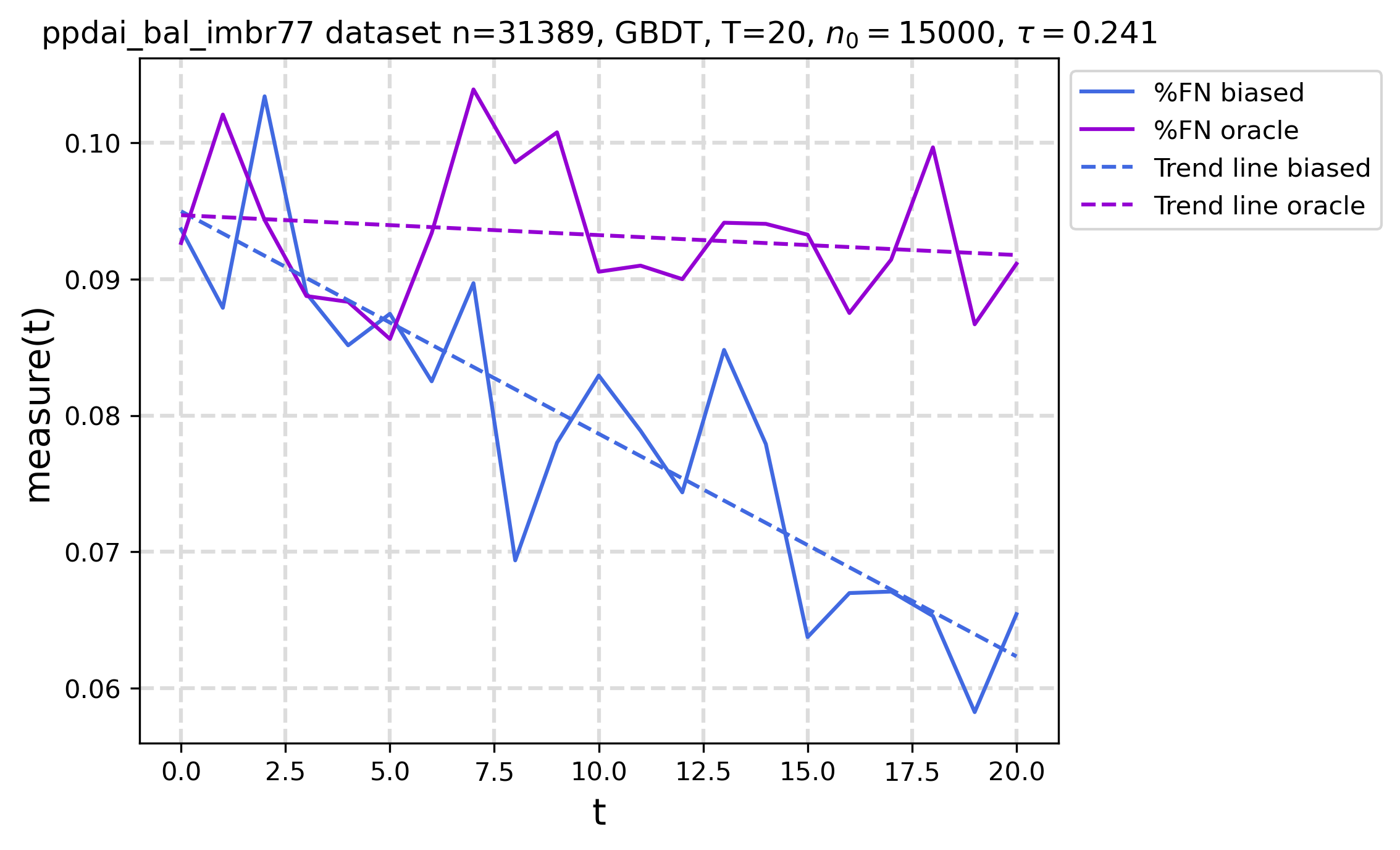} & \includegraphics[scale=.25]{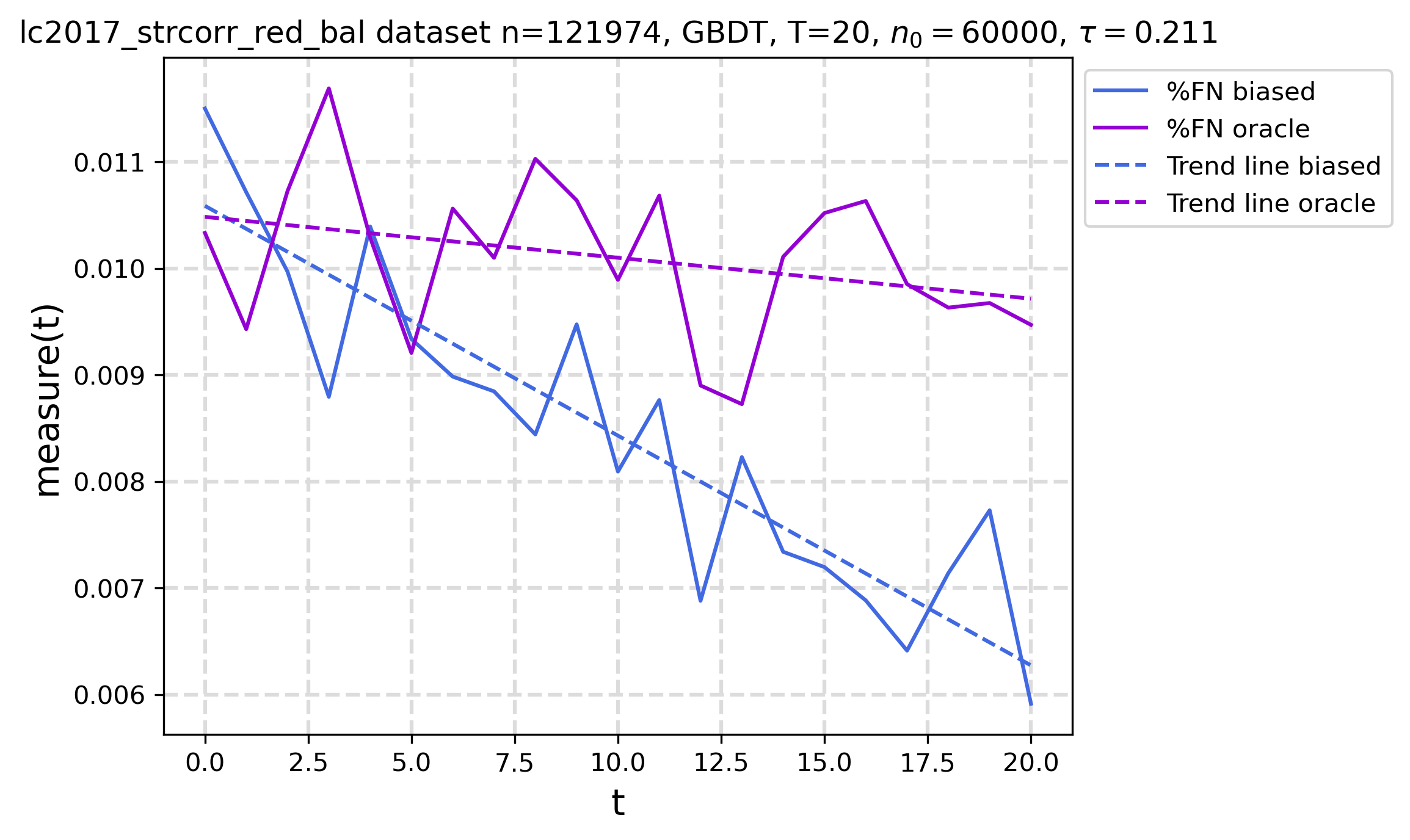}\\
        \includegraphics[scale=.25]{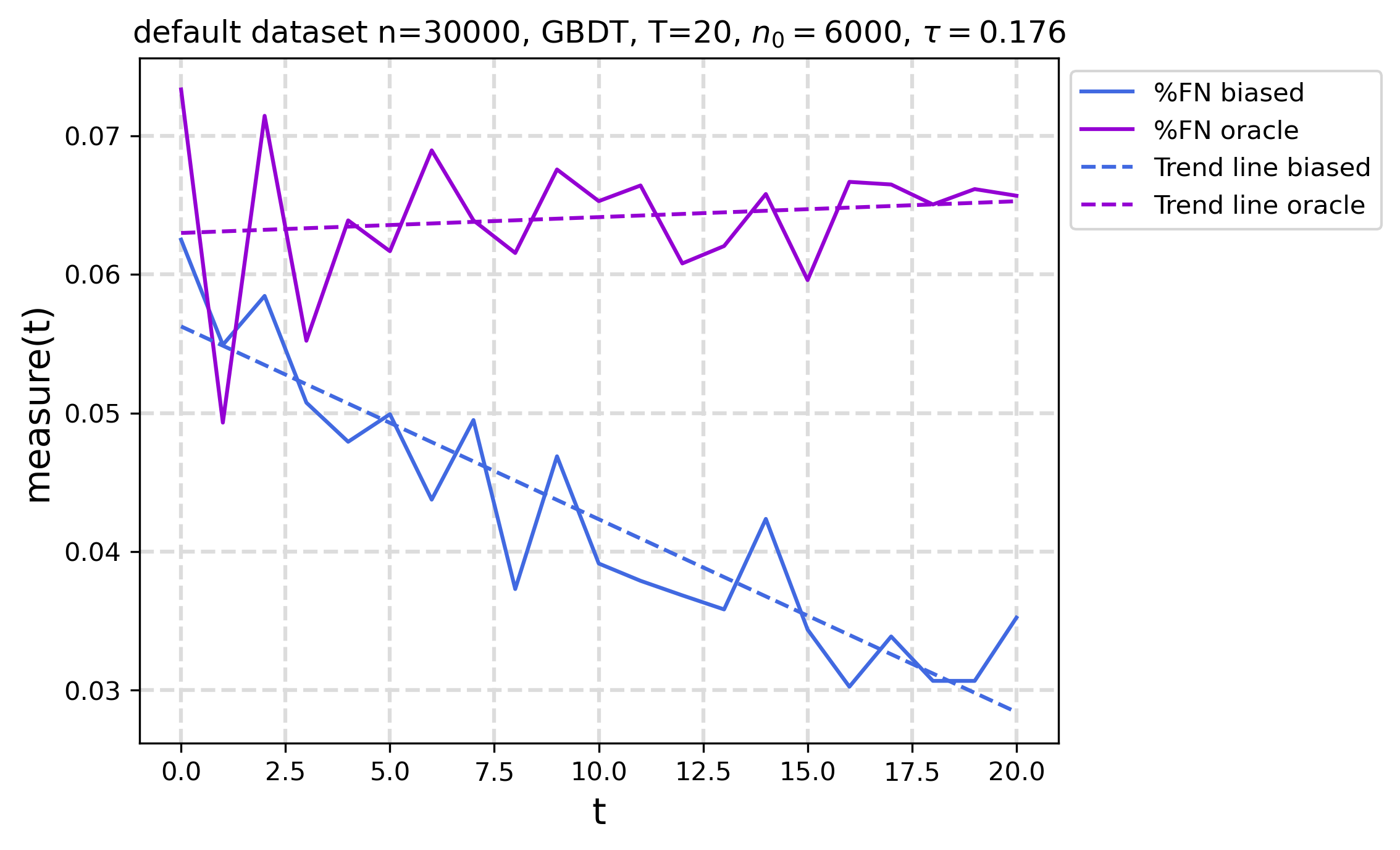} & \includegraphics[scale=.25]{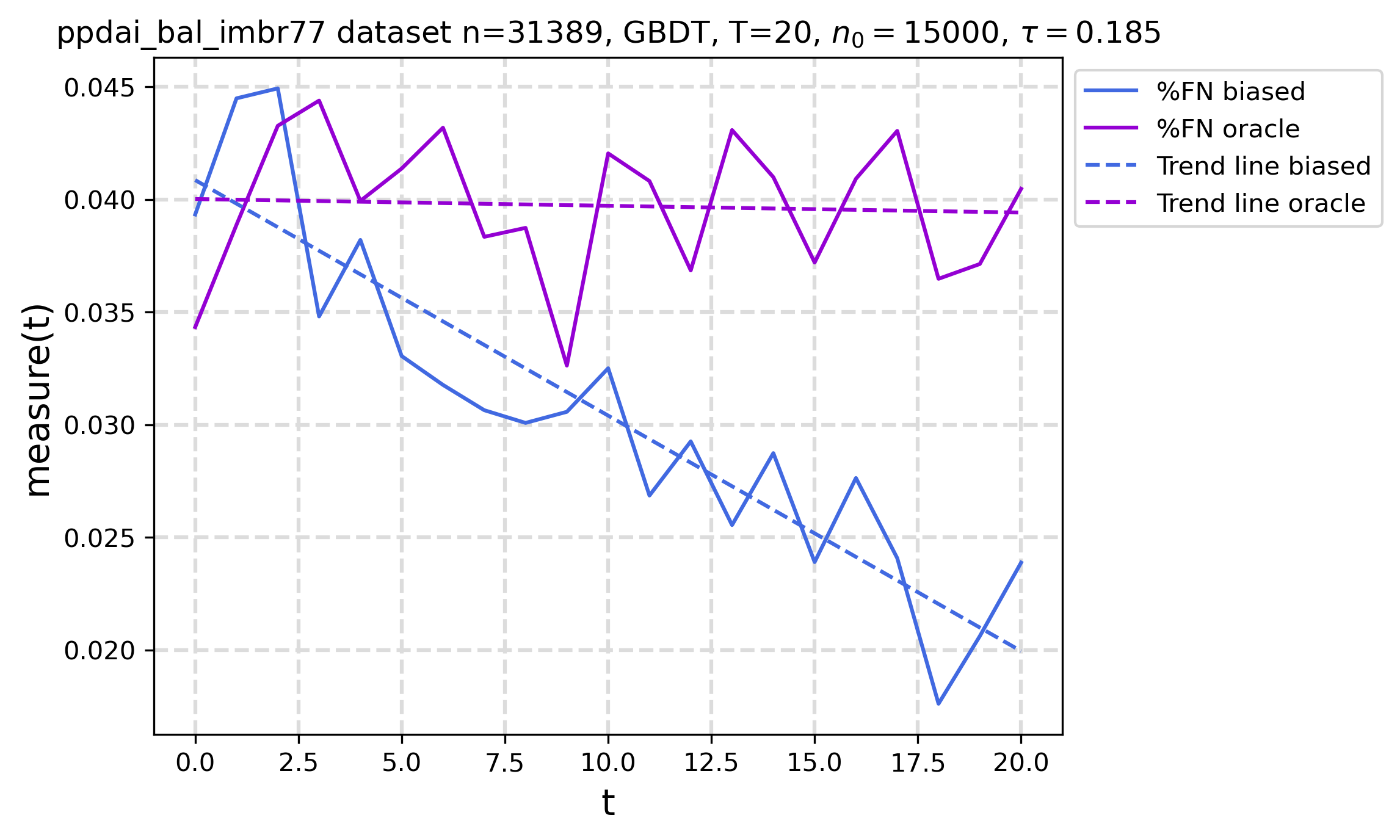} & \includegraphics[scale=.25]{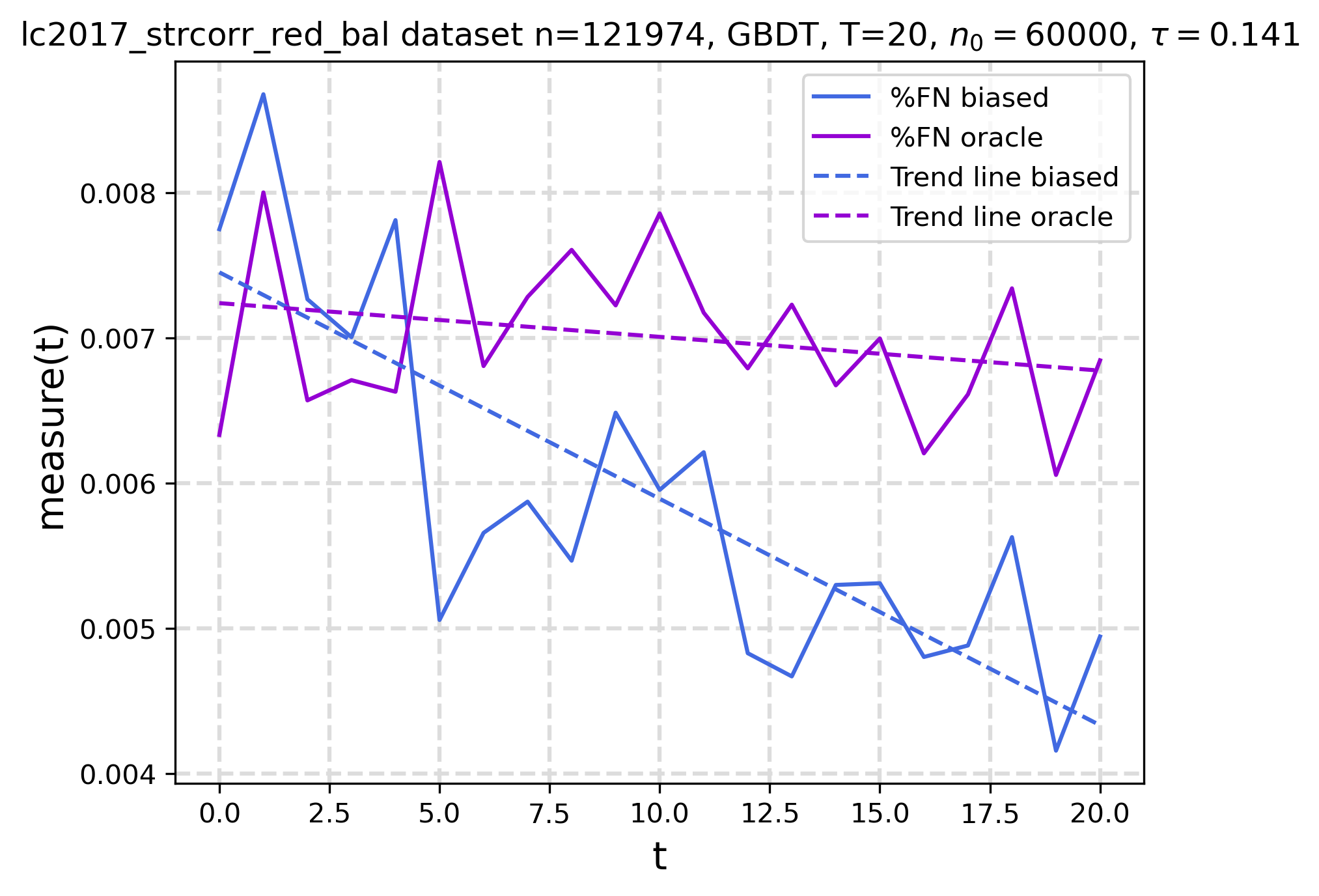}\\
    \end{tabular}
    \caption{Temporal evolution - GBDT - Decision thresholds for $c \in \{1, 3, 5\}$ - Percentage of False Negatives}
    \label{fig:gbdt_tempevol_perFN}
\end{figure*}

\begin{figure*}[t]
    \centering
    \begin{tabular}{ccc}
        Default & ppdai\_bal\_imbr77 & lc17\_bal\_imbr50\\
        \includegraphics[scale=.25]{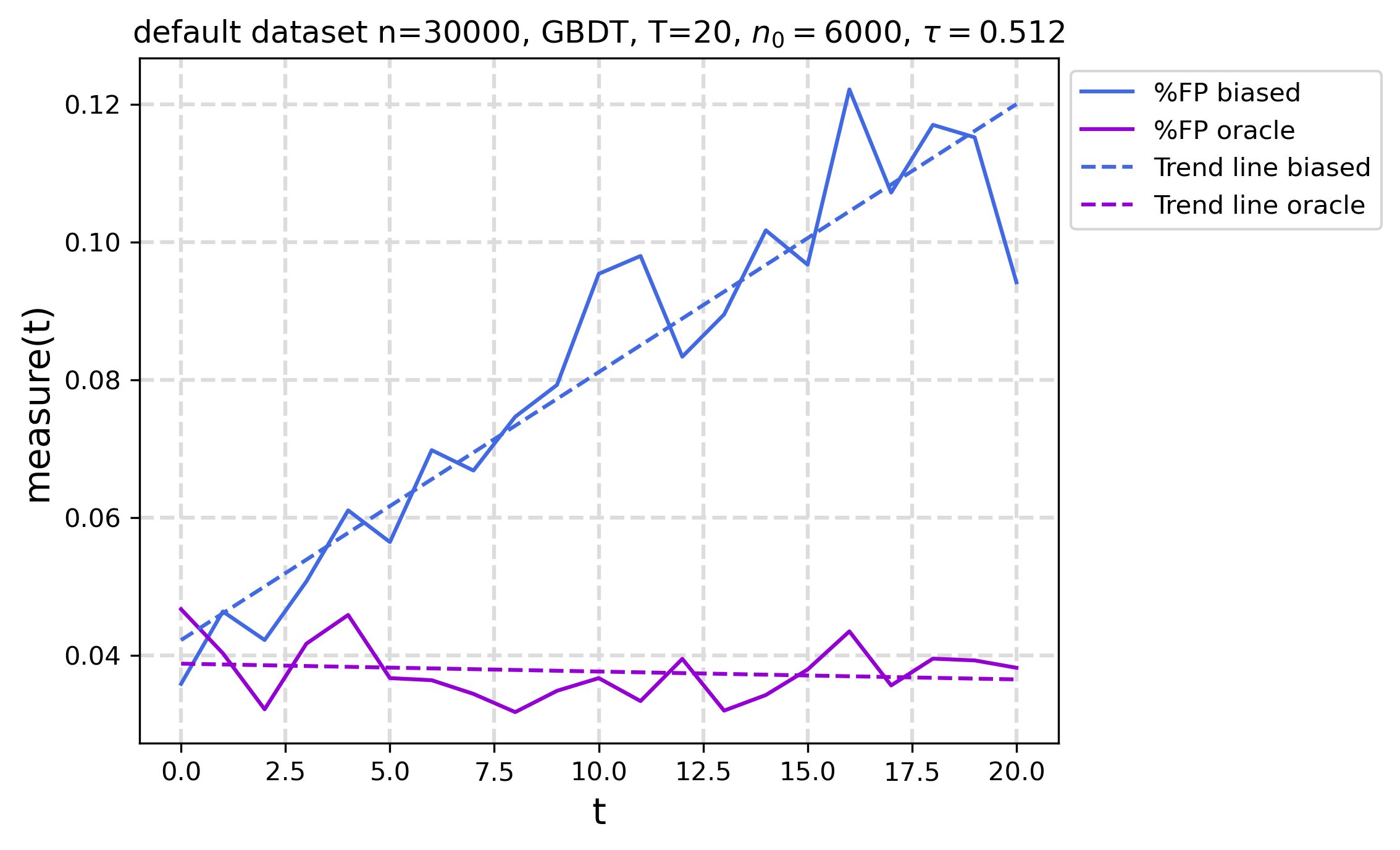} & \includegraphics[scale=.25]{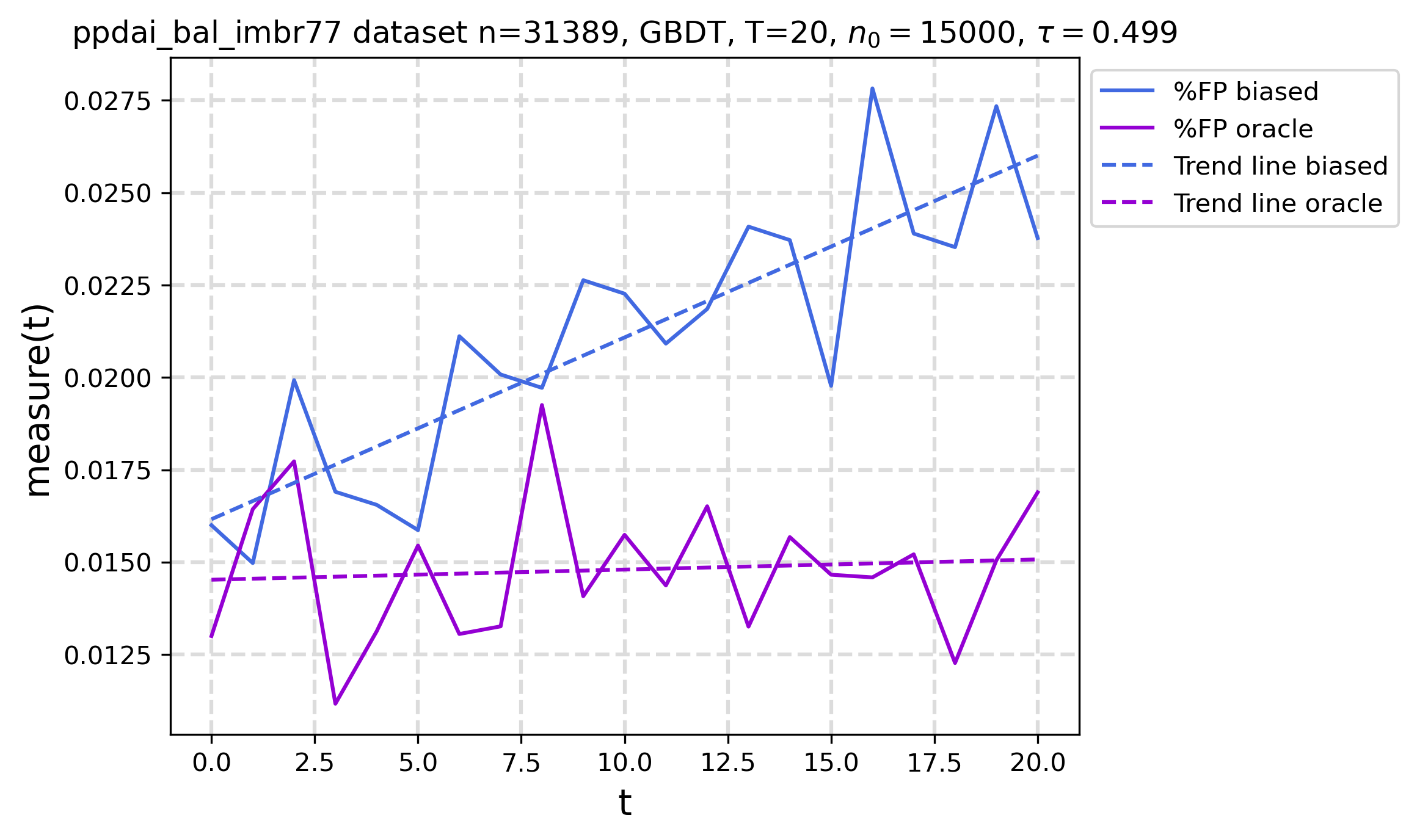} & \includegraphics[scale=.25]{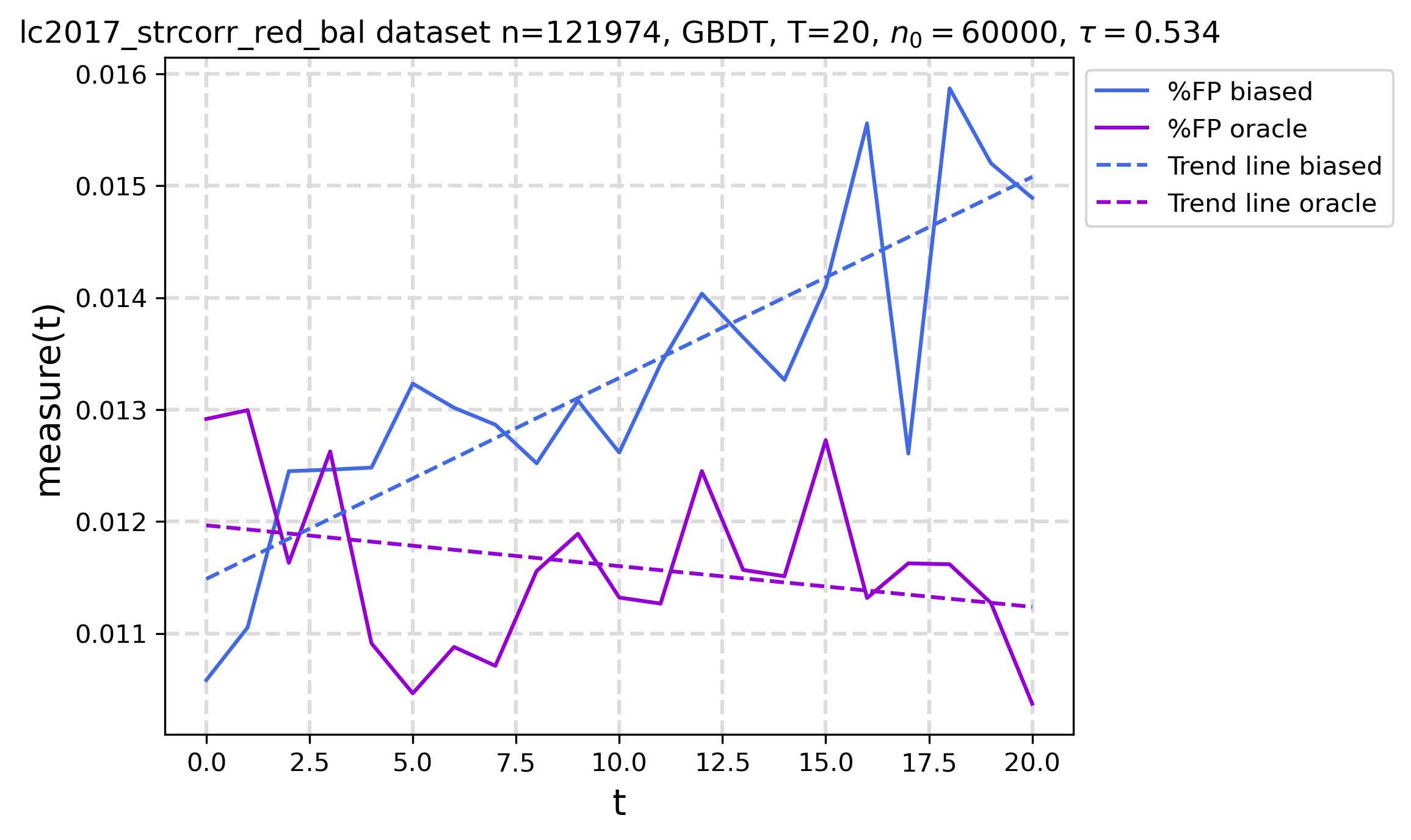}\\
      \includegraphics[scale=.25]{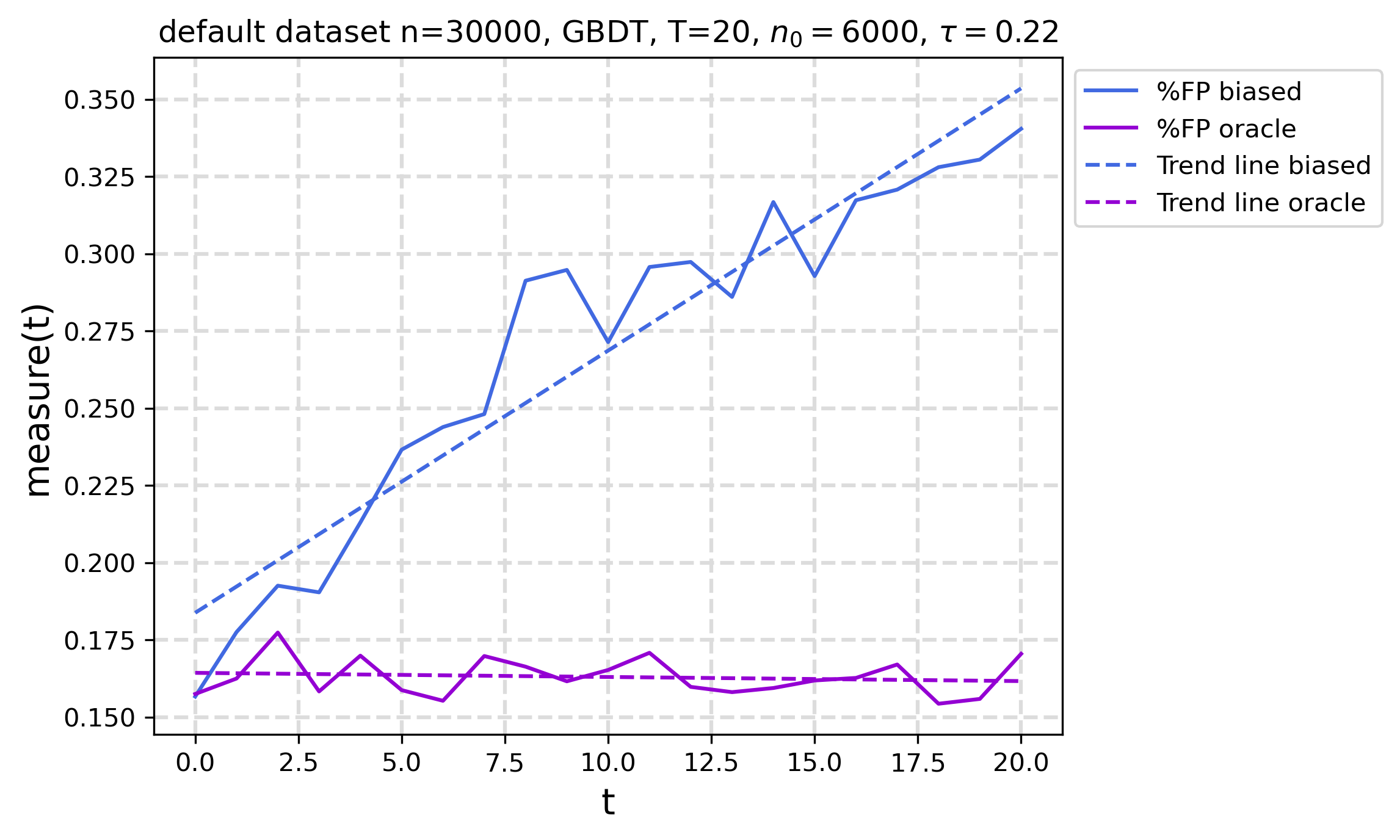} & \includegraphics[scale=.25]{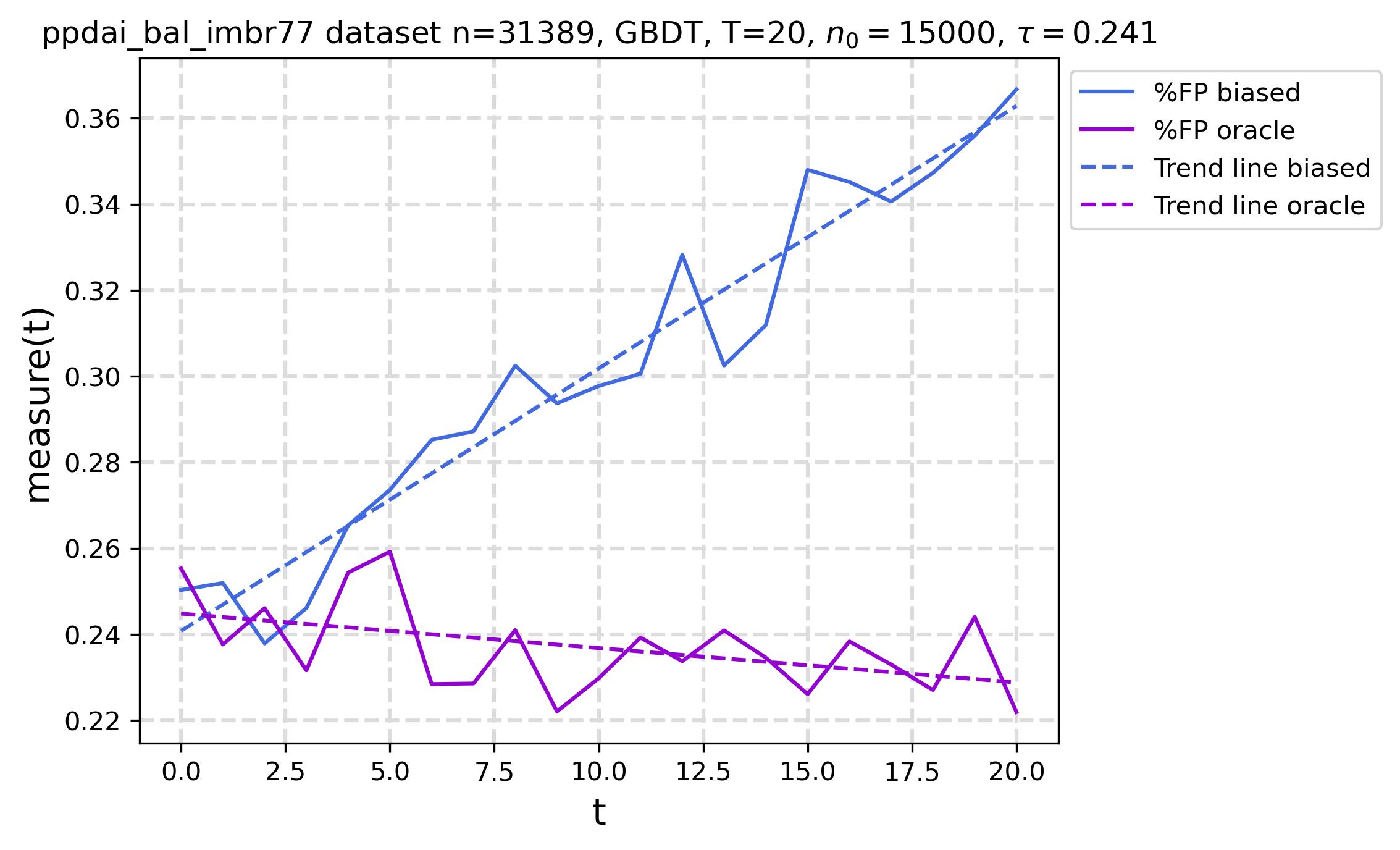} & \includegraphics[scale=.25]{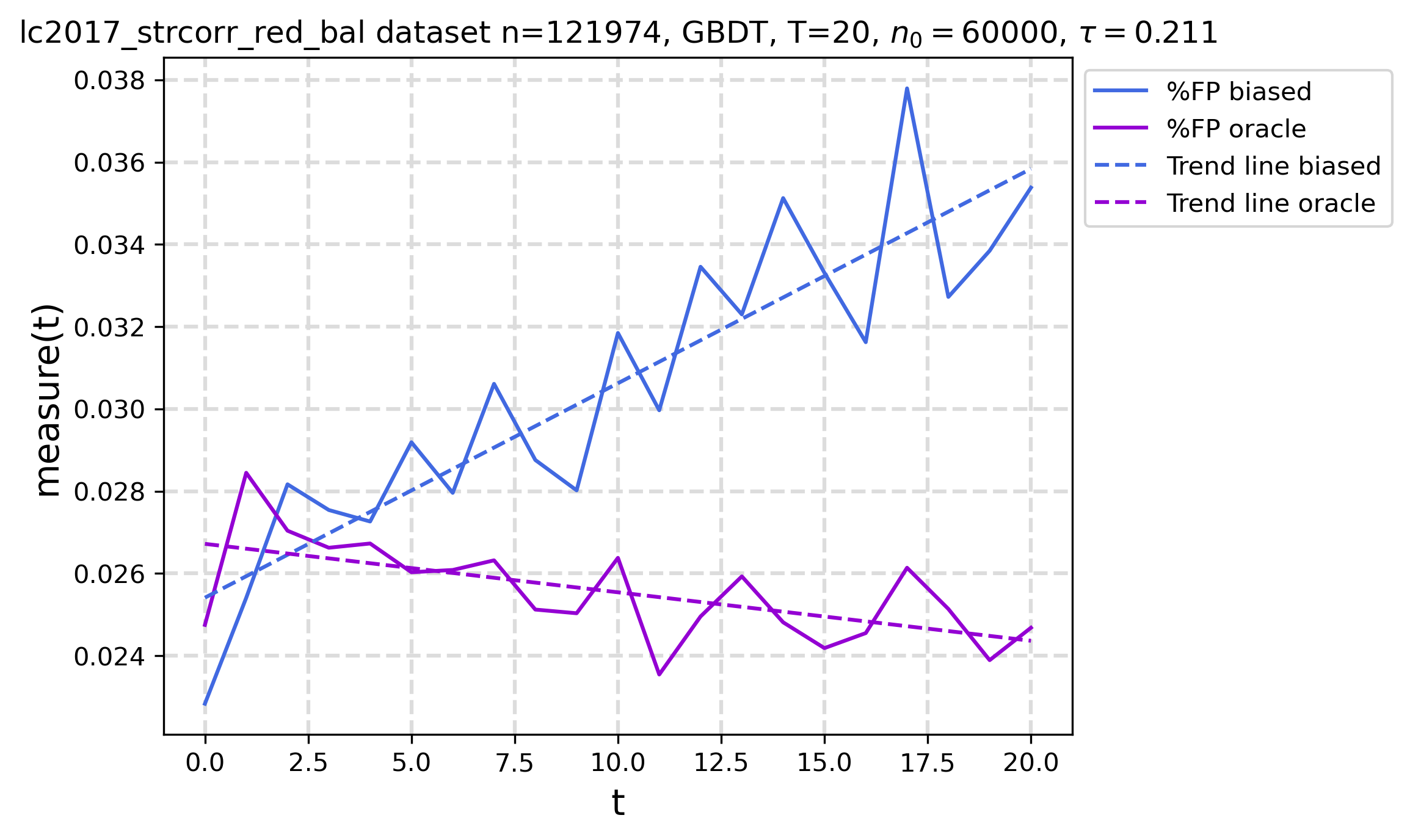}\\
        \includegraphics[scale=.25]{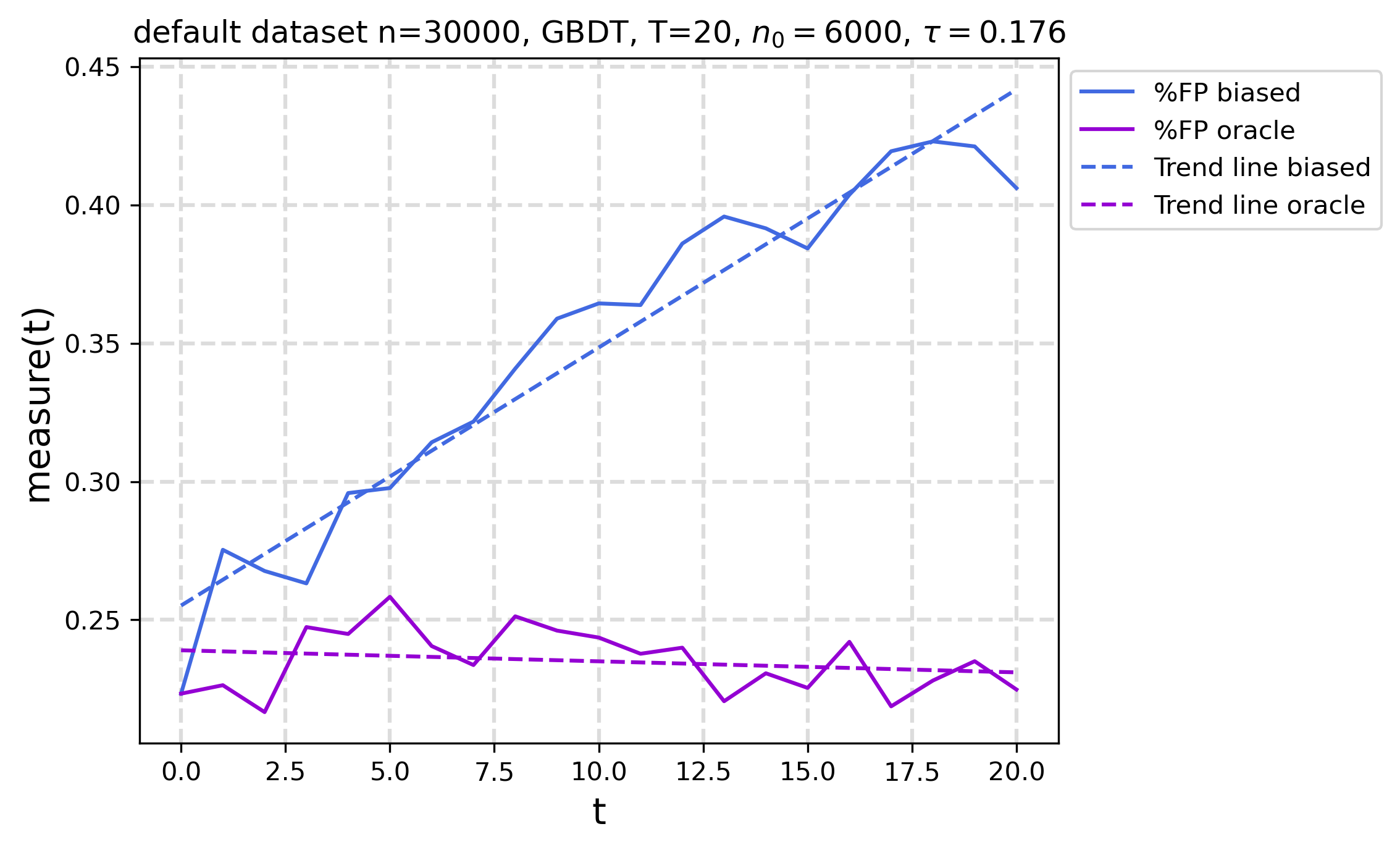} & \includegraphics[scale=.25]{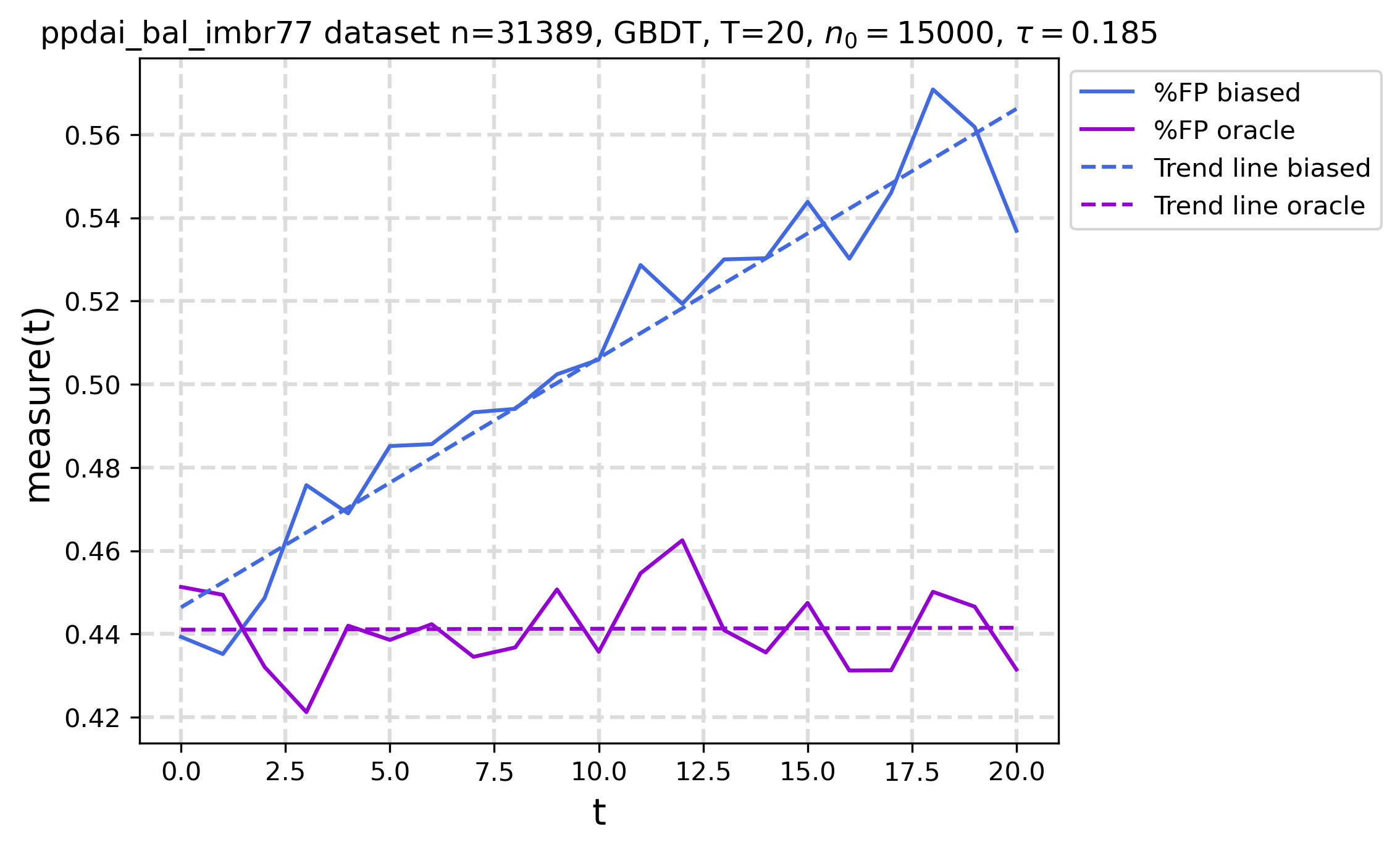} & \includegraphics[scale=.25]{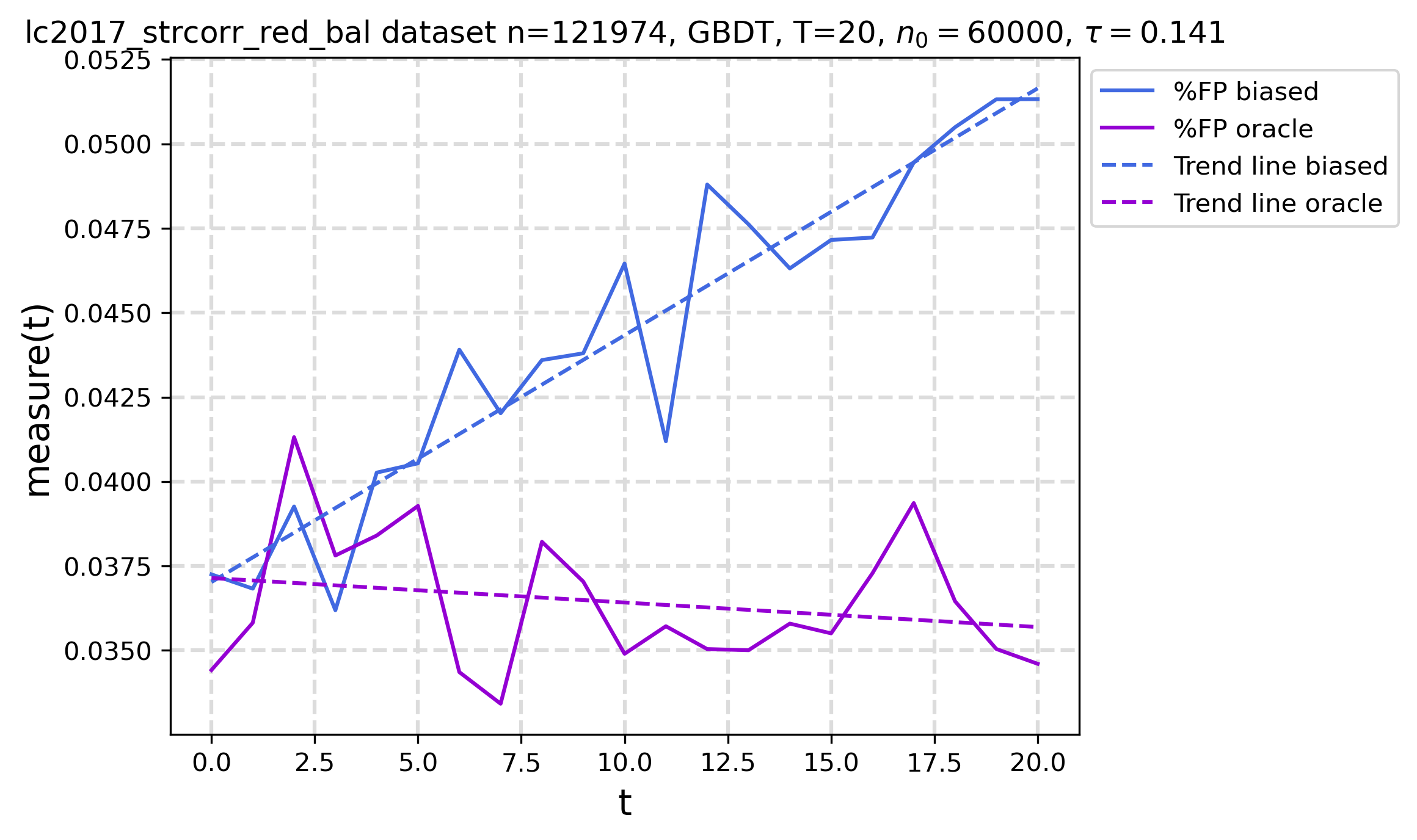}\\
    \end{tabular}
    \caption{Temporal evolution - GBDT - Decision thresholds for $c \in \{1, 3, 5\}$ - Percentage of False Positives}
    \label{fig:gbdt_tempevol_perFP}
\end{figure*}

\begin{figure*}[t]
    \centering
    \begin{minipage}{.4\textwidth}
        \includegraphics[scale=.3]{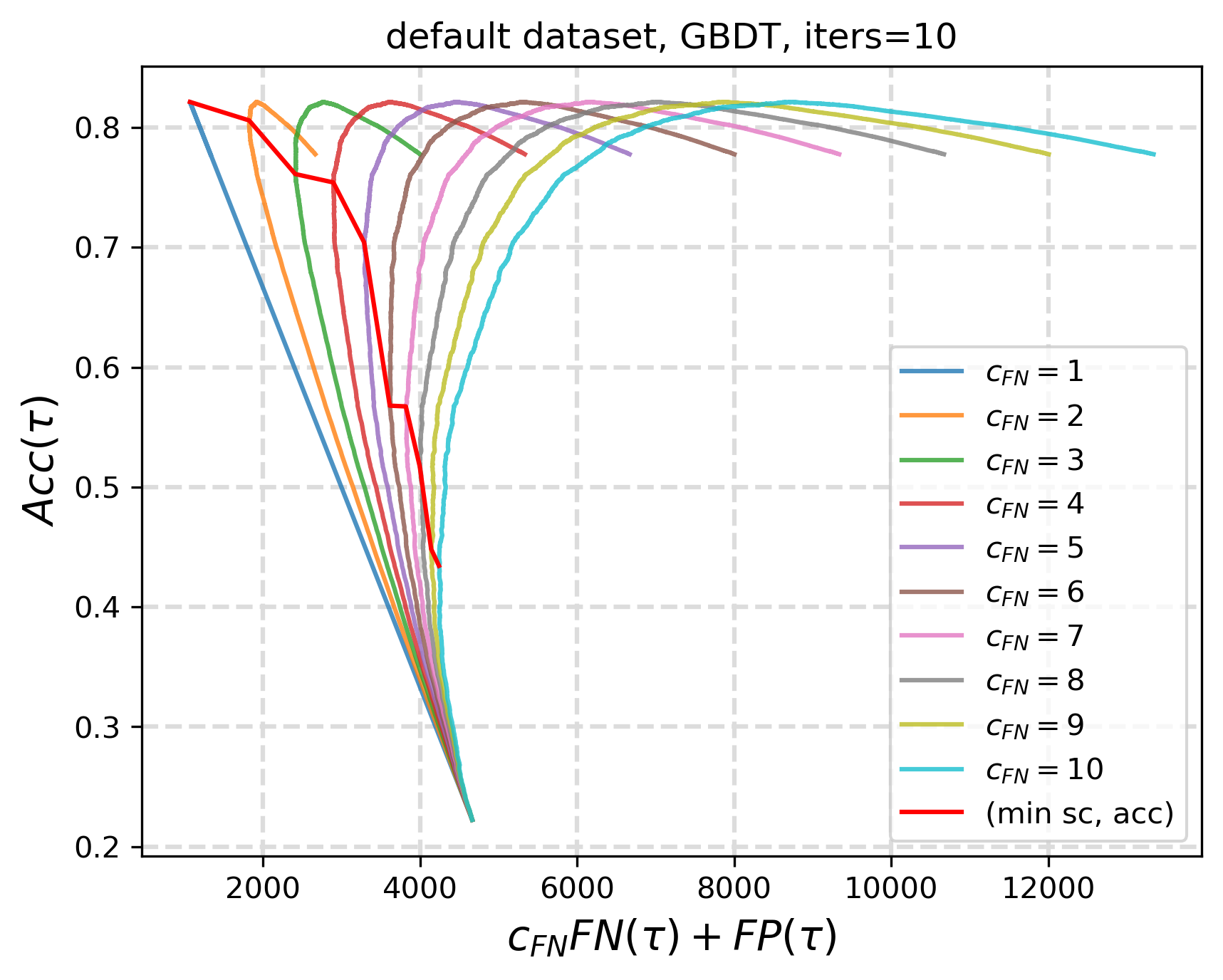}
    \end{minipage}
    \begin{minipage}{.3\textwidth}
        \includegraphics[scale=.3]{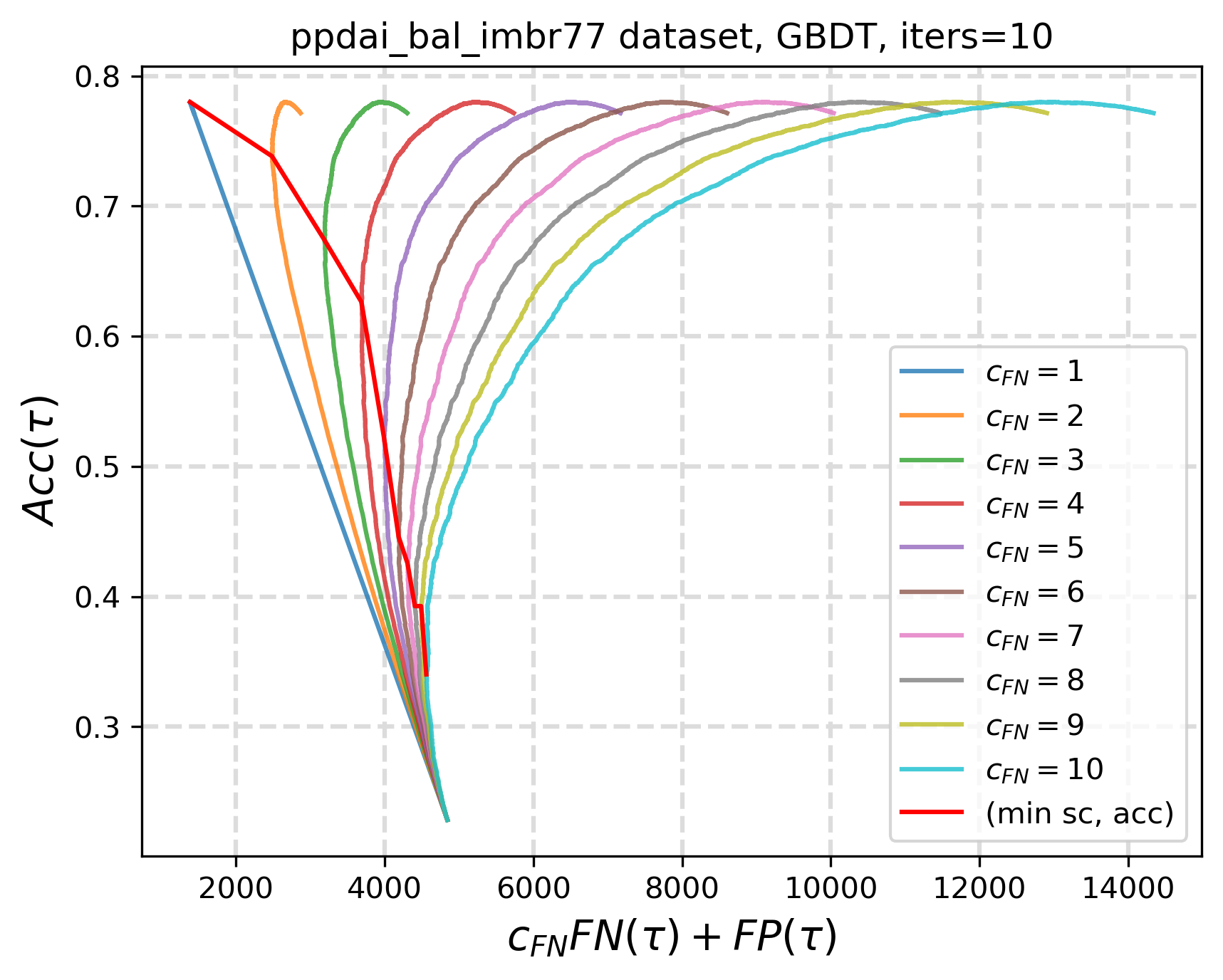}
    \end{minipage}
    \begin{minipage}{.3\textwidth}
        \includegraphics[scale=.3]{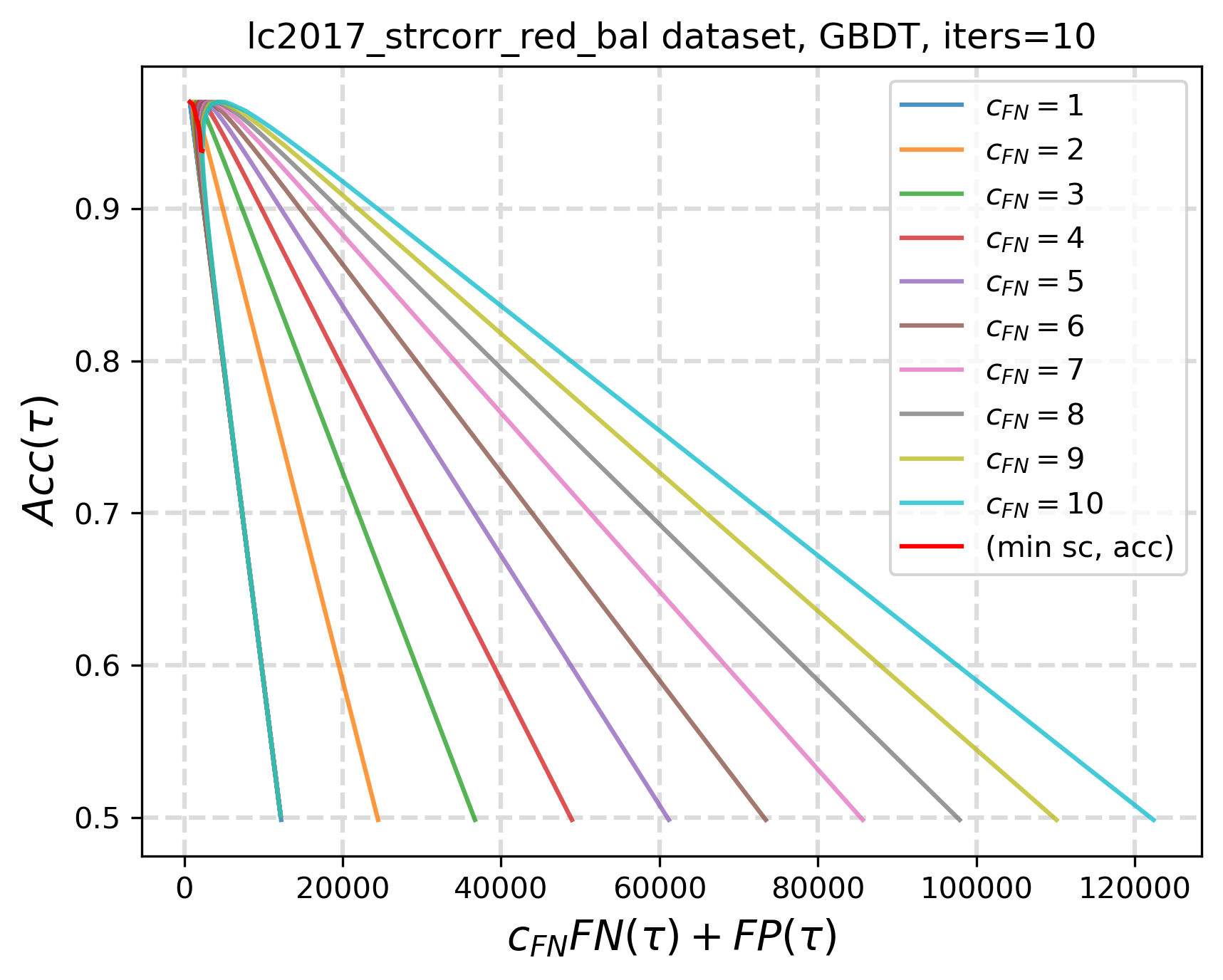}
    \end{minipage}
    \caption{Accuracy vs social cost for $\tau\in [0,1]$ for the GBDT model and all datasets.}
    \label{fig:gbdt_acc_sc_cFN_1_10}
\end{figure*}

\begin{table*}[t]
    \centering
    \begin{tabular}{|c|c|c|c|c|}
        \hline
        $c$ & Instance & $\tau$ & Acc($\tau$) & Relative diff. (\%)\\
        \hline
        $1,\dots,5$ & Max. acc. = Min. s.c. for $c=1$ & 0.500 & 0.782 & 0\\
        \hline
        2 & Min. s.c. & 0.350 & 0.753 & 3.851\\
        \hline
        3 & Min. s.c. & 0.270 & 0.685 & 14.161\\
        \hline
        4 & Min. s.c. & 0.210 & 0.591 & 32.318\\
        \hline
        5 & Min. s.c. & 0.190 & 0.550 & 42.182\\
        \hline
    \end{tabular}
    \caption{Accuracy loss for socially optimal models, for the Random Forest model and the balanced PPDai dataset. Values are rounded to three decimal places. }
    \label{tab:tauacc_tausc_values_rf_ppdai}
\end{table*}

\end{document}